\pdfoutput=1
\documentclass[11pt,twoside,a4paper,cmspaper,final,collab]{cms-tdr}
\def\svnVersion{bdb9b80}\def\svnDate{2026/03/27}\def\cmsCernNoTag{CERN-EP-2025-035}\def\cmsCernDate{\today}\def\cmsMessage{Published in the European Physical Journal C as \href{http://dx.doi.org/10.1140/epjc/s10052-025-14997-y}{\doi{10.1140/epjc/s10052-025-14997-y}.}}
\begin{document}\cmsNoteHeader{SMP-24-003}

\newcommand{\mll}{\ensuremath{m_{\ell\ell}}}
\newcommand{\mttbar}{\ensuremath{m_{\PQt\PAQt}}}
\newcommand{\mgg}{\ensuremath{m_{\PGg\PGg}}}
\newcommand{\rivet}{\textsc{Rivet}\xspace}
\newcommand{\smeftsim}{\textsc{SMEFTsim3}\xspace}
\newcommand{\smeftatnlo}{\textsc{SMEFT@NLO}\xspace}
\newcommand{\fastNLO}{\textsc{fastNLO}\xspace}
\newcommand{\NNLOjet}{\textsc{NNLOjet}\xspace}

\newcommand{\GZ}{\ensuremath{\Gamma_\PZ}}
\newcommand{\sigmahad}{\ensuremath{\sigma_\mathrm{had}^0}}
\newcommand{\Rl}{\ensuremath{R_\ell}}
\newcommand{\Rc}{\ensuremath{R_\PQc}}
\newcommand{\Rb}{\ensuremath{R_\PQb}}
\newcommand{\AlSLC}{\ensuremath{\mathcal{A}_\ell^\mathrm{SLC}}}
\newcommand{\AlLEP}{\ensuremath{\mathcal{A}_\ell^\mathrm{LEP}}}
\newcommand{\Ac}{\ensuremath{\mathcal{A}_\PQc}}
\newcommand{\Ab}{\ensuremath{\mathcal{A}_\PQb}}
\newcommand{\AFBl}{\ensuremath{A_\mathrm{FB}^{0,\ell}}}
\newcommand{\AFBc}{\ensuremath{A_\mathrm{FB}^{0,\PQc}}}
\newcommand{\AFBb}{\ensuremath{A_\mathrm{FB}^{0,\PQb}}}
\newcommand{\GW}{\ensuremath{\Gamma_\PW}}
\newcommand{\BrWhad}{\ensuremath{B_\PW^\mathrm{had}}}
\newcommand{\DeltaAlpha}{\ensuremath{\Delta\alpha}}
\newcommand{\GF}{\ensuremath{G_\text{F}}}

\newcommand{\cllprime}{\ensuremath{c_{\Pl\Pl}^{\prime}}}
\newcommand{\chdd}{\ensuremath{c_{\PH\text{D}}}}
\newcommand{\cbhre}{\ensuremath{\text{Re}(c_{\PQb\PH})}}
\newcommand{\ctbre}{\ensuremath{\text{Re}(c_{\PQt\PB})}}
\newcommand{\ctwre}{\ensuremath{\text{Re}(c_{\PQt\PW})}}
\newcommand{\ctgre}{\ensuremath{\text{Re}(c_{\PQt\mathrm{G}})}}
\newcommand{\cthre}{\ensuremath{\text{Re}(c_{\PQt\PH})}}
\newcommand{\chwb}{\ensuremath{c_{\PH\PW\PB}}}
\newcommand{\chu}{\ensuremath{c_{\PH\PQu}}}
\newcommand{\chd}{\ensuremath{c_{\PH\PQd}}}
\newcommand{\chb}{\ensuremath{c_{\PH\PB}}}
\newcommand{\chbox}{\ensuremath{c_{\PH\Box}}}
\newcommand{\chbq}{\ensuremath{c_{\PH\PQb}}}
\newcommand{\cht}{\ensuremath{c_{\PH\PQt}}}
\newcommand{\che}{\ensuremath{c_{\PH\Pe}}}
\newcommand{\chj}[1]{\ensuremath{c_{\PH\PQq}^{(#1)}}}
\newcommand{\chq}[1]{\ensuremath{c_{\PH\PQQ}^{(#1)}}}
\newcommand{\chl}[1]{\ensuremath{c_{\PH\Pl}^{(#1)}}}
\newcommand{\cw}{\ensuremath{c_{\PW}}}
\newcommand{\cg}{\ensuremath{c_{\mathrm{G}}}}
\newcommand{\clj}[1]{\ensuremath{c_{\Pl\PQq}^{(#1)}}}
\newcommand{\clu}{\ensuremath{c_{\Pl\PQu}}}
\newcommand{\cjj}[2]{\ensuremath{c_{\PQq\PQq}^{(#1,#2)}}}
\newcommand{\chg}{\ensuremath{c_{\PH\mathrm{G}}}}
\newcommand{\chw}{\ensuremath{c_{\PH\PW}}}
\newcommand{\cuu}[1]{\ensuremath{c_{\PQu\PQu}^{(#1)}}}
\newcommand{\cud}[1]{\ensuremath{c_{\PQu\PQd}^{(#1)}}}
\newcommand{\cdd}[1]{\ensuremath{c_{\PQd\PQd}^{(#1)}}}
\newcommand{\cju}[1]{\ensuremath{c_{\PQq\PQu}^{(#1)}}}
\newcommand{\cjd}[1]{\ensuremath{c_{\PQq\PQd}^{(#1)}}}
\newcommand{\cqj}[2]{\ensuremath{c_{\PQQ\PQq}^{(#1,#2)}}}
\newcommand{\cqu}[1]{\ensuremath{c_{\PQQ\PQu}^{(#1)}}}
\newcommand{\cqq}[1]{\ensuremath{c_{\PQQ\PQQ}^{(#1)}}}
\newcommand{\cqd}[1]{\ensuremath{c_{\PQQ\PQd}^{(#1)}}}
\newcommand{\ctj}[1]{\ensuremath{c_{\PQq\PQt}^{(#1)}}}
\newcommand{\ctu}[1]{\ensuremath{c_{\PQt\PQu}^{(#1)}}}
\newcommand{\ctd}[1]{\ensuremath{c_{\PQt\PQd}^{(#1)}}}
\newcommand{\ctt}{\ensuremath{c_{\PQt\PQt}}}
\newcommand{\cte}{\ensuremath{c_{\Pe\PQt}}}
\newcommand{\ctl}{\ensuremath{c_{\Pl\PQt}}}
\newcommand{\cql}[1]{\ensuremath{c_{\Pl\PQQ}^{(#1)}}}
\newcommand{\cqt}[1]{\ensuremath{c_{\PQQ\PQt}^{(#1)}}}

\newcommand{\Qg}{\ensuremath{\mathcal{Q}_{\mathrm{G}}}}
\newcommand{\QgDef}{\ensuremath{f^{abc} G_{\mu}^{a\nu} G_{\nu}^{b\rho} G_{\rho}^{c\mu}}}
\newcommand{\Qw}{\ensuremath{\mathcal{Q}_{\mathrm{W}}}}
\newcommand{\QwDef}{\ensuremath{\varepsilon^{ijk} W_{\mu}^{i\nu} W_{\nu}^{j\rho} W_{\rho}^{k\mu}}}
\newcommand{\Qhbox}{\ensuremath{\mathcal{Q}_{\mathrm{H}\Box}}}
\newcommand{\QhboxDef}{\ensuremath{(H^{\dagger}H)\Box(H^{\dagger}H)}}
\newcommand{\Qhdd}{\ensuremath{\mathcal{Q}_{\mathrm{HD}}}}
\newcommand{\QhddDef}{\ensuremath{(D^{\mu}H^{\dagger}H)(H^{\dagger}D_{\mu}H)}}
\newcommand{\Qhg}{\ensuremath{\mathcal{Q}_{\mathrm{HG}}}}
\newcommand{\QhgDef}{\ensuremath{H^{\dagger}H G_{\mu\nu}^{a}G^{a\mu\nu}}}
\newcommand{\Qhw}{\ensuremath{\mathcal{Q}_{\mathrm{HW}}}}
\newcommand{\QhwDef}{\ensuremath{H^{\dagger}H W_{\mu\nu}^{i}W^{i\mu\nu}}}
\newcommand{\Qhb}{\ensuremath{\mathcal{Q}_{\mathrm{HB}}}}
\newcommand{\QhbDef}{\ensuremath{H^{\dagger}H B_{\mu\nu}B^{\mu\nu}}}
\newcommand{\Qhwb}{\ensuremath{\mathcal{Q}_{\mathrm{HWB}}}}
\newcommand{\QhwbDef}{\ensuremath{H^{\dagger}H W_{\mu\nu}^{i}B^{\mu\nu}}}
\newcommand{\Qth}{\ensuremath{\mathcal{Q}_{\mathrm{tH}}}}
\newcommand{\QthDef}{\ensuremath{(H^{\dagger}H)(\overline{Q}\tilde{H}t)}}
\newcommand{\Qbh}{\ensuremath{\mathcal{Q}_{\mathrm{bH}}}}
\newcommand{\QbhDef}{\ensuremath{(H^{\dagger}H)(\overline{Q}Hb)}}
\newcommand{\Qtw}{\ensuremath{\mathcal{Q}_{\mathrm{tW}}}}
\newcommand{\QtwDef}{\ensuremath{(\overline{Q}\sigma^{\mu\nu}t)\sigma^{i}\tilde{H}W_{\mu\nu}^{i}}}
\newcommand{\Qtb}{\ensuremath{\mathcal{Q}_{\mathrm{tB}}}}
\newcommand{\QtbDef}{\ensuremath{(\overline{Q}\sigma^{\mu\nu}t)\tilde{H}B_{\mu\nu}}}
\newcommand{\Qtg}{\ensuremath{\mathcal{Q}_{\mathrm{tG}}}}
\newcommand{\QtgDef}{\ensuremath{(\overline{Q}\sigma^{\mu\nu}T^{a}t)\tilde{H}G_{\mu\nu}^{a}}}
\newcommand{\QhlOne}{\ensuremath{\mathcal{Q}_{\mathrm{Hl}}^{(1)}}}
\newcommand{\QhlOneDef}{\ensuremath{(H^{\dagger}i\overset\leftrightarrow{D_{\mu}}H)(\overline{l}_{p}\gamma^{\mu}l_{r})}}
\newcommand{\QhlThree}{\ensuremath{\mathcal{Q}_{\mathrm{Hl}}^{(3)}}}
\newcommand{\QhlThreeDef}{\ensuremath{(H^{\dagger}i\overset\leftrightarrow{D_{\mu}^{i}}H)(\overline{l}_{p}\sigma^{i}\gamma^{\mu}l_{r})}}
\newcommand{\Qhe}{\ensuremath{\mathcal{Q}_{\mathrm{He}}}}
\newcommand{\QheDef}{\ensuremath{(H^{\dagger}i\overset\leftrightarrow{D_{\mu}}H)(\overline{e}_{p}\gamma^{\mu}e_{r})}}
\newcommand{\QhjOne}{\ensuremath{\mathcal{Q}_{\mathrm{Hq}}^{(1)}}}
\newcommand{\QhjOneDef}{\ensuremath{(H^{\dagger}i\overset\leftrightarrow{D_{\mu}}H)(\overline{q}\gamma^{\mu}q)}}
\newcommand{\QhjThree}{\ensuremath{\mathcal{Q}_{\mathrm{Hq}}^{(3)}}}
\newcommand{\QhjThreeDef}{\ensuremath{(H^{\dagger}i\overset\leftrightarrow{D_{\mu}^{i}}H)(\overline{q}\sigma^{i}\gamma^{\mu}q)}}
\newcommand{\Qhu}{\ensuremath{\mathcal{Q}_{\mathrm{Hu}}}}
\newcommand{\QhuDef}{\ensuremath{(H^{\dagger}i\overset\leftrightarrow{D_{\mu}}H)(\overline{u}\gamma^{\mu}u)}}
\newcommand{\Qhd}{\ensuremath{\mathcal{Q}_{\mathrm{Hd}}}}
\newcommand{\QhdDef}{\ensuremath{(H^{\dagger}i\overset\leftrightarrow{D_{\mu}}H)(\overline{d}\gamma^{\mu}d)}}
\newcommand{\QhqOne}{\ensuremath{\mathcal{Q}_{\mathrm{HQ}}^{(1)}}}
\newcommand{\QhqOneDef}{\ensuremath{(H^{\dagger}i\overset\leftrightarrow{D_{\mu}}H)(\overline{Q}\gamma^{\mu}Q)}}
\newcommand{\QhqThree}{\ensuremath{\mathcal{Q}_{\mathrm{HQ}}^{(3)}}}
\newcommand{\QhqThreeDef}{\ensuremath{(H^{\dagger}i\overset\leftrightarrow{D_{\mu}^{i}}H)(\overline{Q}\sigma^{i}\gamma^{\mu}Q)}}
\newcommand{\Qht}{\ensuremath{\mathcal{Q}_{\mathrm{Ht}}}}
\newcommand{\QhtDef}{\ensuremath{(H^{\dagger}i\overset\leftrightarrow{D_{\mu}}H)(\overline{t}\gamma^{\mu}t)}}
\newcommand{\Qhbq}{\ensuremath{\mathcal{Q}_{\mathrm{Hb}}}}
\newcommand{\QhbqDef}{\ensuremath{(H^{\dagger}i\overset\leftrightarrow{D_{\mu}}H)(\overline{b}\gamma^{\mu}b)}}
\newcommand{\QljOne}{\ensuremath{\mathcal{Q}_{\mathrm{lq}}^{(1)}}}
\newcommand{\QljOneDef}{\ensuremath{(\overline{l}_{p}\gamma_{\mu}l_{r})(\overline{q}\gamma^{\mu}q)}}
\newcommand{\QljThree}{\ensuremath{\mathcal{Q}_{\mathrm{lq}}^{(3)}}}
\newcommand{\QljThreeDef}{\ensuremath{(\overline{l}_{p}\sigma^{i}\gamma_{\mu}l_{r})(\overline{q}\sigma^{i}\gamma^{\mu}q)}}
\newcommand{\QllPrime}{\ensuremath{\mathcal{Q}_{\mathrm{ll}}^{\prime}}}
\newcommand{\QllPrimeDef}{\ensuremath{(\overline{l}_{p}\gamma_{\mu}l_{r})(\overline{l}_{r}\gamma^{\mu}l_{p})}}
\newcommand{\QqlOne}{\ensuremath{\mathcal{Q}_{\mathrm{lQ}}^{(1)}}}
\newcommand{\QqlOneDef}{\ensuremath{(\overline{l}_{p}\gamma_{\mu}l_{r})(\overline{Q}\gamma^{\mu}Q)}}
\newcommand{\QqlThree}{\ensuremath{\mathcal{Q}_{\mathrm{lQ}}^{(3)}}}
\newcommand{\QqlThreeDef}{\ensuremath{(\overline{l}_{p}\sigma^{i}\gamma_{\mu}l_{r})(\overline{Q}\sigma^{i}\gamma^{\mu}Q)}}
\newcommand{\QqqOne}{\ensuremath{\mathcal{Q}_{\mathrm{QQ}}^{(1)}}}
\newcommand{\QqqOneDef}{\ensuremath{(\overline{Q}\gamma_{\mu}Q)(\overline{Q}\gamma^{\mu}Q)}}
\newcommand{\QjjOneOne}{\ensuremath{\mathcal{Q}_{\mathrm{qq}}^{(1,1)}}}
\newcommand{\QjjOneOneDef}{\ensuremath{(\overline{q}\gamma_{\mu}q)(\overline{q}\gamma^{\mu}q)}}
\newcommand{\QjjOneEight}{\ensuremath{\mathcal{Q}_{\mathrm{qq}}^{(1,8)}}}
\newcommand{\QjjOneEightDef}{\ensuremath{(\overline{q}T^{a}\gamma_{\mu}q)(\overline{q}T^{a}\gamma^{\mu}q)}}
\newcommand{\QjjThreeOne}{\ensuremath{\mathcal{Q}_{\mathrm{qq}}^{(3,1)}}}
\newcommand{\QjjThreeOneDef}{\ensuremath{(\overline{q}\sigma^{i}\gamma_{\mu}q)(\overline{q}\sigma^{i}\gamma^{\mu}q)}}
\newcommand{\QjjThreeEight}{\ensuremath{\mathcal{Q}_{\mathrm{qq}}^{(3,8)}}}
\newcommand{\QjjThreeEightDef}{\ensuremath{(\overline{q}\sigma^{i}T^{a}\gamma_{\mu}q)(\overline{q}\sigma^{i}T^{a}\gamma^{\mu}q)}}
\newcommand{\QqjOneOne}{\ensuremath{\mathcal{Q}_{\mathrm{Qq}}^{(1,1)}}}
\newcommand{\QqjOneOneDef}{\ensuremath{(\overline{Q}\gamma_{\mu}Q)(\overline{q}\gamma^{\mu}q)}}
\newcommand{\QqjOneEight}{\ensuremath{\mathcal{Q}_{\mathrm{Qq}}^{(1,8)}}}
\newcommand{\QqjOneEightDef}{\ensuremath{(\overline{Q}T^{a}\gamma_{\mu}Q)(\overline{q}T^{a}\gamma^{\mu}q)}}
\newcommand{\QqjThreeOne}{\ensuremath{\mathcal{Q}_{\mathrm{Qq}}^{(3,1)}}}
\newcommand{\QqjThreeOneDef}{\ensuremath{(\overline{Q}\sigma^{i}\gamma_{\mu}Q)(\overline{q}\sigma^{i}\gamma^{\mu}q)}}
\newcommand{\QqjThreeEight}{\ensuremath{\mathcal{Q}_{\mathrm{Qq}}^{(3,8)}}}
\newcommand{\QqjThreeEightDef}{\ensuremath{(\overline{Q}\sigma^{i}T^{a}\gamma_{\mu}Q)(\overline{q}\sigma^{i}T^{a}\gamma^{\mu}q)}}
\newcommand{\Qte}{\ensuremath{\mathcal{Q}_{\mathrm{et}}}}
\newcommand{\QteDef}{\ensuremath{(\overline{e}_{p}\gamma_{\mu}e_{r})(\overline{t}\gamma^{\mu}t)}}
\newcommand{\Qtt}{\ensuremath{\mathcal{Q}_{\mathrm{tt}}}}
\newcommand{\QttDef}{\ensuremath{(\overline{t}\gamma_{\mu}t)(\overline{t}\gamma^{\mu}t)}}
\newcommand{\QuuOne}{\ensuremath{\mathcal{Q}_{\mathrm{uu}}^{(1)}}}
\newcommand{\QuuOneDef}{\ensuremath{(\overline{u}\gamma_{\mu}u)(\overline{u}\gamma^{\mu}u)}}
\newcommand{\QuuEight}{\ensuremath{\mathcal{Q}_{\mathrm{uu}}^{(8)}}}
\newcommand{\QuuEightDef}{\ensuremath{(\overline{u}T^{a}\gamma_{\mu}u)(\overline{u}T^{a}\gamma^{\mu}u)}}
\newcommand{\QtuOne}{\ensuremath{\mathcal{Q}_{\mathrm{tu}}^{(1)}}}
\newcommand{\QtuOneDef}{\ensuremath{(\overline{t}\gamma_{\mu}t)(\overline{u}\gamma^{\mu}u)}}
\newcommand{\QtuEight}{\ensuremath{\mathcal{Q}_{\mathrm{tu}}^{(8)}}}
\newcommand{\QtuEightDef}{\ensuremath{(\overline{t}T^{a}\gamma_{\mu}t)(\overline{u}T^{a}\gamma^{\mu}u)}}
\newcommand{\QddOne}{\ensuremath{\mathcal{Q}_{\mathrm{dd}}^{(1)}}}
\newcommand{\QddOneDef}{\ensuremath{(\overline{d}\gamma_{\mu}d)(\overline{d}\gamma^{\mu}d)}}
\newcommand{\QddEight}{\ensuremath{\mathcal{Q}_{\mathrm{dd}}^{(8)}}}
\newcommand{\QddEightDef}{\ensuremath{(\overline{d}T^{a}\gamma_{\mu}d)(\overline{d}T^{a}\gamma^{\mu}d)}}
\newcommand{\QudOne}{\ensuremath{\mathcal{Q}_{\mathrm{ud}}^{(1)}}}
\newcommand{\QudOneDef}{\ensuremath{(\overline{u}\gamma_{\mu}u)(\overline{d}\gamma^{\mu}d)}}
\newcommand{\QudEight}{\ensuremath{\mathcal{Q}_{\mathrm{ud}}^{(8)}}}
\newcommand{\QudEightDef}{\ensuremath{(\overline{u}T^{a}\gamma_{\mu}u)(\overline{d}T^{a}\gamma^{\mu}d)}}
\newcommand{\QtdOne}{\ensuremath{\mathcal{Q}_{\mathrm{td}}^{(1)}}}
\newcommand{\QtdOneDef}{\ensuremath{(\overline{t}\gamma_{\mu}t)(\overline{d}\gamma^{\mu}d)}}
\newcommand{\QtdEight}{\ensuremath{\mathcal{Q}_{\mathrm{td}}^{(8)}}}
\newcommand{\QtdEightDef}{\ensuremath{(\overline{t}T^{a}\gamma_{\mu}t)(\overline{d}T^{a}\gamma^{\mu}d)}}
\newcommand{\Qlu}{\ensuremath{\mathcal{Q}_{\mathrm{lu}}}}
\newcommand{\QluDef}{\ensuremath{(\overline{l}_{p}\gamma_{\mu}l_{r})(\overline{u}\gamma^{\mu}u)}}
\newcommand{\Qtl}{\ensuremath{\mathcal{Q}_{\mathrm{lt}}}}
\newcommand{\QtlDef}{\ensuremath{(\overline{l}_{p}\gamma_{\mu}l_{r})(\overline{t}\gamma^{\mu}t)}}
\newcommand{\QquOne}{\ensuremath{\mathcal{Q}_{\mathrm{Qu}}^{(1)}}}
\newcommand{\QquOneDef}{\ensuremath{(\overline{Q}\gamma_{\mu}Q)(\overline{u}\gamma^{\mu}u)}}
\newcommand{\QquEight}{\ensuremath{\mathcal{Q}_{\mathrm{Qu}}^{(8)}}}
\newcommand{\QquEightDef}{\ensuremath{(\overline{Q}T^{a}\gamma_{\mu}Q)(\overline{u}T^{a}\gamma^{\mu}u)}}
\newcommand{\QtjOne}{\ensuremath{\mathcal{Q}_{\mathrm{qt}}^{(1)}}}
\newcommand{\QtjOneDef}{\ensuremath{(\overline{q}\gamma_{\mu}q)(\overline{t}\gamma^{\mu}t)}}
\newcommand{\QtjEight}{\ensuremath{\mathcal{Q}_{\mathrm{qt}}^{(8)}}}
\newcommand{\QtjEightDef}{\ensuremath{(\overline{q}T^{a}\gamma_{\mu}q)(\overline{t}T^{a}\gamma^{\mu}t)}}
\newcommand{\QqtOne}{\ensuremath{\mathcal{Q}_{\mathrm{Qt}}^{(1)}}}
\newcommand{\QqtOneDef}{\ensuremath{(\overline{Q}\gamma_{\mu}Q)(\overline{t}\gamma^{\mu}t)}}
\newcommand{\QqtEight}{\ensuremath{\mathcal{Q}_{\mathrm{Qt}}^{(8)}}}
\newcommand{\QqtEightDef}{\ensuremath{(\overline{Q}T^{a}\gamma_{\mu}Q)(\overline{t}T^{a}\gamma^{\mu}t)}}
\newcommand{\QqdOne}{\ensuremath{\mathcal{Q}_{\mathrm{Qd}}^{(1)}}}
\newcommand{\QqdOneDef}{\ensuremath{(\overline{Q}\gamma_{\mu}Q)(\overline{d}\gamma^{\mu}d)}}
\newcommand{\QqdEight}{\ensuremath{\mathcal{Q}_{\mathrm{Qd}}^{(8)}}}
\newcommand{\QqdEightDef}{\ensuremath{(\overline{Q}T^{a}\gamma_{\mu}Q)(\overline{d}T^{a}\gamma^{\mu}d)}}
\newcommand{\QjuOne}{\ensuremath{\mathcal{Q}_{\mathrm{qu}}^{(1)}}}
\newcommand{\QjuOneDef}{\ensuremath{(\overline{q}\gamma_{\mu}q)(\overline{u}\gamma^{\mu}u)}}
\newcommand{\QjuEight}{\ensuremath{\mathcal{Q}_{\mathrm{qu}}^{(8)}}}
\newcommand{\QjuEightDef}{\ensuremath{(\overline{q}T^{a}\gamma_{\mu}q)(\overline{u}T^{a}\gamma^{\mu}u)}}
\newcommand{\QjdOne}{\ensuremath{\mathcal{Q}_{\mathrm{qd}}^{(1)}}}
\newcommand{\QjdOneDef}{\ensuremath{(\overline{q}\gamma_{\mu}q)(\overline{d}\gamma^{\mu}d)}}
\newcommand{\QjdEight}{\ensuremath{\mathcal{Q}_{\mathrm{qd}}^{(8)}}}
\newcommand{\QjdEightDef}{\ensuremath{(\overline{q}T^{a}\gamma_{\mu}q)(\overline{d}T^{a}\gamma^{\mu}d)}}

\newcommand{\Uthree}{\ensuremath{\mathrm{U}(3)}}
\newcommand{\Utwo}{\ensuremath{\mathrm{U}(2)}}
\newcommand{\topUtL}{\ensuremath{\Uthree_\mathrm{l} \times \Uthree_\Pe \times \Utwo_\PQq \times \Utwo_\PQu \times \Utwo_\PQd}}

\newcommand{\ttX}{\ensuremath{\PQt(\PAQt)\PX}\xspace}
\newcommand{\ttH}{\ensuremath{\ttbar\PH}\xspace}
\newcommand{\ttll}{\ensuremath{\ttbar\Pell\overline{\Pell}}\xspace}
\newcommand{\ttlnu}{\ensuremath{\ttbar\Pell\PGn}\xspace}
\newcommand{\tllq}{\ensuremath{\PQt\Pell\overline{\Pell}\PQq}\xspace}
\newcommand{\tHq}{\ensuremath{\PQt\PH\PQq}\xspace}
\newcommand{\tttt}{\ensuremath{\ttbar\ttbar}\xspace}

\newcommand{\nWCsConstr}{64}
\newcommand{\nEVsConstr}{43}

\newcommand{\HtoGamGam}{\ensuremath{\PH\to\PGg\PGg}\xspace}
\newcommand{\WGam}{\ensuremath{\PW\PGg}\xspace}
\newcommand{\ZtoNuNu}{\ensuremath{\PZ\to\PGn\PGn}\xspace}
\newcommand{\WW}{\ensuremath{\PW\PW}\xspace}

\newlength\cmsTabSkip\setlength{\cmsTabSkip}{2ex}
\providecommand{\cmsTable}[1]{\resizebox{\textwidth}{!}{#1}}

\ifthenelse{\boolean{cms@external}}{\providecommand{\maybeCmsTable}[1]{\resizebox{\columnwidth}{!}{#1}}}{\providecommand{\maybeCmsTable}{}}

\ifthenelse{\boolean{cms@external}}{\providecommand{\specialTableA}[1]{\resizebox*{\columnwidth}{!}{#1}}}{\providecommand{\specialTableA[1]}{\resizebox*{!}{0.87\textheight}{#1}}}

\makeatletter
\newcommand{\doublehat}[1]{
\begingroup
  \let\macc@kerna\z@
  \let\macc@kernb\z@
  \let\macc@nucleus\@empty
  \hat{\mathchoice
    {\raisebox{.3ex}{\vphantom{\ensuremath{\displaystyle #1}}}}
    {\raisebox{.3ex}{\vphantom{\ensuremath{\textstyle #1}}}}
    {\raisebox{.16ex}{\vphantom{\ensuremath{\scriptstyle #1}}}}
    {\raisebox{.14ex}{\vphantom{\ensuremath{\scriptscriptstyle #1}}}}
    \smash{\hat{#1}}}
\endgroup
}
\makeatother

\cmsNoteHeader{SMP-24-003}

\title{Combined effective field theory interpretation of Higgs boson, electroweak vector boson, top quark, and multijet measurements}
\titlerunning{Combined EFT interpretation of Higgs boson, electroweak vector boson, top quark, and multijet measurements}
\date{\today}

\abstract{
Constraints on Wilson coefficients (WCs) corresponding to dimension-6 operators of the standard model effective field theory (SMEFT) are determined from a simultaneous fit to seven sets of CMS measurements probing Higgs boson, electroweak vector boson, top quark, and multijet production. Measurements of electroweak precision observables are also included and provide complementary constraints to those from the CMS experiment. The CMS measurements, using LHC proton-proton collision data at $\sqrt{s}=13\TeV$, corresponding to integrated luminosities of 36.3 or 138\fbinv, are chosen to provide sensitivity to a broad set of operators, for which consistent SMEFT predictions can be derived.
These are primarily measurements of differential cross sections which are parameterized as functions of the WCs.
In measurements targeting \ttX production, SMEFT effects are modelled at the detector level.
Individual constraints on 64 WCs, and constraints on 43 linear combinations of WCs, are obtained.
}

\hypersetup{
pdfauthor={CMS Collaboration},
pdftitle={Combined effective field theory interpretation of Higgs boson, electroweak vector boson, top quark, and multi-jet measurements},
pdfsubject={CMS},
pdfkeywords={CMS, EFT, Higgs boson, Standard Model, Top}}

\maketitle

\section{Introduction}\label{sec:introduction}

The advent of the LHC era has allowed an extensive exploration of the standard model (SM) and beyond across a broad energy range, from the discovery of the Higgs boson (\PH) at a mass of 125\GeV by the ATLAS and CMS Collaborations in 2012~\cite{Aad:2012tfa,Chatrchyan:2012xdj,Chatrchyan:2013lba} to searches for the direct production of heavy new particles at the \TeVns{} scale~\cite{CMS:2024phk,ATLAS:2024lda,ATLAS:2024fdw}. The discovery of new particles would provide unambiguous evidence of physics beyond the SM~(BSM).
To date, no BSM particles have been found, which motivates a complementary strategy to look for indirect evidence of BSM physics via deviations from theoretical predictions in known SM processes.

{\tolerance=1200
The SM effective field theory (SMEFT) provides a framework for such indirect searches~\cite{BRIVIO20191}.
It characterizes deviations caused by new particles at an energy scale $\Lambda$, assumed to be much higher than the electroweak scale, without depending on the realization of any specific BSM model.
Such an approach is sensitive to scales $\Lambda$ beyond the maximum energy reach of the LHC, which is important if BSM particles are too heavy to be produced on-shell.
The SM Lagrangian, $\mathcal{L}_{\text{SM}}$, is treated as the lowest order term in an expansion in powers of $1/\Lambda$,
\begin{equation}
   \label{eqn:eft_lagrangian}
   \mathcal{L}_{\text{SMEFT}} = \mathcal{L}_{\text{SM}} + \sum_{d,j} \frac{c_j^{(d)}}{\Lambda^{d-4}}\mathcal{Q}_j^{(d)},
\end{equation}
where $\mathcal{Q}_j^{(d)}$ are operators of mass dimension $d \geq 5$, and the $c_j^{(d)}$ are Wilson coefficients (WCs) parameterizing the strength of the interaction introduced by each SMEFT operator.
In this paper, we focus only on dimension-6 operators.
These are the lowest dimension operators beyond the SM when ignoring odd-dimensional operators, which violate lepton or baryon number. These dimension-6 operators are generally expected to give the leading BSM contribution to any process measured at the LHC, with higher-dimensional operators suppressed by factors proportional to $1/\Lambda^4$ or higher.
\par}

There are 2499 dimension-6 operators that together form an independent basis~\cite{Grzadkowski:2010es}.
It is currently not feasible to constrain this many operators simultaneously. However, the imposition of flavour symmetries can reduce this number significantly.
We adopt the \topUtL~(``\texttt{topU3l}'') symmetry of Ref.~\cite{smeftsim:theory,smeftsim:guide}, which treats the first- and second-generation quarks as one set of fields, and the third generation quarks as another independent set.
This reduces the basis to 182 operators.
Of these, 53 have both a Charge-Parity (CP)-conserving and a CP-violating variant.
The latter are not considered here, since the sets of measurements included in this combined interpretation do not make use of observables that can distinguish between CP-conserving and CP-violating effects. This leaves a total of 129 operators.

\begin{figure*}[hbtp]
   \centering
   \includegraphics[width=0.25\textwidth]{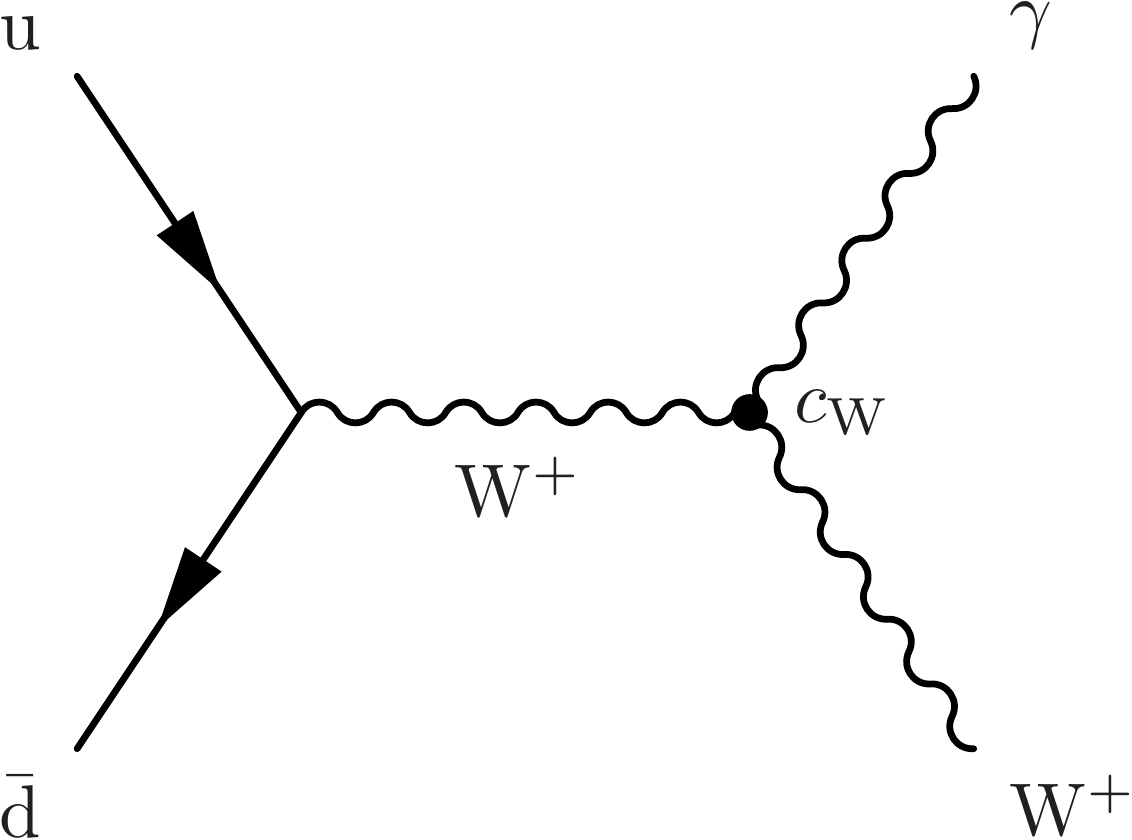}\quad\quad
   \includegraphics[width=0.25\textwidth]{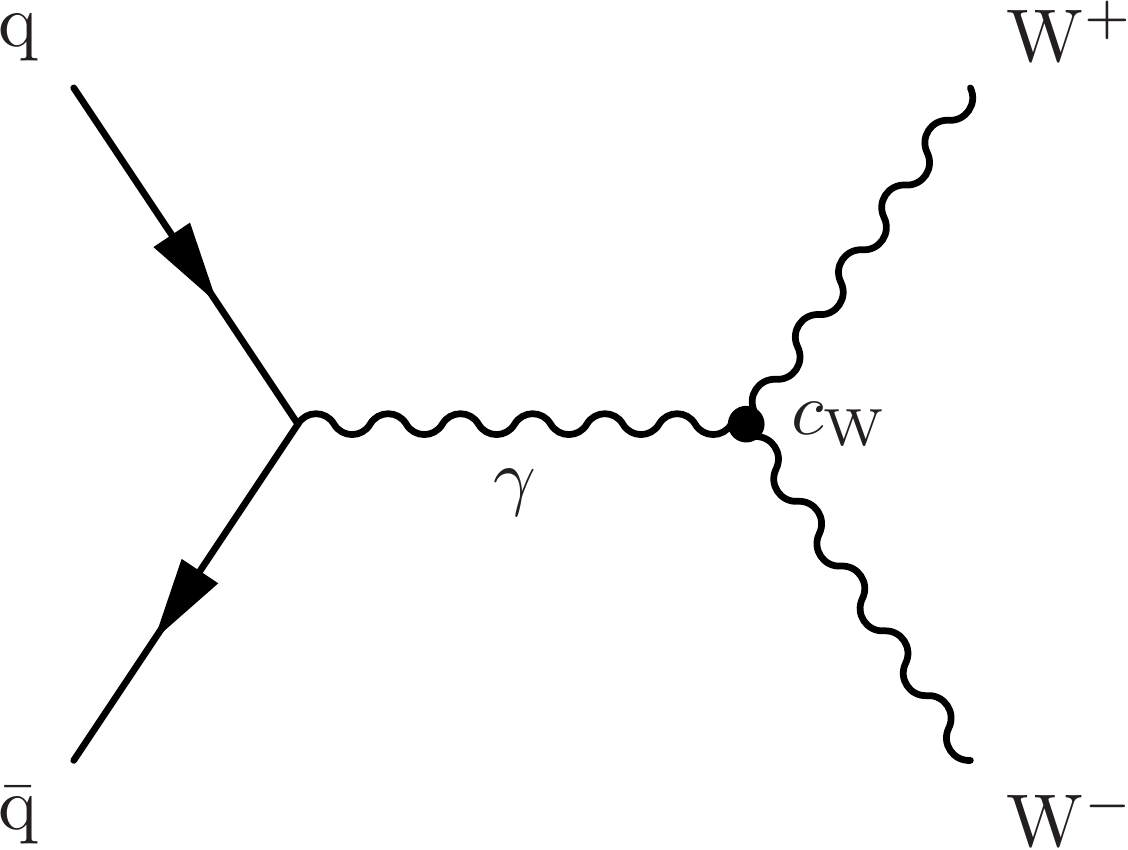}\quad\quad
   \includegraphics[width=0.30\textwidth]{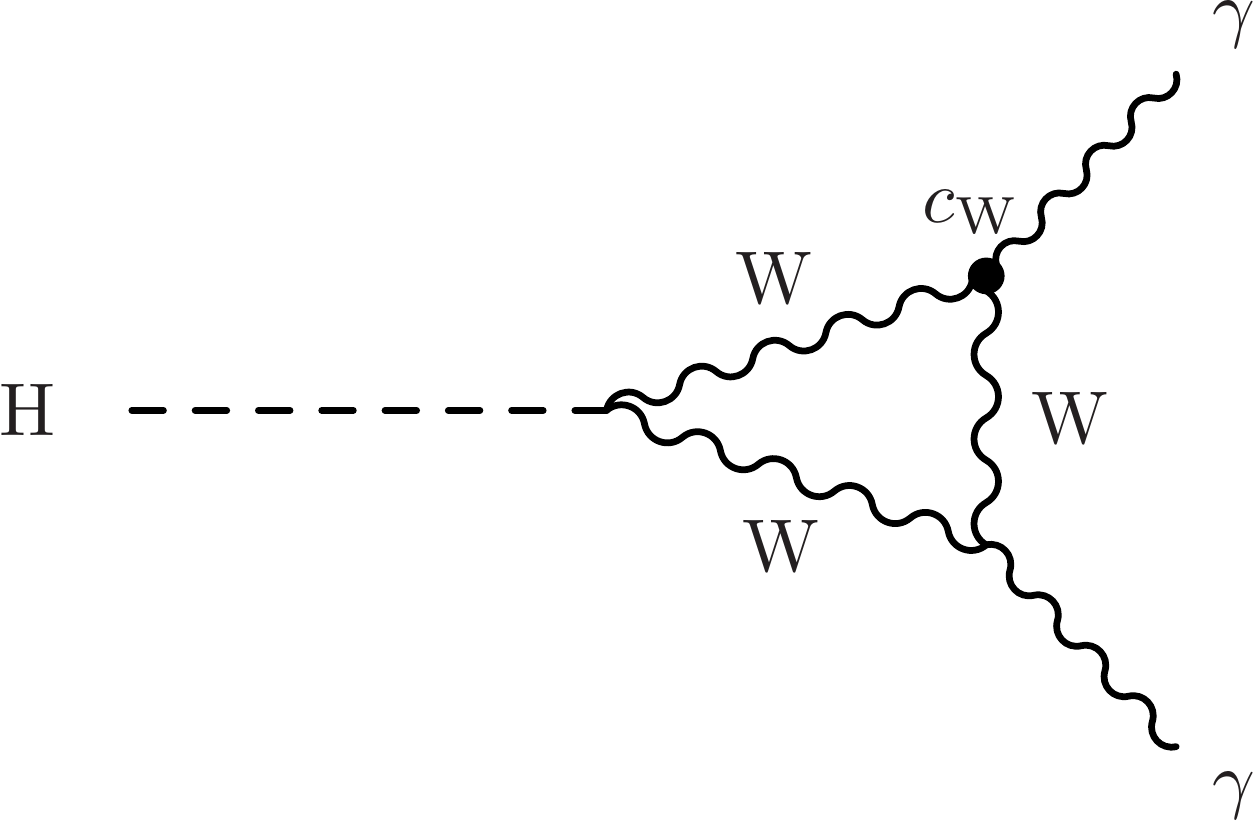}
   \caption{Example Feynman diagrams of modifications of SM processes by the SMEFT operator $\Qw$: $\WGam$ production (left), $\WW$ production (centre), $\HtoGamGam$ decay (right). The WC $\cw$ controls the strength of the interaction.}\label{fig:feynman_diagrams}
\end{figure*}

Each operator will typically impact multiple processes measured at the LHC, and any process will be sensitive to multiple operators. Figure~\ref{fig:feynman_diagrams} shows examples of SM processes modified by the operator $\Qw = \QwDef$, where $\varepsilon$ is the Levi-Civita symbol and $W$ denotes a \PW boson field strength tensor.
Constraints can be determined by a single measurement, but depending on the number of observables measured, this typically requires the assumption that the other WCs are zero. For example, different WCs might affect the same observable, and it is not always possible to include additional observables in the measurement to break this degeneracy.
However, it is expected that the presence of BSM physics would manifest itself through modifications of multiple operators at the same time, which highlights the interest in constraining multiple WCs simultaneously.
Measurements of individual SM processes have been interpreted in EFTs by both the ATLAS and CMS Collaborations.
A selection of these results is reported in Refs.~\cite{CMS:2022uox,ATLAS:2020fcp,ATLAS:2020rej,CMS:2021klw,CMS:2022hjj,CMS:2021aly,ATLAS:2024hmk,ATLAS:2022waa,CMS:2020gtj,CMS:2021icx,ATLAS:2023sua,ATLAS:2023dkz,CMS:2024ksn}.

The most general EFT interpretation requires constraints to be set simultaneously on all WCs, using a global set of measurements as input~\cite{Castro:2809469}.
Such constraints have already been set by several global fit collaborations, using publicly available measurements from experiments at the LHC, and beyond, as input~\cite{Ellis_2021,bartocci2023global,Brivio:2021alv,Celada:2024mcf}.
Additionally, EFT interpretations based on measurements of a range of processes in the top quark physics sector have been performed by the CMS Collaboration~\cite{CMS:2023xyc}, and a combined EFT interpretation of Higgs boson measurements~\cite{ATLAS:2024lyh} has been performed by the ATLAS Collaboration.

In this paper, a combined SMEFT interpretation of CMS measurements covering multiple sectors of the SM via a simultaneous likelihood fit is presented.
Combining multiple analyses requires careful consideration of the statistical overlap between events entering signal regions and the impact of SMEFT operators on the background models used in each analysis. For this combination, measurements for which a consistent SMEFT prediction is available or can be derived are considered, and the event samples selected by each analysis must be statistically independent. Analyses are also required to have small backgrounds, or estimate backgrounds from data if their contribution is significant. Out of the measurements satisfying these criteria, a selection is made that aims to provide complementary sensitivity to a broad set of operators. As only a subset of all CMS measurements is considered, this combined measurement cannot provide sensitivity to all possible operators. Because the selected analyses typically have small backgrounds, or estimate them from data otherwise, the impact of the SMEFT operators on the background processes is not taken into account.

The analyses included are measurements of (i) Higgs boson production in the \HtoGamGam decay channel~\cite{CMS:2021kom}; (ii) top quark-antiquark pair (\ttbar) production in the lepton plus jets final state~\cite{PhysRevD.104.092013}; (iii)
\ttH, \ttll, \ttlnu, \tllq, \tHq, and \tttt production~\cite{CMS:2023xyc} that are collectively referred to in the following as ``\ttX''; (iv)
$\WW(\Pell\PGn\Pell\PGn)$~\cite{CMS:2020mxy} and (v) $\PW(\Pell\PGn)\PGg$ production~\cite{CMS:2021cxr}, where $\Pell = \Pe,\PGm$; (vi)
\ZtoNuNu~\cite{CMS:2020hkd}; and (vii) inclusive jet production~\cite{CMS:2021yzl}.
Measurements of electroweak precision observables (EWPO)~\cite{EWPO:ex1,EWPO:ex2,ewpd4lhc} are also included and provide complementary constraints.

In total, the effects of \nWCsConstr{} operators, listed in Table~\ref{tab:smeftops_other}, are studied. The operators involving three vector boson field strength tensors ($X^3$) are mainly constrained by the \HtoGamGam, \WGam, and inclusive jet production measurements.
Operators connecting two Higgs fields and two vector boson field strength tensors ($X^2 H^2$) are primarily constrained by the \HtoGamGam measurement, while the strongest constraints on operators involving fermion fields and Higgs fields, but no covariant derivatives ($\psi^2 H^3$, $\psi^2 XH$), come from the \HtoGamGam, \ttbar, and \ttX measurements.
The constraints on operators in the classes combining Higgs fields and covariant derivatives ($H^4 D^2$, $\psi^2 H^2 D$) primarily arise from the EWPO measurements.
The inclusive jet, \ttbar, and \ttX measurements provide the strongest constraints on four-fermion operators ($\psi^4$).

\begin{table*}[hp]
	\topcaption{The SMEFT operators studied in this analysis, following the definitions of Ref.~\cite{smeftsim:theory,smeftsim:guide}, where $(q,u,d)$ denote quark fields of the first two generations, $(Q,t,b)$ quark fields of the third generation, and $(l,e,\nu)$ lepton fields of all three generations. The Higgs doublet field is indicated by $H$; $D$ represents a covariant derivative; $\Box$ is the d'Alembert operator; $X = G, W, B$ denotes a vector boson field strength tensor; $p,r$ are flavour indices. Fermion fields are represented by $\psi$, with $L$ and $R$ indicating left- and right-handed fermion fields.}\label{tab:smeftops_other}
    \centering
    \resizebox*{\textwidth}{0.84\textheight}{
    \begin{tabular}{p{7mm}p{42mm}p{6mm}p{46mm}p{6mm}p{41mm}}
    	\multicolumn{6}{c}{$\boldsymbol{X^{3}}$} \\
        \Qg & $= \QgDef$ & \Qw & $= \QwDef$ & & \\
        [\cmsTabSkip]
        \multicolumn{6}{c}{$\boldsymbol{H^{4} D^{2}}$} \\
		\Qhbox & $= \QhboxDef$ & \Qhdd & $= \QhddDef$ & & \\
		[\cmsTabSkip]
		\multicolumn{6}{c}{$\boldsymbol{X^{2} H^{2}}$} \\
		\Qhg & $= \QhgDef$ & \Qhw & $= \QhwDef$ & \Qhb & $= \QhbDef$ \\
		\Qhwb & $= \QhwbDef$ & & & & \\
		[\cmsTabSkip]
		\multicolumn{6}{c}{$\boldsymbol{\psi^{2} H^{3}}$} \\
		\Qth & $= \QthDef$ & \Qbh & $= \QbhDef$ & & \\
		[\cmsTabSkip]
		\multicolumn{6}{c}{$\boldsymbol{\psi^{2} X H}$} \\
		\Qtw & $= \QtwDef$ & \Qtb & $= \QtbDef$ & \Qtg & $= \QtgDef$ \\
		[\cmsTabSkip]
		\multicolumn{6}{c}{$\boldsymbol{\psi^{2} H^{2} D}$} \\
		\QhlOne & $= \QhlOneDef$ & \QhlThree & $= \QhlThreeDef$ & \Qhe & $= \QheDef$ \\
		\QhjOne & $= \QhjOneDef$ & \QhjThree & $= \QhjThreeDef$ & \Qhu & $= \QhuDef$ \\
		\Qhd & $= \QhdDef$ & \QhqOne & $= \QhqOneDef$ & \QhqThree & $= \QhqThreeDef$ \\
		\Qht & $= \QhtDef$ & \Qhbq & $= \QhbqDef$ & &
		\\[\cmsTabSkip]
		\multicolumn{6}{c}{$\boldsymbol{\psi^{4}}\text{\textbf{, }}\mathbf{(\overline{L}L)(\overline{L}L)}$} \\
		\QljOne & $= \QljOneDef$ & \QljThree & $= \QljThreeDef$ & \QqlOne & $= \QqlOneDef$ \\
		\QqlThree & $= \QqlThreeDef$ & \QqqOne & $= \QqqOneDef$ & \QllPrime & $= \QllPrimeDef$ \\
		\QjjOneOne & $= \QjjOneOneDef$ & \QjjOneEight & $= \QjjOneEightDef$ & \QjjThreeOne & $= \QjjThreeOneDef$ \\
		\QjjThreeEight & $= \QjjThreeEightDef$ & \QqjOneOne & $= \QqjOneOneDef$ & \QqjOneEight & $= \QqjOneEightDef$ \\
		\QqjThreeOne & $= \QqjThreeOneDef$ & \QqjThreeEight & $= \QqjThreeEightDef$ & &
		\\[\cmsTabSkip]
		\multicolumn{6}{c}{$\boldsymbol{\psi^{4}}\text{\textbf{, }}\mathbf{(\overline{R}R)(\overline{R}R)}$} \\
		\Qte & $= \QteDef$ & \Qtt & $= \QttDef$ & \QuuOne & $= \QuuOneDef$ \\
		\QuuEight & $= \QuuEightDef$ & \QtuOne & $= \QtuOneDef$ & \QtuEight & $= \QtuEightDef$ \\
		\QddOne & $= \QddOneDef$ & \QddEight & $= \QddEightDef$ & \QudOne & $= \QudOneDef$ \\
		\QudEight & $= \QudEightDef$ & \QtdOne & $= \QtdOneDef$ & \QtdEight & $= \QtdEightDef$
		\\[\cmsTabSkip]
		\multicolumn{6}{c}{$\boldsymbol{\psi^{4}}\text{\textbf{, }} \mathbf{(\overline{L}L)(\overline{R}R)}$} \\
		\Qlu & $= \QluDef$ & \Qtl & $= \QtlDef$ & \QjuOne & $= \QjuOneDef$ \\
		\QjuEight & $= \QjuEightDef$ & \QquOne & $= \QquOneDef$ & \QquEight & $= \QquEightDef$ \\
		\QtjOne & $= \QtjOneDef$ & \QtjEight & $= \QtjEightDef$ & \QqtOne & $= \QqtOneDef$ \\
		\QqtEight & $= \QqtEightDef$ & \QjdOne & $= \QjdOneDef$ & \QjdEight & $= \QjdEightDef$ \\
		\QqdOne & $= \QqdOneDef$ & \QqdEight & $= \QqdEightDef$ & & \\
	\end{tabular}}
\end{table*}

{\tolerance=800
The majority of the analyses included in the combined interpretation are measurements in which cross sections are reinterpreted as constraints on the WCs, while the \ttX measurement uses the ``direct'' approach, in which detector-level predictions are obtained by incorporating the EFT effects through event weights applied to the simulated signal samples.
An additional set of results, excluding the \ttX measurement, is produced via a simplified likelihood (defined in Section~\ref{sec:comb_procedure}) constructed from publicly available information, and validated against the experimental likelihood model. This additional set of results can, therefore, serve as a basis for future reinterpretation.
It is also fast to evaluate, facilitating statistical tests of the model that would otherwise be computationally prohibitive.
\par}

The effect of the SMEFT operators on the signal processes consists of a part that is linear in the WCs, and a part that is quadratic in them. The linear part represents the interference between the SM and the BSM effect described by the SMEFT operators, while the quadratic part is purely related to the BSM contribution. In both cases, only diagrams containing a single SMEFT contribution are considered. Most of the constraints presented in this analysis are set taking into account only the contributions that are linear in the WCs.

This paper is organized as follows. The CMS detector and event reconstruction are introduced in Section~\ref{sec:cmsdetector}. The analyses included in this combined interpretation are summarized in Section~\ref{sec:input_analyses} and the modifications made to those analyses are discussed in Section~\ref{sec:modifications_inputs}. The SMEFT parameterizations and the combination procedure are discussed in Sections~\ref{sec:parameterization} and~\ref{sec:comb_procedure}. Section~\ref{sec:results} presents the results, and a summary is given in Section~\ref{sec:summary}. Tabulated results, including the SMEFT parameterizations, are provided in the HEPData record for this analysis~\cite{HEPData}.

\section{The CMS detector and event reconstruction}\label{sec:cmsdetector}

The CMS apparatus~\cite{CMS:2008xjf,CMS:2023gfb} is a multipurpose, nearly hermetic detector, designed to trigger on~\cite{CMS:2020cmk,CMS:2016ngn,CMS:2024aqx} and identify electrons, muons, photons, and hadrons~\cite{CMS:2020uim,CMS:2018rym,CMS:2014pgm}. Its central feature  is a superconducting solenoid of 6\unit{m} internal diameter, providing a magnetic field of 3.8\unit{T}.
Within the solenoid volume are a silicon pixel and strip tracker, a lead tungstate crystal electromagnetic calorimeter (ECAL), and a brass and scintillator hadron calorimeter (HCAL), each composed of a barrel and two endcap sections.
Forward calorimeters extend the pseudorapidity ($\eta$) coverage provided by the barrel and endcap detectors.
Muons are measured in gas-ionization detectors embedded in the steel flux-return yoke outside the solenoid.
More detailed descriptions of the CMS detector, together with a definition of the coordinate system used and the relevant kinematic variables, can be found in Refs.~\cite{CMS:2008xjf,CMS:2023gfb}.

A particle-flow algorithm~\cite{CMS:2017yfk} aims to reconstruct and identify each individual particle in an event, with an optimized combination of information from the various elements of the CMS detector. The energy of photons is obtained from the ECAL measurement. The energy of electrons is determined from a combination of the electron momentum at the primary interaction vertex as determined by the tracker, the energy of the corresponding ECAL cluster, and the energy sum of all bremsstrahlung photons spatially compatible with originating from the electron track. The energy of muons is obtained from the curvature of the corresponding track. The energy of charged hadrons is determined from a combination of their momentum measured in the tracker and the matching ECAL and HCAL energy deposits, corrected for the response function of the calorimeters to hadronic showers. Finally, the energy of neutral hadrons is obtained from the corresponding corrected ECAL and HCAL energies.

Jets are reconstructed by clustering particle flow candidates using the anti-\kt algorithm~\cite{Cacciari:2008gp, Cacciari:2011ma} with a distance parameter $R = 0.4$, 0.7, or 0.8. Jets originating from \PQb quarks are identified (\PQb~tagged) using multivariate algorithms~\cite{BTV-16-002,Bols:2020bkb,CMS-DP-2018-058}. Boosted hadronically decaying top quarks are reconstructed as anti-\kt jets with $R = 0.8$ and identified using neural networks.

The missing transverse momentum vector \ptvecmiss is computed as the negative vector sum of the transverse momenta of all the particle flow candidates in an event, and its magnitude is denoted as \ptmiss~\cite{CMS:2019ctu}.

\section{Analyses included in the combination}\label{sec:input_analyses}
In this section, we briefly describe the analyses included in the combined interpretation. For each of the studied processes, we have incorporated the most recent CMS measurement. Where a measurement that only analyzes data collected in 2016 is used, a corresponding measurement using the 2016--2018 data set is not available. A summary of analysis type, whether the experimental statistical model was available, and the observables that were used, is given in Table~\ref{tab:input_summary}.

\begin{table*}[h]
   \topcaption{Summary of input analysis characteristics. The observables are defined in the following sections and the experimental likelihood is defined in Section~\ref{sec:comb_procedure}.}\label{tab:input_summary}
   \centering
   \maybeCmsTable{
   \begin{tabular}{p{2.1cm}p{5.7cm}p{4.6cm}p{2.0cm}}
       Analysis & Type of measurement & Observables used & Experimental likelihood\\
       \hline
       $\HtoGamGam$ & Differential cross sections & STXS bins~\cite{Berger:2019wnu} & \checkmark \\
       $\WGam$ & Fiducial differential cross sections & $\pt^{\PGg} \times \abs{\phi_f}$~\cite{CMS:2021cxr} & \checkmark \\
       $\ZtoNuNu$ & Fiducial differential cross sections & $\pt^{\PZ}$ & \checkmark \\
       $\WW$ & Fiducial differential cross sections & $\mll$ &\checkmark \\
       \ttbar & Fiducial differential cross sections & $\mttbar$ & $\times$\\
       \ttX & Direct EFT & Yields in regions of interest & \checkmark \\
       Inclusive jet & Fiducial differential cross sections & $\pt^{\text{jet}}\times \abs{y ^{\text{jet}}}$ & $\times$\\
       EWPO & Pseudo-observables & \GZ, \sigmahad, \Rl, \Rc, \Rb, \AlSLC, \AlLEP, \Ac, \Ab, \AFBl, \AFBc, \AFBb, \GW, \BrWhad, \DeltaAlpha~\cite{ewpd4lhc} & $\times$\\
   \end{tabular}}
\end{table*}

\subsection{\texorpdfstring{Measurement of $\PH\to\boldsymbol{\gamma\gamma}$}{Measurement of H to gamma gamma}}\label{sec:inputs_Hgg}
In the \HtoGamGam analysis, described in more detail in Ref.~\cite{CMS:2021kom}, the major Higgs boson production modes are studied, including gluon-gluon fusion (ggH), vector boson fusion (VBF), vector boson associated production (VH), production associated with a top quark-antiquark pair (\ttH), and production in association with a single top quark (tH). The analysis considers final states with at least two photons, plus additional objects (leptons, jets) for categories targeting specific production modes. The two photons with highest transverse momentum are required to have $\pt>\mgg/3$ and $\pt>\mgg/4$, respectively, with an invariant mass $\mgg$ in the range $100<\mgg<180\GeV$.
The analysis is performed on proton-proton ($\Pp\Pp$) collision data collected between 2016 and 2018, corresponding to an integrated luminosity of 138\fbinv.
The Higgs boson signal is extracted by measuring a narrow peak in the diphoton invariant mass spectrum, on top of a smoothly falling background.

{\tolerance=4800
The MC samples for each Higgs boson production mode are generated with \MGvATNLO 2.4.2 \cite{MG5} at NLO accuracy in QCD. Events produced via the ggH production mode are weighted as a function of Higgs boson \pt and number of jets, to match the prediction from the \textsc{NNLOPS} program \cite{Hamilton_2013}. The NNPDF\,3.0 \cite{Ball_2015} (3.1 \cite{Ball_2017}) parton distribution function (PDF) set is used for event samples compatible with the 2016 (2017 and 2018) data. Parton showering and hadronization are simulated with \PYTHIA 8.230 \cite{pythia82}. The samples are normalized to the production cross sections and branching fractions recommended in Ref.~\cite{LHCHiggsCrossSectionWorkingGroup:2016ypw}.
\par}

The \HtoGamGam analysis performs measurements of the simplified template cross sections (STXS) \cite{Berger:2019wnu}, a set of fiducial bins for different kinematic variables defined for each Higgs boson production mode.
Stage 1.2 of the STXS is used. For ggH events, bins in Higgs boson \pt and number of jets are defined. For events with at least two jets, there is an additional binning in dijet mass and the \pt of the Higgs boson plus dijet system. Events in which Higgs bosons are produced via VBF or in association with a hadronically decaying vector boson have a binning in the number of jets, the Higgs boson \pt, dijet mass, and the \pt of the Higgs boson plus dijet system. For Higgs bosons produced in association with a leptonically decaying vector boson, a binning in the type of associated vector boson, the number of jets, and the \pt of the vector bosons is used. For \ttH production, the \pt of the Higgs boson is used to define a binning.  Not all of the STXS bins can be measured in the analysis; some of these bins are merged, while others are fixed to their SM prediction.

\subsection{\texorpdfstring{Measurement of $\PW\boldsymbol{\gamma}$}{Measurement of W gamma}}\label{sec:inputs_Wgamma}
The \WGam production process is studied in Ref.~\cite{CMS:2021cxr} using the $\Pp\Pp$ collision data set collected by the CMS experiment between 2016 and 2018, corresponding to an integrated luminosity of 138\fbinv.
Leptonic decays of the \PW boson to an electron or muon and a corresponding neutrino are considered.

\sloppypar{
Signal samples are generated with \MGvATNLO~2.6.5 at NLO in QCD, interfaced with \PYTHIA~8.230 to simulate parton showering and hadronization. The NNPDF\,3.1 PDF set is used in the signal simulation. }

Several differential cross sections are measured, and a dedicated EFT analysis is performed, studying only one operator, $\Qw$ (referred to as $\mathcal{O}_{3\PW}$ in Ref.~\cite{CMS:2021cxr}, defined in Table~\ref{tab:smeftops_other}), which modifies the triple and quartic gauge couplings.
For this, a simultaneous measurement of the photon transverse momentum ($\pt^{\PGg}$) and the azimuthal angle ($\abs{\phi_{f}}$) of the charged lepton in the centre-of-mass frame of the $\WGam$ system is performed.
The latter corresponds to the angle between the \PW boson decay plane and the plane spanned by the \PW boson momentum and $\WGam$ boost vectors.
This two-dimensional approach provides increased sensitivity to the interference between the SM and the $\Qw$ contribution than using the transverse momentum alone~\cite{PANICO2018473,Azatov2017}. In the combined interpretation presented here, we therefore select the two-dimensional measurement of $\pt^{\PGg}$ and $\abs{\phi_{f}}$ as the input.

\subsection{\texorpdfstring{Measurement of $\PZ\to\boldsymbol{\nu\nu}$}{Measurement of Z to nu nu}}\label{sec:inputs_Znn}
The $\ZtoNuNu$ analysis measures total and fiducial differential cross sections for the production of a \PZ boson decaying into two neutrinos~\cite{CMS:2020hkd}. The $\PGn\PGn$ final state is characterized by large missing transverse momentum, $\ptmiss>250\GeV$.
Events must also contain at least one jet with $\pt>100\GeV$ and $\abs{\eta}<2.4$. Events with additional photons, leptons, and \PQb-tagged jets are rejected.
The measurement analyzes $\Pp\Pp$ collision data collected in 2016, corresponding to an integrated luminosity of 36.3\fbinv.

{\tolerance=800
Simulated MC signal samples are produced at NLO in QCD using \MGvATNLO 2.2.2, and corrected by \PZ boson \pt dependent higher order electroweak terms. Parton showering and hadronization are simulated with \PYTHIA 8.212 and PDFs are taken from NNPDF\,3.0.
\par}

The signal is extracted through a binned maximum likelihood fit to the missing transverse momentum spectrum, with the major backgrounds estimated using control regions in data, and further minor backgrounds estimated from simulated samples of events.

\subsection{\texorpdfstring{Measurement of $\WW$}{Measurement of WW}}\label{sec:inputs_WW}
In the analysis described in Ref.~\cite{CMS:2020mxy}, fiducial and differential cross section measurements of the production of a $\PWp\PWm$ pair are performed using the $\Pp\Pp$ collision data set collected in 2016, corresponding to an integrated luminosity of 36.3\fbinv.

\WW production via \qqbar annihilation ($\qqbar \to \WW$) is generated with \POWHEG 2.0 \cite{Nason:2004rx,Frixione:2007vw,Alioli:2010xd,Melia:2011tj,Nason:2013ydw} at NLO precision and \WW production via gluon-gluon fusion ($\Pg\Pg \to \WW$) is generated at LO precision with \MCFM 7.0 \cite{Campbell_2010}. The signal samples are corrected to reproduce the \pt spectrum of the \WW system calculated at NNLO precision in perturbative QCD \cite{Gehrmann_2014}. All samples use \PYTHIA 8.212 and the NNPDF\,2.3 \cite{Ball_2013} PDF set.

{\tolerance=800
Leptonic final states of the \PW bosons are analyzed, considering only electrons and muons.
The two leptons must have opposite electric charge, and both the same-flavour ($\Pem\Pep$, $\Pgmm\Pgmp$) and different-flavour ($\Pem\Pgmp$, $\Pep\Pgmm$) lepton channels are analyzed.
The leading (subleading) lepton is required to have a minimum \pt of 25\,(20)\GeV.
In the combination presented here, we use the different-flavour lepton decay channels, because they have smaller Drell-Yan contamination.
\par}

Measurements of several observables are provided, such as lepton transverse momentum, angular separation between the two leptons, and invariant mass of the two leptons, $\mll$. For the interpretation presented in this paper, we choose $\mll$ as the observable. The reconstructed $\mll$ distribution is used in the original analysis to set constraints on dimension-6 WCs in the anomalous triple gauge coupling framework, supporting the choice of observable for the present interpretation.

\subsection{\texorpdfstring{Measurement of \ttbar}{Measurement of ttbar}}\label{sec:inputs_TT}
The \ttbar analysis in the single-lepton plus jets channel~\cite{PhysRevD.104.092013} measures the production cross section of top quark pairs differentially and double-differentially, analyzing pp collision data collected in 2016--2018 and corresponding to an integrated luminosity of 138\fbinv.

{\tolerance=1200
The production of \ttbar events is simulated with \POWHEG~2.0~\cite{Nason:2004rx,Frixione:2007vw,Alioli:2010xd,Frixione:2007nw} at NLO accuracy in QCD. The NNPDF\,3.0 (3.1) PDF set is used for event samples compatible with the 2016 (2017 and 2018) data. Parton showering and hadronization are simulated with \PYTHIA~8.2. The \ttbar samples are normalized to the inclusive cross section calculated with \textsc{Top++}\,v2~\cite{Czakon_2014} at NNLO accuracy in QCD. 
\par}

Both measurements performed at the parton level and at the particle level are provided. At the parton level, where the \ttbar pair is considered before its decay, all effects related to top quark decays, hadronization, and detector acceptance are corrected based on theoretical assumptions. At the particle level, the \ttbar pair is defined based on jets and leptons that can be directly observed in the detector. This reduces extrapolation uncertainties and results in fewer bin-to-bin migrations. In the interpretation presented in this paper, we therefore make use of the particle-level measurements.

The analysis selects events with exactly one electron or muon with $\pt>30\GeV$ and $\abs{\eta}<2.4$. Events with additional electrons or muons with $\pt>15\GeV$ and $\abs{\eta}<2.4$ are rejected. The analysis selection requires the presence of a candidate for a boosted hadronically decaying top quark, or at least four jets with $\pt>30\GeV$ and $\abs{\eta}<2.4$. In the latter case, at least two of the jets must be \PQb tagged.

{\tolerance=800
Measurements of the following observables are performed: leptonically decaying top quark \pt and rapidity;  hadronically decaying top quark \pt and rapidity; mass, rapidity, and \pt of the \ttbar system; leading and subleading top quark \pt. Several double-differential measurements in mass, rapidity, and \pt are also performed.
For the SMEFT interpretation presented here, we tested the different measured observables for optimal sensitivity, by analyzing (a) the trace and (b) the product of eigenvalues greater than or equal to unity of the Hessian matrix of the measurement, parameterized in terms of the WCs. These criteria maximize the number of directions in the SMEFT parameter space to which the measurement is sensitive. The \pt of the subleading top quark, $\pt^{t_\text{low}}$, and the invariant mass of the \ttbar system, $\mttbar$, were the most sensitive observables. Since there are known issues with the modelling of top quark \pt distributions~\cite{Catani:2019hip}, we choose $\mttbar$ as input for the SMEFT interpretation.
\par}

\subsection{\texorpdfstring{Measurement of \ttX}{Measurement of ttX}}\label{sec:inputs_ttX}
{\tolerance=800
A search for new physics in the production of top quarks associated with additional leptons, documented in Ref.~\cite{CMS:2023xyc}, is included in the combined interpretation presented here. This search uses the $\Pp\Pp$ collision data set collected by the CMS experiment between 2016 and 2018, corresponding to an integrated luminosity of 138\fbinv, and is sensitive to operators affecting the production of \ttH, \ttll, \ttlnu, \tllq, \tHq, and \tttt.
\par}

{\tolerance=800
Independent measurements of such processes exist, but cannot easily be reinterpreted in terms of constraints on SMEFT operators, as the event selections typically overlap between these measurements. A single consistent analysis is thus needed to avoid using the same events multiple times in the interpretation. Therefore, the \ttX analysis implements an approach designed to target the effect of dimension-6 EFT operators directly, relying on detector-level observables: the number of events in different regions of interest, defined by the multiplicities of final state objects, and additional kinematical variables. These observables are parameterized as a function of the WCs by incorporating the effect of dimension-6 EFT operators in the event weights of the simulated signal samples. This approach was developed in Ref.~\cite{CMS:2020lrr}, which contains a more detailed description.
\par}

The signal samples are generated at LO accuracy in QCD with \MGvATNLO 2.6.5, using the NNPDF\,3.1 PDF set. \PYTHIA 8.240 is used to simulate partons showering, hadronization, and the decays of Higgs bosons and top quarks. All signal samples are normalized to inclusive cross sections with higher order QCD and electroweak corrections. The \tllq~\cite{MG5} and \tHq~\cite{LHCHiggsCrossSectionWorkingGroup:2016ypw} cross sections are calculated with NLO QCD corrections, and the \ttH~\cite{LHCHiggsCrossSectionWorkingGroup:2016ypw}, \ttll~\cite{LHCHiggsCrossSectionWorkingGroup:2016ypw}, \ttlnu~\cite{Frederix_2021}, and \tttt~\cite{vanbeekveld2022} cross sections with NLO QCD and electroweak corrections.

\subsection{Measurement of inclusive jet production}\label{sec:inputs_QCD}

A measurement of the double-differential cross section in \pt and rapidity $y$ for inclusive jet production is performed in Ref.~\cite{CMS:2021yzl} using $\Pp\Pp$ collision data collected during 2016 and corresponding to an integrated luminosity of 36.3\fbinv.
The jet \pt is measured in up to 22 bins from 97\GeV to 3.1\TeV, and $\abs{y}$ is measured in four bins up to $\abs{y} = 2$. There are a total of 78 bins in the ($\pt,\abs{y}$) parameter space.

Results are given for particle-flow jets clustered using the anti-\kt algorithm~\cite{Cacciari:2008gp} with $R = 0.4$ or 0.7. A quantum chromodynamics (QCD) analysis is also performed using state-of-the-art next-to-next-to-leading order (NNLO) QCD predictions, including an interpretation for four-fermion SMEFT operators, where the cross section dependence is computed at next-to-leading order (NLO).
In the combination presented here, we use the $R=0.7$ measurements as input, as in the QCD analysis of Ref.~\cite{CMS:2021yzl}. The larger radius reduces the modelling impact from out-of-cone radiation effects.

\subsection{Electroweak precision observables}\label{sec:inputs_ewpo}
{\tolerance=800
The interpretation presented here incorporates electroweak precision observables sensitive to the couplings of electroweak vector bosons to fermions. Their implementation is based on the \textsc{ewpd4lhc} tool~\cite{ewpd4lhc}. A short description of the included observables is given in this section.
\par}

The \PZ pole observables, \GZ, \sigmahad, \Rl, \Rc, \Rb, \AlSLC, \Ac, \Ab, \AlLEP, \AFBl, \AFBc, and \AFBb, were measured at LEP and SLC~\cite{EWPO:ex1}. The \PZ boson total width, \GZ, the hadronic pole cross section, \sigmahad, and the ratios \Rl, \Rc, and \Rb, constrain the \PZ boson couplings to left- and right-handed fermions. They are defined as
\ifthenelse{\boolean{cms@external}}{
\begin{equation}
   \begin{aligned}
	\sigmahad &= \frac{12\pi}{m_\PZ^2} \frac{\Gamma_{\Pe\Pe} \Gamma_\text{had}}{\GZ^2}, & \Rl &= \frac{\Gamma_\text{had}}{\Gamma_{\Pell\Pell}}, \\
	\Rc &= \frac{\Gamma_{\PQc\PQc}}{\Gamma_\text{had}}, 	& \Rb &= \frac{\Gamma_{\PQb\PQb}}{\Gamma_\text{had}},
   \end{aligned}
\end{equation}
}
{
\begin{equation}
	\sigmahad = \frac{12\pi}{m_\PZ^2} \frac{\Gamma_{\Pe\Pe} \Gamma_\text{had}}{\GZ^2}, \quad
	\Rl = \frac{\Gamma_\text{had}}{\Gamma_{\Pell\Pell}}, \quad
	\Rc = \frac{\Gamma_{\PQc\PQc}}{\Gamma_\text{had}}, \quad
	\Rb = \frac{\Gamma_{\PQb\PQb}}{\Gamma_\text{had}},
\end{equation}
}
where $\Gamma_{\mathrm{ff}}$ denotes the partial decay width of the \PZ boson to a fermion-antifermion pair, and $\Gamma_\mathrm{had} = \Gamma_{\PQu\PQu} + \Gamma_{\PQd\PQd} + \Gamma_{\PQc\PQc} + \Gamma_{\PQs\PQs} + \Gamma_{\PQb\PQb}$ is the hadronic \PZ boson partial decay width.

The asymmetry parameters \AlSLC, \Ac, and \Ab, determined using polarized electron beams at SLC, and \AlLEP, determined from \PGt polarization measurements at LEP, as well as the forward-backward asymmetries, \AFBl, \AFBc, \AFBb, measured at LEP, distinguish between the \PZ boson couplings to left- and right-handed fermions. The forward-backward asymmetries are defined as
\begin{equation}
	A_\mathrm{FB} = \frac{N_\mathrm{F} - N_\mathrm{B}}{N_\mathrm{F} + N_\mathrm{B}}.
\end{equation}
In this expression, $N_\mathrm{F}$ ($N_\mathrm{B}$) is the number of events where the charged lepton, \PQc quark, or \PQb quark is produced in the direction of the electron beam (positron beam).

Measurements of the \PW boson total width, \GW, and the hadronic \PW boson branching fraction, \BrWhad, constrain the \PW boson couplings to left-handed fermions \cite{EWPO:ex2}. \BrWhad~was measured at LEP, while the most precise measurement of \GW~is obtained from a combination of LEP and Tevatron data~\cite{EWPO:ex2,EWPO:ex3}.

{\tolerance=800
The running of the electromagnetic coupling,
\ifthenelse{\boolean{cms@external}}
{
\begin{multline}
   \DeltaAlpha(m_\PZ) = \DeltaAlpha_\mathrm{had}(m_\PZ) + \DeltaAlpha_\mathrm{lep}(m_\PZ) \\ = 1 - \frac{\alpha\bigl(Q^2=0\bigr)}{\alpha\bigl(Q^2=m_{\PZ}^2\bigr)},
\end{multline}
}
{
\begin{equation}
	\DeltaAlpha(m_\PZ) = \DeltaAlpha_\mathrm{had}(m_\PZ) + \DeltaAlpha_\mathrm{lep}(m_\PZ) = 1 - \frac{\alpha\bigl(Q^2=0\bigr)}{\alpha\bigl(Q^2=m_{\PZ}^2\bigr)},
\end{equation}
}
is also treated as an observable. We use the value of $\DeltaAlpha_\mathrm{had}(m_\PZ)$ from Ref.~\cite{Davier_2020} in combination with the theoretical calculation of the leptonic contributions $\DeltaAlpha_\mathrm{lep}(m_\PZ)$ at four-loop order~\cite{STURM2013698}.
\par}

Theoretical predictions for the electroweak precision observables, in the SM and the SMEFT, are calculated with \textsc{ewpd4lhc}. The SM predictions are calculated, at two-loop accuracy or higher, using the interpolation formulas of Refs.~\cite{Dubovyk_2019,Awramik_2006,Awramik_2004,Cho_2011}. SMEFT corrections are evaluated with \smeftsim~\cite{smeftsim:theory,smeftsim:guide}. The input parameters used to calculate the SM predictions are given in Table~\ref{tab:ewpo1}. The predicted and measured values of the 15 observables included in the fit are summarized in Table~\ref{tab:ewpo2}.

\begin{table*}[h]
   \topcaption{Input parameters used to calculate SM predictions for the electroweak precision observables. The values of $m_{\PW}$, $m_{\PZ}$, $m_{\PH}$, $m_{\PQt}$, and \GF~are set to the PDG averages~\cite{pdg}, with an additional uncertainty of $0.5\GeV$ on $m_{\PQt}$ to account for ambiguities in the definition of the top quark mass~\cite{Hoang_2020}. The strong coupling $\alpha_s$ is set to the average of the Flavour Lattice Averaging Group (FLAG)~\cite{flag}, as it is more robust against SMEFT effects than the PDG value~\cite{trott2023}.}\label{tab:ewpo1}
   \centering
   \begin{tabular}{lrcllllrcll}
       Parameter & \multicolumn{3}{c}{Value} & Ref. & & Parameter & \multicolumn{3}{c}{Value} & Ref. \\
       \hline
		$m_{\PW}$ (\GeVns{}) & 80.377 & \hspace{-3mm}$\pm$ & \hspace{-3mm}0.012 & \cite{pdg} & & $m_{\PQt}$ (\GeVns{}) & 172.69 & \hspace{-3mm}$\pm$ & \hspace{-3mm}0.58 & \cite{pdg,Hoang_2020} \\
		$m_{\PZ}$ (\GeVns{}) & 91.1876 & \hspace{-3mm}$\pm$ & \hspace{-3mm}0.0021 & \cite{pdg} & & \GF~($10^{-12}\GeV^{-2}$) & 11663788 & \hspace{-3mm}$\pm$ & \hspace{-3mm}6 & \cite{pdg} \\
		$m_{\PH}$ (\GeVns{}) & 125.25 & \hspace{-3mm}$\pm$ & \hspace{-3mm}0.17 & \cite{pdg} & & $\alpha_s$ & 0.1183 & \hspace{-3mm}$\pm$ & \hspace{-3mm}0.0007 & \cite{flag} \\
   \end{tabular}
\end{table*}

\begin{table*}[h]
   \topcaption{Predicted and measured values of the 15 electroweak precision observables included in the fit. $\DeltaAlpha(m_\PZ)$ is the sum of $\DeltaAlpha_\mathrm{had}(m_\PZ)=0.02753\pm0.00010$~\cite{Davier_2020} and $\DeltaAlpha_\mathrm{lep}(m_\PZ)=0.0314979\pm0.0000002$~\cite{STURM2013698}. The measurements of \GZ~and \sigmahad~include corrections to Ref.~\cite{EWPO:ex1}, accounting for an underestimation of the integrated luminosity and using an updated Bhabha scattering cross section, as recommended by the PDG~\cite{pdg}. The SM predictions are calculated in the $\{m_\PW, m_\PZ, \GF\}$ input parameter scheme with the input parameters of Table~\ref{tab:ewpo1}, using \textsc{ewpd4lhc}~\cite{ewpd4lhc}.}\label{tab:ewpo2}
   \centering
   \begin{tabular}{llrcllrclll}
    Observable & & \multicolumn{3}{c}{Prediction} & & \multicolumn{3}{c}{Measurement} & & Ref. \\
   \hline
	\GZ~(\GeVns{}) & & 2.49561 & \hspace{-3mm}$\pm$ & \hspace{-3mm}0.00084 & & 2.4955  & \hspace{-3mm}$\pm$ & \hspace{-3mm}0.0023  & & \cite{EWPO:ex1,pdg} \\
	\sigmahad~(pb) & & 41488.2 & \hspace{-3mm}$\pm$ & \hspace{-3mm}7.2     & & 41480.2 & \hspace{-3mm}$\pm$ & \hspace{-3mm}32.5    & & \cite{EWPO:ex1,pdg} \\
	\Rl            & & 20.7578 & \hspace{-3mm}$\pm$ & \hspace{-3mm}0.0086  & & 20.767  & \hspace{-3mm}$\pm$ & \hspace{-3mm}0.025   & & \cite{EWPO:ex1} \\
	\Rc            & & 0.17224 & \hspace{-3mm}$\pm$ & \hspace{-3mm}0.00005 & & 0.1721  & \hspace{-3mm}$\pm$ & \hspace{-3mm}0.0030  & & \cite{EWPO:ex1} \\
	\Rb            & & 0.21586 & \hspace{-3mm}$\pm$ & \hspace{-3mm}0.00010 & & 0.21629 & \hspace{-3mm}$\pm$ & \hspace{-3mm}0.00066 & & \cite{EWPO:ex1} \\
	\AlSLC         & & 0.1503  & \hspace{-3mm}$\pm$ & \hspace{-3mm}0.0019  & & 0.1513  & \hspace{-3mm}$\pm$ & \hspace{-3mm}0.0021  & & \cite{EWPO:ex1} \\
	\AlLEP         & & 0.1503  & \hspace{-3mm}$\pm$ & \hspace{-3mm}0.0019  & & 0.1465  & \hspace{-3mm}$\pm$ & \hspace{-3mm}0.0033  & & \cite{EWPO:ex1} \\
	\Ac            & & 0.6692  & \hspace{-3mm}$\pm$ & \hspace{-3mm}0.0008  & & 0.670   & \hspace{-3mm}$\pm$ & \hspace{-3mm}0.027   & & \cite{EWPO:ex1} \\
	\Ab            & & 0.9350  & \hspace{-3mm}$\pm$ & \hspace{-3mm}0.0002  & & 0.923   & \hspace{-3mm}$\pm$ & \hspace{-3mm}0.020   & & \cite{EWPO:ex1} \\
	\AFBl          & & 0.01695 & \hspace{-3mm}$\pm$ & \hspace{-3mm}0.00043 & & 0.0171  & \hspace{-3mm}$\pm$ & \hspace{-3mm}0.0010  & & \cite{EWPO:ex1} \\
	\AFBc          & & 0.0755  & \hspace{-3mm}$\pm$ & \hspace{-3mm}0.0011  & & 0.0707  & \hspace{-3mm}$\pm$ & \hspace{-3mm}0.0035  & & \cite{EWPO:ex1} \\
	\AFBb          & & 0.1054  & \hspace{-3mm}$\pm$ & \hspace{-3mm}0.0014  & & 0.0992  & \hspace{-3mm}$\pm$ & \hspace{-3mm}0.0016  & & \cite{EWPO:ex1} \\
	\GW~(\GeVns{}) & & 2.0918  & \hspace{-3mm}$\pm$ & \hspace{-3mm}0.0010  & & 2.085   & \hspace{-3mm}$\pm$ & \hspace{-3mm}0.042   & & \cite{EWPO:ex2,EWPO:ex3,pdg} \\
	\BrWhad        & & 0.6754  & \hspace{-3mm}$\pm$ & \hspace{-3mm}0.0000  & & 0.6741  & \hspace{-3mm}$\pm$ & \hspace{-3mm}0.0027  & & \cite{EWPO:ex2} \\
	$\DeltaAlpha(m_\PZ)$ & & 0.05793 & \hspace{-3mm}$\pm$ & \hspace{-3mm}0.00074 & & 0.05903 & \hspace{-3mm}$\pm$ & \hspace{-3mm}0.00010 & & \cite{Davier_2020,STURM2013698}
   \end{tabular}
\end{table*}

\section{Modifications to the input analyses}\label{sec:modifications_inputs}

Modifications to some of the analyses included in this combination are made with respect to their original publications. These changes are made to ensure a consistent treatment of theoretical uncertainties in the interpretation, the use of state-of-the-art SM predictions in the inclusive jet measurement, and a consistent use of the integrated luminosity calibrations and uncertainties.

When cross section measurements are reported, theoretical uncertainties in the cross sections are properties of the prediction with which the measurement is compared, and not uncertainties in the measured cross sections themselves.
However, for the purpose of interpreting those cross sections in the context of an EFT, the theoretical uncertainties must be taken into account. This is achieved by repeating the cross section measurements, where the theoretical uncertainties in the cross sections are included in the total uncertainty in the measurement. The theoretical uncertainties that are considered in all the analyses are uncertainties in the PDF and in the QCD factorization and renormalization scale.

For the \HtoGamGam and the \WGam measurements, theoretical uncertainties in the SM predictions are incorporated in the statistical model and the cross section measurements are repeated, allowing the corresponding nuisance parameters to vary in the model.
These uncertainties were already available in the likelihood constructed for the original measurements, but were kept fixed to their nominal values in that case. The theoretical uncertainties in the predictions for the \HtoGamGam measurement include uncertainties due to missing higher orders and PDF uncertainties following the recommendations from Ref.~\cite{LHCHiggsCrossSectionWorkingGroup:2016ypw}, as well as uncertainties specific to the STXS measurements, described in more detail in Ref.~\cite{CMS:2021kom}. 
Theoretical uncertainties in the \WGam predictions due to missing higher orders in the cross section calculation are obtained from the variation of renormalization and factorization scales. PDF uncertainties are evaluated following the \textsc{PDF4LHC} prescription~\cite{pdf4lhc}.

For the inclusive jet measurement, the NNLO QCD cross sections are extracted, using the \fastNLO tool \cite{Britzger:2012bs,Wobisch:2011ij}, from interpolation grids~\cite{Britzger:2022lbf} that were derived from \NNLOjet~\cite{Currie:2016bfm,Currie:2018xkj,Gehrmann:2018szu} predictions. These cross sections are updated for this combination to use the CT18~\cite{Hou:2019efy} PDF set at NNLO, which is expected to give better agreement with CMS jet data than the CT14~\cite{Dulat:2015mca} set used previously. Multiplicative corrections for nonperturbative and NLO electroweak effects are applied, and are the same as those used in Ref.~\cite{CMS:2021yzl}. Several sources of uncertainty in the final prediction are evaluated. The PDF uncertainty considers the 28 independent CT18 PDF eigenvector variations. For the missing higher orders in QCD, the uncertainty is treated as uncorrelated between different $\abs{y}$ bins, but correlated between the \pt bins of a given $\abs{y}$ bin.  Also included are an uncertainty in the nonperturbative correction, from a comparison of \PYTHIA and \HERWIGpp  tunes, and the statistical uncertainties originating from the interpolation grids.

In the \ZtoNuNu, \WW, and \ttbar measurements, theoretical uncertainties are evaluated by passing samples of simulated events through the \rivet codes~\cite{rivet} that replicate the phase space selections for these analyses. 
The \WW and \ttbar samples are generated at NLO in QCD with the \POWHEG~2.0 event generator, and \ZtoNuNu events are generated at NLO in QCD with the \MGvATNLO~2.6.5 generator. All samples are interfaced with \PYTHIA~8.240 with the CP5 underlying event tune~\cite{CMS:2019csb} for parton showering and hadronization, and in the samples produced with \MGvATNLO, the FxFx jet merging scheme is used~\cite{Frederix:2012ps}. The NNPDF\,3.1 PDF set at NNLO in QCD is incorporated in these samples.
The samples include PDF as well as renormalization and factorization scale weights. The factorization and renormalization scale uncertainties are considered as two independent sources of systematic uncertainty, to provide more degrees of freedom in the fit than taking the envelope of the two would.

The uncertainties are incorporated as nuisance parameters in the likelihood fit, or, for the analyses for which the experimental likelihood description is not available, we construct a covariance matrix of theoretical uncertainties. This theoretical uncertainty covariance matrix is added to the experimental covariance matrix that was provided in the measurements.

{\tolerance=800
The \ttX analysis originally used the \textsc{dim6top} EFT description~\cite{AguilarSaavedra:2018nen}, and this has been translated to SMEFT as necessary. Additional operators that were not considered in the original analysis are added by post-generation reweighting of the signal samples~\cite{10.21468/SciPostPhysCommRep.4}. This is achieved by generating external matrix element libraries with \MGvATNLO. Both linear and quadratic terms are accounted for. The $\Qhbox = \QhboxDef$ operator (rate only, defined in Table~\ref{tab:smeftops_other}) and two-heavy-two-light quark operators (differential) are incorporated in the \ttX analysis for the purpose of this combination. Other operators were also studied and the sensitivity was found to be minimal.
\par}

For the analyses incorporated in the interpretations with their experimental likelihood, the PDF uncertainties are considered to be correlated. The factorization and renormalization scale uncertainties are not correlated between the different processes, as they reflect the missing higher orders in the perturbative QCD calculation of each specific scattering process. In addition to theoretical uncertainties, experimental uncertainties related to the integrated luminosity, the jet energy scale and resolution, the missing transverse energy unclustered energy scale (MET unclustered), the trigger inefficiency caused by the gradual timing shift in the ECAL trigger inputs in the $\abs{\eta} > 2.4$ region (L1 Prefiring), and the pileup modelling are correlated between measurements where possible. When the uncertainty schemes used in the analyses are not compatible, they are treated as uncorrelated. The results are not significantly impacted by the correlation of systematic uncertainties between measurements, as their contribution to the total uncertainty in each WC is small. A summary of the correlation scheme is given in Table~\ref{tab:correlations}. 

The integrated luminosity uncertainties are correlated between measurements that use the same data set and for which the experimental likelihood is available. The integrated luminosity calibrations and uncertainties were updated to the latest available values~\cite{CMS-LUM-17-003,CMS-PAS-LUM-17-004,CMS-PAS-LUM-18-002}. The majority of the luminosity uncertainties are correlated between the different data-taking years, with uncorrelated contributions to the total luminosity uncertainty of 0.26\%, 0.60\%, and 0.65\% for 2016, 2017, and 2018, respectively.

\begin{table}[h]
   \topcaption{Correlation scheme of the systematic uncertainties.}\label{tab:correlations}
   \centering
   \maybeCmsTable{
   \begin{tabular}{lccccc}
	Uncertainty source 	  & \HtoGamGam & \WGam 	  	& \ZtoNuNu   & \WW 		  & \ttX \\
    \hline
    PDF 				  & 		   & \checkmark & \checkmark & 			  & \\
    Luminosity 			  & \checkmark & \checkmark & \checkmark & \checkmark & \checkmark \\
    Jet energy scale 	  & \checkmark & 			& \checkmark & 		      & \\
    Jet energy resolution & \checkmark & \checkmark & \checkmark & 			  & \checkmark \\
    MET unclustered		  & \checkmark & 			&			 & \checkmark & \\
    L1 Prefiring 		  & 		   &			& \checkmark & \checkmark & \checkmark \\
    Pileup				  & 		   & \checkmark & \checkmark & \checkmark & \checkmark
   \end{tabular}}
\end{table}

\section{The SMEFT parameterization}\label{sec:parameterization}

To interpret the measured cross sections or signal yields as constraints on the contributions from BSM physics, these quantities need to be parameterized as functions of the WCs. The scattering cross section is proportional to the square of the matrix element, $\abs{\mathcal{M}}^2$, of the process. In the presence of new interactions introduced by dimension-6 SMEFT operators, the matrix element can be written as
\begin{equation}
   \label{eqn:matrixelement}
   \mathcal{M} = \mathcal{M}_{\text{SM}} + \sum_j \frac{c_j}{\Lambda^2}\mathcal{M}_j + \mathcal{O}\left(\frac{c_j^2}{\Lambda^4}\right),
\end{equation}
where the SM matrix element is given by $\mathcal{M}_{\text{SM}}$ and the $\mathcal{M}_j$ describe the matrix elements corresponding to the new physics interactions. In the following, we restrict SMEFT contributions to diagrams with a single insertion of a new physics interaction, therefore dropping the $\mathcal{O}(c_j^2/\Lambda^4)$ term.

Squaring this expression to obtain a cross section gives
\ifthenelse{\boolean{cms@external}}
{
\begin{multline}
   \label{eqn:matrixelementsq}
   \sigma \propto \abs{\mathcal{M}}^2 = \abs{\mathcal{M}_{\text{SM}}}^2 + 2\sum_j \frac{c_j}{\Lambda^2}\text{Re}\left(\mathcal{M}_j\mathcal{M}^{*}_{\text{SM}}\right) \\ +\sum_{j,k}\frac{c_{j}c_{k}}{\Lambda^4}\text{Re}\left(\mathcal{M}_j\mathcal{M}^{*}_k\right) .
\end{multline}
}
{
\begin{equation}
   \label{eqn:matrixelementsq}
   \sigma \propto \abs{\mathcal{M}}^2 = \abs{\mathcal{M}_{\text{SM}}}^2 + 2\sum_j \frac{c_j}{\Lambda^2}\text{Re}\left(\mathcal{M}_j\mathcal{M}^{*}_{\text{SM}}\right)+\sum_{j,k}\frac{c_{j}c_{k}}{\Lambda^4}\text{Re}\left(\mathcal{M}_j\mathcal{M}^{*}_k\right) .
\end{equation}
}
This means that the cross section of a process $\alpha$ in a kinematic bin $i$ (for example, a specific \pt range) can be written as
\ifthenelse{\boolean{cms@external}}
{
\begin{multline}
   \label{eqn:crosssec_EFT}
   \sigma^{i}_{\alpha,\text{SMEFT}} = \sigma^{i}_{\alpha,\text{SM}} + \sigma^i_{\alpha,\text{int}}(\vec{c}) + \sigma^i_{\alpha,\text{BSM}}(\vec{c}) \\ = \sigma^i_{\alpha,\text{SM}} \left( 1 + \sum_j A_{\alpha,j}^i\frac{c_j}{\Lambda^2} + \sum_{j,k} B_{\alpha,jk}^i\frac{c_{j}c_{k}}{\Lambda^4} \right),
\end{multline}
}
{
\begin{equation}
   \label{eqn:crosssec_EFT}
   \sigma^{i}_{\alpha,\text{SMEFT}} = \sigma^{i}_{\alpha,\text{SM}} + \sigma^i_{\alpha,\text{int}}(\vec{c}) + \sigma^i_{\alpha,\text{BSM}}(\vec{c}) = \sigma^i_{\alpha,\text{SM}} \left( 1 + \sum_j A_{\alpha,j}^i\frac{c_j}{\Lambda^2} + \sum_{j,k} B_{\alpha,jk}^i\frac{c_{j}c_{k}}{\Lambda^4} \right),
\end{equation}
}
where $\vec{c}$ is the full set of WCs.
Here, the symbol $\sigma$ incorporates both the production cross section and any relevant branching fractions for the decays of unstable particles. The subscript `int' denotes the SM-BSM interference part of the cross section, and the subscript `BSM' the purely BSM-related part.

{\tolerance=1200
Thus, the cross sections can be parameterized into a part that is linear in the WCs and a part that is quadratic in them.
The strengths of the linear and quadratic contributions are described by the constants $A_{\alpha,j}^i$ and $B_{\alpha,jk}^i$, respectively.
These constants are computed by generating events at leading order (LO), with extra parton emissions to partially capture NLO QCD effects, using \MGvATNLO~2.0.16 interfaced with \PYTHIA~8.3~\cite{pythia83} to simulate parton showering and hadronization. The MLM jet merging scheme~\cite{Alwall:2007fs} is employed in these samples. The effects of the SMEFT operators on the generated processes are modelled using \smeftsim, following the procedures outlined in Ref.~\cite{10.21468/SciPostPhysCommRep.4}. For the loop-induced processes $\Pg\Pg\to\PH$ and $\Pg\Pg\to\PZ\PH$, \smeftatnlo~\cite{smeftatnlo} is used. Analytic calculations are used for the $\PH \to \PGg\PGg$ decay~\cite{hyyparam}.
\par}

As analytic calculations are used for the \HtoGamGam decay, but not for its production, the production and decay for this process are parameterized separately. This parameterization also depends on the total Higgs boson width, and therefore the parameterization involves a division by a sum over $c_j$ and $c_{j}c_{k}$. This leads to a parameterization with terms of $\mathcal{O}(c_j^3/\Lambda^6)$ and above. The parameterization is truncated at $\mathcal{O}(c_j^2/\Lambda^4)$ or $\mathcal{O}(c_j/\Lambda^2)$, depending on the results presented.

The corrections from the SMEFT do not only modify interaction vertices; the masses and decay widths of intermediate particles can also be modified. Propagator corrections are incorporated in the parameterizations, and are evaluated with \smeftsim. These corrections are not well defined at orders $\mathcal{O}(c_j^2/\Lambda^4)$ or above, therefore they are only considered up to linear order. Propagator corrections and vertex corrections are considered simultaneously, while ensuring that double insertions---the inclusion of multiple new physics interactions---in a single diagram are avoided.

The NNPDF\,3.1 NNLO PDF set is used when computing the parameterizations. The SM masses, widths, and couplings are set to the values in Table~\ref{tab:param_card}. All parameterizations use the $\{m_{\PW}, m_{\PZ}, \GF \}$ input parameter scheme~\cite{Brivio:2792440} and the \texttt{topU3l} flavour symmetry.

\begin{table}[h]
   \topcaption{The SM parameters used in the event generation to derive the SMEFT parameterizations~\cite{pdg}.}\label{tab:param_card}
   \centering
   \maybeCmsTable{
   \begin{tabular}{cccc}
     Parameter & Value & Parameter & Value \\
     \hline
     $m_{\PW}$  & 80.377\GeV  & $\Gamma_{\PW}$  & 2.085\GeV \\
     $m_{\PZ}$  & 91.1876\GeV & $\Gamma_{\PZ}$  & 2.4955\GeV \\
     $m_{\PH}$  & 125.25\GeV  & $\Gamma_{\PH}$  & 3.2\MeV \\
     $\overline{m}_{\PQt}$ & 172.69\GeV  & $\Gamma_{\PQt}$ & 1.42\GeV \\
     $\overline{m}_{\PQb}$ & 4.18\GeV    & $\GF$ & $1.166379 \times 10^{5} \GeV^{-2}$ \\
   \end{tabular}}
\end{table}

The phase space selections for each kinematic measurement bin are reproduced using \rivet 3.1.9~\cite{rivet}.

For the \ZtoNuNu, \WW, \WGam, \ttbar, and inclusive jet analyses, multiple samples of events, with orthogonal phase space selections, are used to ensure a sufficient number of events are available to derive the parameterization for all measurement bins.

As discussed in Section~\ref{sec:inputs_ttX}, the \ttX analysis directly incorporates the effects of dimension-6 EFT operators in the weights of the simulated signal events. Like the cross sections in Eq.~(\ref{eqn:crosssec_EFT}), the weight function for each event $\beta$ is parameterized by a polynomial of second order in the WCs,
 \begin{equation}
   \label{eqn:ttX_eventweight}
    w^\beta(\vec{c},\vec{\nu}) = u_0^\beta(\vec{\nu}) + \sum_j \frac{c_j}{\Lambda^2} u_{1j}^\beta(\vec{\nu}) + \sum_{j,k} \frac{c_j c_k}{\Lambda^4} u_{2jk}^\beta(\vec{\nu}),
\end{equation}
where $\vec{\nu}$ are nuisance parameters corresponding to systematic uncertainties, as will be described in Section~\ref{sec:comb_procedure}.
The predicted yield for a given analysis bin $i$, as a function of the WCs and the nuisance parameters, is calculated by summing the weight functions of each event that passes the selection criteria of that bin,
\ifthenelse{\boolean{cms@external}}
{
\begin{multline}
   \label{eqn:ttX_yield}
    e^i(\vec{c},\vec{\nu}) = \sum_\beta w^\beta(\vec{c},\vec{\nu}) = U_0^i(\vec{\nu}) \\ + \sum_j \frac{c_j}{\Lambda^2} U_{1j}^i(\vec{\nu}) + \sum_{j,k} \frac{c_j c_k}{\Lambda^4} U_{2jk}^i(\vec{\nu}).
\end{multline}
}
{
\begin{equation}
   \label{eqn:ttX_yield}
    e^i(\vec{c},\vec{\nu}) = \sum_\beta w^\beta(\vec{c},\vec{\nu}) = U_0^i(\vec{\nu}) + \sum_j \frac{c_j}{\Lambda^2} U_{1j}^i(\vec{\nu}) + \sum_{j,k} \frac{c_j c_k}{\Lambda^4} U_{2jk}^i(\vec{\nu}).
\end{equation}
}
In this equation, the coefficients of the yield parameterization are the sums of the coefficients of the weight functions, \eg $U_{1j}^i(\vec{\nu}) = \sum_\beta u_{1j}^\beta(\vec{\nu})$. With this approach, detector-level predictions can be obtained at any arbitrary point in the EFT parameter space. In the combined interpretation presented here, we reuse the EFT predictions of the original \ttX analysis in Ref.~\cite{CMS:2023xyc}, with the modifications discussed in Section~\ref{sec:modifications_inputs}.

Although parameterizations of the SMEFT corrections are computed up to quadratic order, the majority of the results we report use parameterizations truncated at $\mathcal{O}(c_j/\Lambda^2)$. As visible in Eq.~(\ref{eqn:crosssec_EFT}), parameterization terms quadratic in $c_j$ enter with a factor $1/\Lambda^4$, the same order in $1/\Lambda$ as the linear terms of a parameterization containing dimension-8 operators. To avoid the inconsistency of considering only some contributions at order $1/\Lambda^4$, the main results that are  computed thus only use parameterizations at $\mathcal{O}(c_j/\Lambda^2)$. However, a comparison of the constraints using the linear-only and the linear-plus-quadratic parameterization is provided. This gives an indication of how much the inclusion of orders $1/\Lambda^4$ could change the sensitivity of the results.

\begin{figure*}[hp]
   	\centering
      \rotatebox{-90}{
      \begin{minipage}{\textheight}
   		\includegraphics[width=\textwidth]{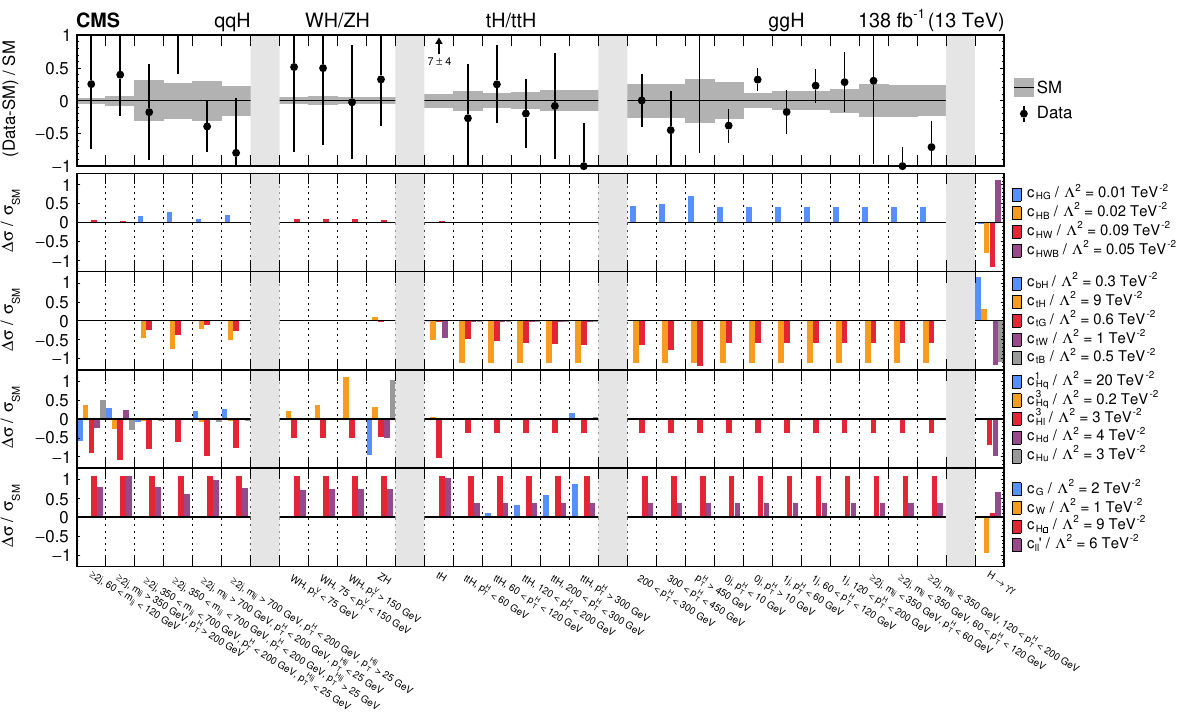}
   		\caption{Relative effect of the linear SMEFT terms for the WCs that affect the Higgs STXS cross sections and the \HtoGamGam branching fraction. The parameters $c_j/\Lambda^2$ are set to different values to ensure the effect of all WCs can be visualized on the same $y$ axis scale. The upper panel shows the measured values and their uncertainties relative to the predictions in the SM. As these are measurements of the cross sections times branching fraction, no measurement is displayed in the rightmost bin (labelled ``\HtoGamGam'').}\label{fig:param_effect1}
      \end{minipage}}
\end{figure*}

\begin{figure*}[hp]
   	\centering
      \rotatebox{-90}{
      \begin{minipage}{\textheight}
   		\includegraphics[width=\textwidth]{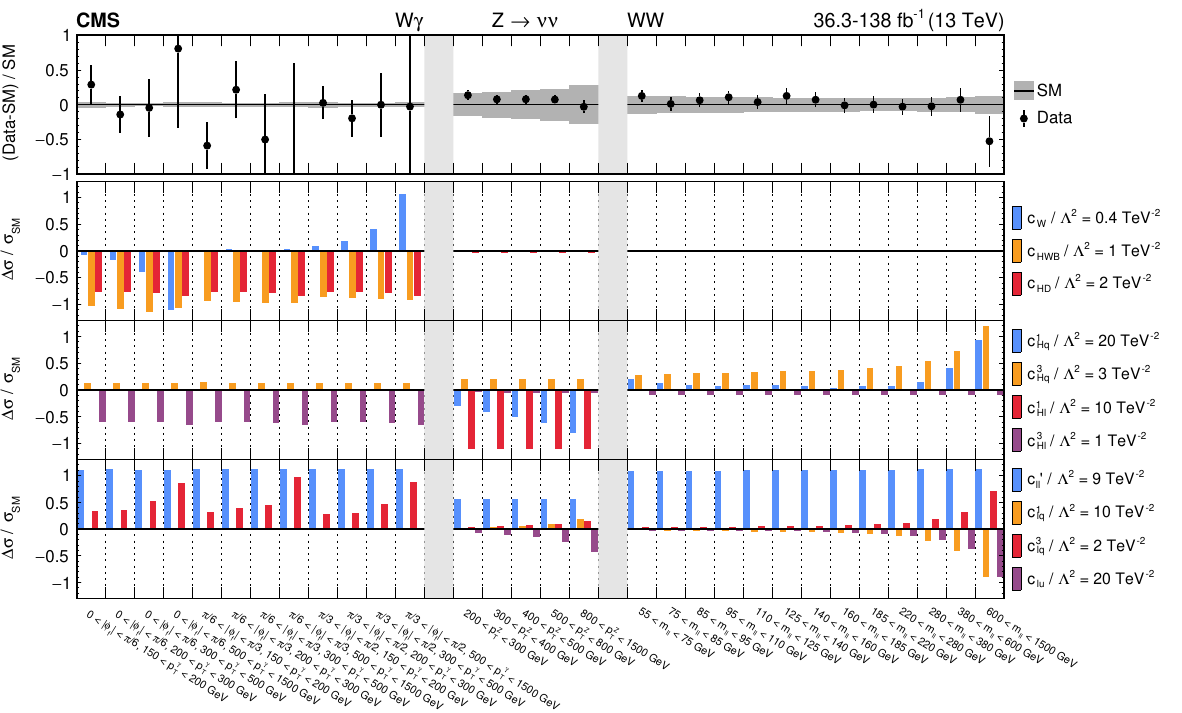}
   		\caption{Relative effect of the linear SMEFT terms for the WCs that affect the $\WGam$, $\ZtoNuNu$, and $\WW$ differential cross sections. The parameters $c_j/\Lambda^2$ are set to different values to ensure the effect of all WCs can be visualized on the same $y$ axis scale. The upper panel shows the measured values and their uncertainties relative to the predictions in the SM.}\label{fig:param_effect2}
      \end{minipage}}
\end{figure*}

\begin{figure*}[hp]
   	\centering
      \rotatebox{-90}{
      \begin{minipage}{\textheight}
   		\includegraphics[width=\textwidth]{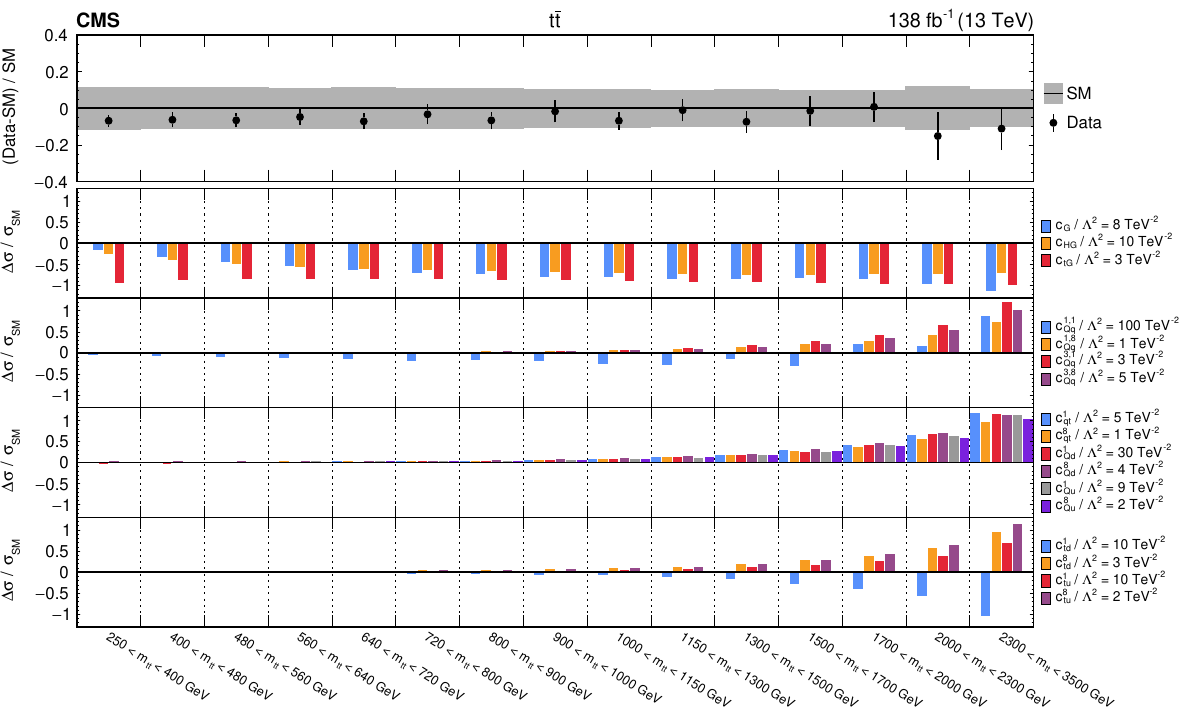}
   		\caption{Relative effect of the linear SMEFT terms for the WCs that affect the \ttbar differential cross sections. The parameters $c_j/\Lambda^2$ are set to different values to ensure the effect of all WCs can be visualized on the same $y$ axis scale. The upper panel shows the measured values and their uncertainties relative to the predictions in the SM.}\label{fig:param_effect3}
      \end{minipage}}
\end{figure*}

\begin{figure*}[hp]
   	\centering
      \rotatebox{-90}{
      \begin{minipage}{\textheight}
   		\includegraphics[width=\textwidth]{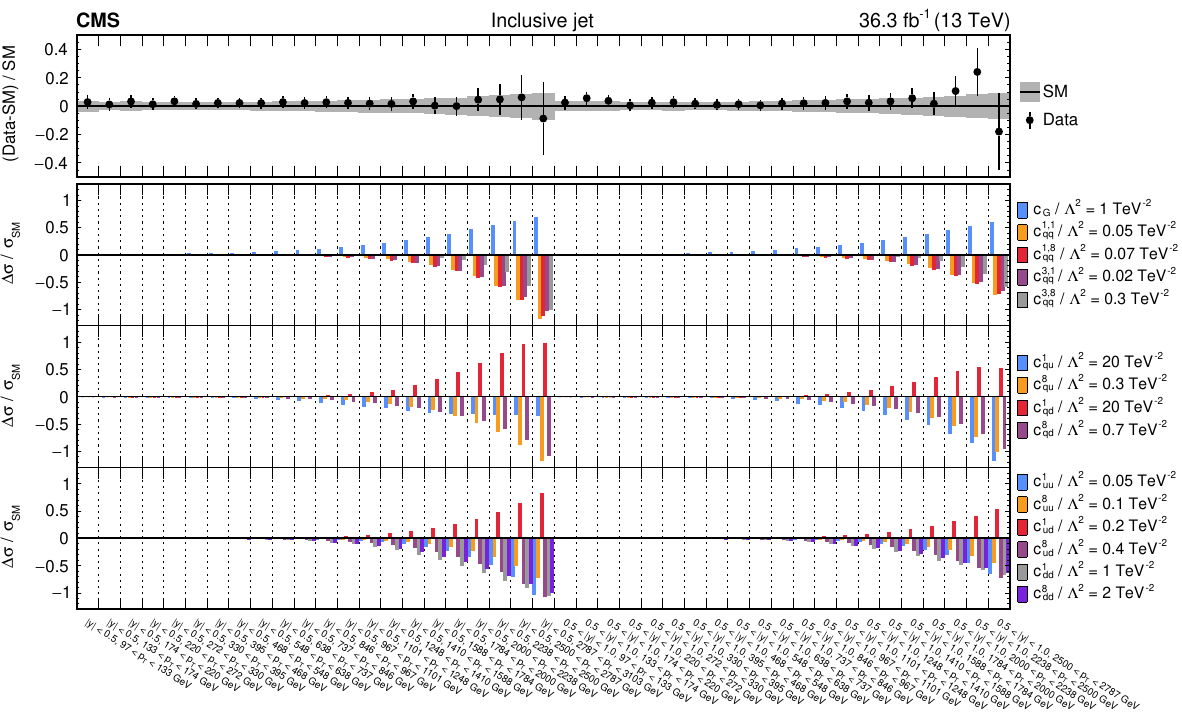}
   		\caption{Relative effect of the linear SMEFT terms for the WCs that affect the inclusive jet differential cross sections in the rapidity bins $(0, 0.5)$ and $(0.5, 1)$. The parameters $c_j/\Lambda^2$ are set to different values to ensure the effect of all WCs can be visualized on the same $y$ axis scale. The upper panel shows the measured values and their uncertainties relative to the predictions in the SM.}\label{fig:param_effect4}
      \end{minipage}}
\end{figure*}

\begin{figure*}[hp]
   	\centering
      \rotatebox{-90}{
      \begin{minipage}{\textheight}
   		\includegraphics[width=\textwidth]{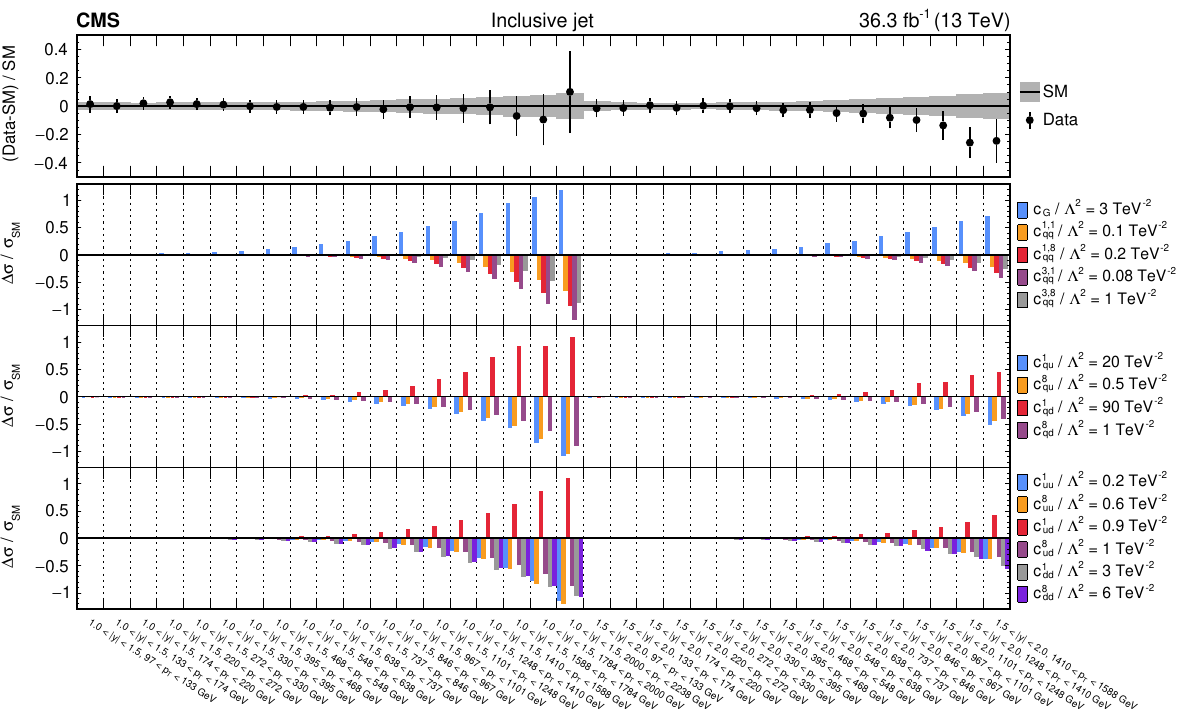}
   		\caption{Relative effect of the linear SMEFT terms for the WCs that affect the inclusive jet differential cross sections in the rapidity bins $(1, 1.5)$ and $(1.5, 2)$. The parameters $c_j/\Lambda^2$ are set to different values to ensure the effect of all WCs can be visualized on the same $y$ axis scale. The upper panel shows the measured values and their uncertainties relative to the predictions in the SM.}\label{fig:param_effect5}
      \end{minipage}}
\end{figure*}

\begin{figure*}[hp]
   	\centering
      \rotatebox{-90}{
      \begin{minipage}{\textheight}
   		\includegraphics[width=\textwidth]{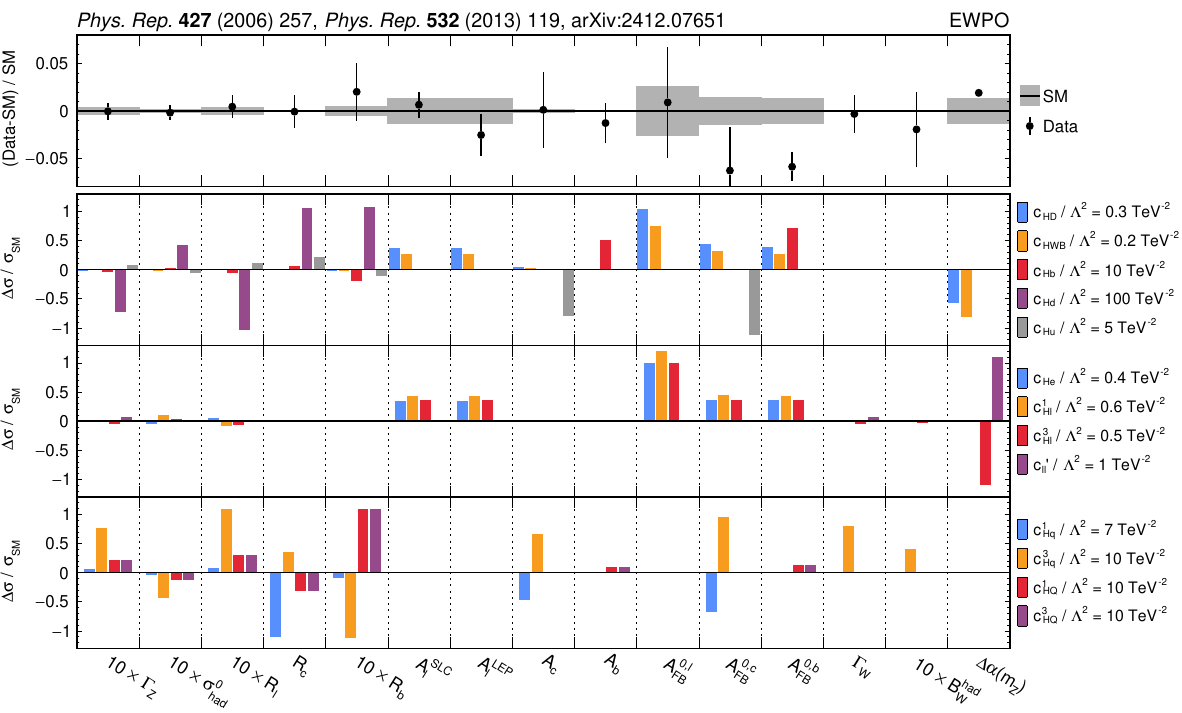}
   		\caption{Relative effect of the linear SMEFT terms for the WCs that affect the EWPO~\cite{EWPO:ex1,EWPO:ex2,ewpd4lhc}. The parameters $c_j/\Lambda^2$ are set to different values to ensure the effect of all WCs can be visualized on the same $y$ axis scale. The upper panel shows the measured values and their uncertainties relative to the predictions in the SM.}\label{fig:param_effect6}
      \end{minipage}}
\end{figure*}

The relative effect of the linear part of the parameterizations on the \HtoGamGam, \WGam, \ZtoNuNu, \WW, \ttbar, and inclusive jet  cross sections and the EWPO is shown in Figs.~\ref{fig:param_effect1}--\ref{fig:param_effect6}. This quantity is computed as the change in the cross section, relative to the SM expectation, for the parameter values indicated in the legend. For $c_j/\Lambda^2=1\TeV^{-2}$ this corresponds to the constant $A_{\alpha,j}^i$ in Eq.~(\ref{eqn:crosssec_EFT}). The upper panels in these figures show the measured values of the cross sections with respect to the SM predictions. The corresponding figures in the rotated basis, described in Section~\ref{sec:results}, are given in Appendix~\ref{sec:ev_impacts}.

\section{Combination procedure}\label{sec:comb_procedure}
This combination follows the statistical methodology described in Ref.~\cite{CMS-NOTE-2011-005}, as implemented in the \textsc{Combine}~\cite{CMS:2024onh} tool, which is based on the \textsc{RooFit} and \textsc{RooStats}~\cite{Verkerke:2003ir} frameworks.
The results are constraints on WCs and their linear combinations, in the form of 68\% and 95\% confidence intervals, evaluated using a profile likelihood ratio test statistic,
\begin{equation}
   \label{eqn:plr}
   q(\vec{c}) = -2\ln\left({\frac{\mathcal{L}(\vec{c},\doublehat{\vec{\nu}}(\vec{c}))}{\mathcal{L}(\hat{\vec{c}},\hat{\vec{\nu}})}}\right),
\end{equation}
where $\vec{c}$ represent the parameters of interest (POIs), which in this case are the WCs or their linear combinations, and $\vec{\nu}$ indicate nuisance parameters that encode the effects of theoretical and experimental uncertainties.
The quantities $\hat{\vec{c}}$ and $\hat{\vec{\nu}}$ describe unconditional maximum likelihood estimates of the parameters. The conditional maximum likelihood estimate of the nuisance parameters for fixed values of the POIs, $\vec{c}$, is given by $\doublehat{\vec{\nu}} (\vec{c})$. The best fit parameter values of the POIs are taken to be the unconditional maximum likelihood estimate $\hat{\vec{c}}$. The 68\% and 95\% confidence intervals are constructed as the union of intervals for which $q(\vec{c})<1$ and $q(\vec{c})<3.84$, respectively.

The likelihood in this combination is expressed as
\begin{equation}
\label{eqn:likelihood_full}
   \mathcal{L} \left( \text{data} \,;\, \vec{c},\vec{\nu} \right) = \mathcal{L}^{\text{expt}} \left( \vec{c},\vec{\nu} \right)  \mathcal{L}^{\text{simpl}} \left( \vec{c} \right),
\end{equation}
where
\ifthenelse{\boolean{cms@external}}
{
   \begin{align}
   \mathcal{L}^{\text{expt}} \left(\vec{c},\vec{\nu} \right)
   &= \prod_i\text{Poisson} \left( n_i \,;\,\sum_j\mu'^j(\vec{c})s_i^j(\vec{\nu}) + b_i(\vec{\nu}) \right) \nonumber \\  &
   \hspace{0.4\columnwidth}\prod_k p_k \left( y_k \,;\, \nu_k \right); \label{eqn:likelihood_expt} \\
   \mathcal{L}^{\text{simpl}} \left(\vec{c} \right)
   &= \frac{\exp \left( -\frac{1}{2} \left(\vec{\mu}(\vec{c}) - \hat{\vec{\mu}}\right)^T V^{-1} \left(\vec{\mu}(\vec{c}) - \hat{\vec{\mu}}\right) \right)}{\sqrt{(2\pi)^m \text{det}(V)}}. \label{eqn:likelihood_simpl}
\end{align}
}
{
\begin{align}
   \mathcal{L}^{\text{expt}} \left(\vec{c},\vec{\nu} \right)
   &= \prod_i\text{Poisson} \left( n_i \,;\,\sum_j\mu'^j(\vec{c})s_i^j(\vec{\nu}) + b_i(\vec{\nu}) \right)
   \, \prod_k p_k \left( y_k \,;\, \nu_k \right); \label{eqn:likelihood_expt} \\
   \mathcal{L}^{\text{simpl}} \left(\vec{c} \right)
   &= \frac{\exp \left( -\frac{1}{2} \left(\vec{\mu}(\vec{c}) - \hat{\vec{\mu}}\right)^T V^{-1} \left(\vec{\mu}(\vec{c}) - \hat{\vec{\mu}}\right) \right)}{\sqrt{(2\pi)^m \text{det}(V)}}. \label{eqn:likelihood_simpl}
\end{align}
}
The first term in Eq.~(\ref{eqn:likelihood_full}), $\mathcal{L}^{\text{expt}}$, covers the measurements, listed in Table~\ref{tab:input_summary}, for which an experimental likelihood is available.
The index $j$ corresponds to the signal processes in the different measurement bins, and the index $i$ runs over all the reconstruction-level bins in the distributions that are being fitted. The parameters $\mu'^j$ are estimators of the cross sections for process (measurement bin) $j$, relative to the SM expectation, and are parameterized in terms of the POIs.
The parameters $n$, $s$, and $b$ represent the observed number of events, the expected number of signal events, and the expected number of background events, respectively. Systematic uncertainties are incorporated as nuisance parameters $\vec{\nu}$, which enter the likelihood paired with auxiliary observables $\vec{y}$.
The factors $p_k \left( y_k \,;\, \nu_k \right)$ represent the probability densities of these auxiliary observables for a given value of the nuisance parameter $y_k$.
In the \ttX measurement, the expected yields in the different analysis bins are directly parameterized in terms of the WCs, without the intermediate step of measuring cross sections.
The construction of the likelihood for this analysis is similar to Eq.~(\ref{eqn:likelihood_expt}). However, instead of $\sum_j\mu'^j(\vec{c})s_i^j(\vec{\nu})+b_i({\vec{\nu}})$, the expected yield in bin $i$ takes the form $e_i(\vec{\nu},\vec{c})$.

The second term in Eq.~(\ref{eqn:likelihood_full}), $\mathcal{L}^{\text{simpl}}$, includes the remaining measurements. It represents a simplified likelihood modelled as a multivariate Gaussian, where $\vec{\mu}(\vec{c})$ represents the cross sections or pseudo-observables relative to their SM expectation, parameterized in terms of the POI, and $\hat{\vec{\mu}}$ represents the corresponding measured best fit values. This model contains no nuisance parameters; instead, experimental and theoretical uncertainties are included in the $m{\times}m$ covariance matrix, $V = V_\text{ex} + V_\text{th}$, where $m$ is the number of measurement bins. As the measurements are assumed not to be correlated with each other, this is a block-diagonal covariance matrix, which considers only correlations between the parameters of each individual measurement. The theoretical uncertainty covariance matrix $V_\text{th}$ is constructed taking into account expected correlations between measurement bins. The consistency of the simplified likelihood approach with the experimental likelihood approach is validated in Appendix~\ref{sec:validation_simplified}.

In what follows, the likelihood given by Eq.~(\ref{eqn:likelihood_full}) is referred to as a `hybrid' likelihood.
A second, simplified, likelihood is defined in Appendix~\ref{sec:simplified}. This likelihood incorporates all input measurements, apart from the \ttX measurement, into $\mathcal{L}^{\text{simpl}}$. For measurements that were originally performed following an experimental likelihood description, we extract the covariance matrix including all theoretical uncertainties to build this simplified likelihood.
As the \ttX measurement does not measure cross sections, but directly targets SMEFT effects, it can not be included in this simplified likelihood construction.

\begin{figure*}[htbp]
   \centering
   \includegraphics[width=\textwidth]{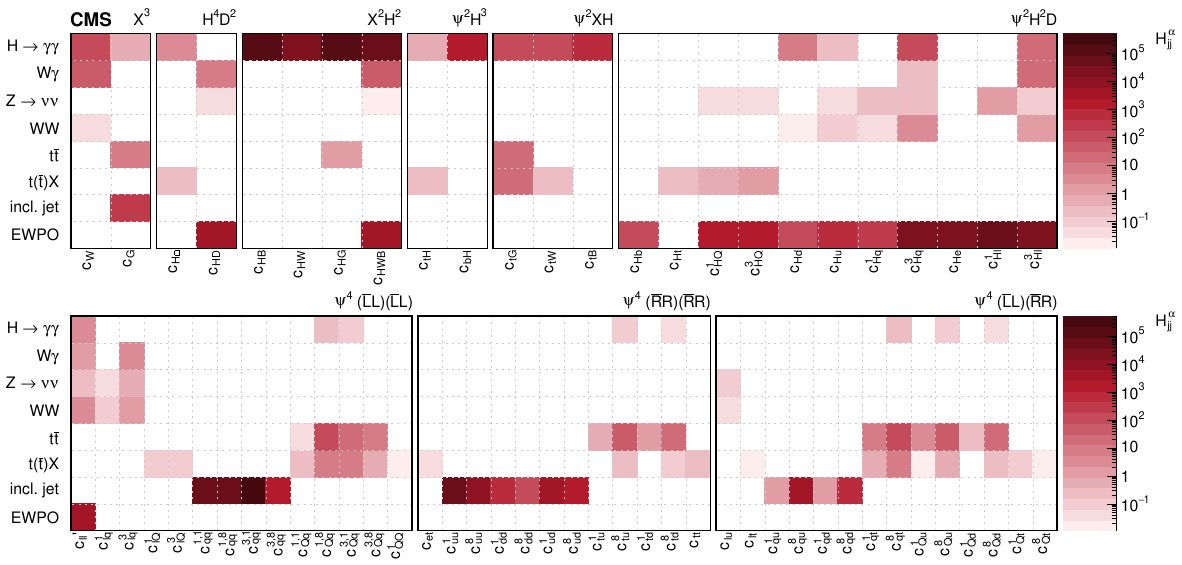}
   \caption{Diagonal entries $H_{jj}^{\alpha}$ of the Hessian matrix evaluated for each input channel. These indicate which of the input channels are expected to be the most sensitive to any given operator. Larger values of $H_{jj}^{\alpha}$ correspond to higher sensitivity.}\label{fig:diag_hessian}
\end{figure*}

In the combined measurement it is not possible to simultaneously constrain all the WCs that affect the processes of interest, because some of the WCs have (nearly) degenerate effects on the distributions of the input measurements. Therefore, in addition to constraining WCs one-by-one---setting constraints on a specific WC while fixing the others to their SM expectations---linear combinations of WCs are also constrained in a simultaneous fit.
These linear combinations are determined by performing a principal component analysis (PCA).
We extract the Hessian matrix $H$ of the combined measurement, parameterized in terms of the WCs, retaining only linear terms. This is thus the matrix of second derivatives of the log-likelihood with respect to the WCs. The diagonal entries $H_{jj}^{\alpha}$ of the Hessian matrix evaluated for each input channel $\alpha$ are shown in Fig.~\ref{fig:diag_hessian}; they give an indication of the sensitivity of each channel $\alpha$ to each operator $\mathcal{Q}_j$. The quantity $1/\sqrt{H_{jj}^{\alpha}}$ is an estimate of half the expected 68\% confidence interval in $c_j/\Lambda^2$, evaluated with input channel $\alpha$.
An eigendecomposition $H = \mathcal{R}^{T} \Lambda \mathcal{R}$ of this matrix is performed, which breaks the matrix down into a matrix $\mathcal{R}$, consisting of eigenvectors of the input Hessian matrix, and a diagonal matrix $\Lambda$ of the corresponding eigenvalues.
The eigenvectors contained in the matrix $\mathcal{R}$ are used to define a set of linear combinations of WCs, $\mathrm{EV}_j = \mathcal{R}^{\,jk} c_k$, that are orthogonal to each other. The likelihood is parameterized in terms of the linear combinations, instead of the WCs, using the transpose of the rotation matrix $\mathcal{R}$.

The PCA returns as many eigenvectors as there are WCs. The quantity $1/\sqrt{\lambda}$, where $\lambda$ represents the eigenvalue for a given eigenvector, gives an estimate of half the expected 68\% confidence interval for that eigenvector. Eigenvectors for which this quantity is greater than five are not taken into account in the analysis. The choice to use a cutoff of five is arbitrary. We consider the analysis to be insensitive to these directions in the SMEFT parameter space, and fix them to their SM value (0) in the combined fit.

\section{Results}\label{sec:results}

Results are presented for the combination of all input measurements, using the experimental likelihood approach where possible (`hybrid').
As discussed in Section~\ref{sec:parameterization}, constraints on the individual WCs are provided for both linear-only and linear-plus-quadratic parameterizations, following some of the recommendations outlined in Ref.~\cite{Brivio:2798082}. It should be noted that deriving the constraints using the linear and quadratic parts of the parameterization is only feasible for the constraints on the individual WCs, obtained by varying the parameters one at a time. Introducing the parameterization terms that are quadratic in the WCs means the eigenvectors obtained via the approach described in Section~\ref{sec:comb_procedure} are no longer guaranteed to be orthogonal. In addition, the introduction of quadratic terms can lead to the likelihood fit converging to a local minimum when allowing multiple parameters to vary simultaneously.

\begin{figure*}[htbp]
   	\centering
      \rotatebox{-90}{
      \begin{minipage}{\textheight}
   		\includegraphics[width=\textwidth]{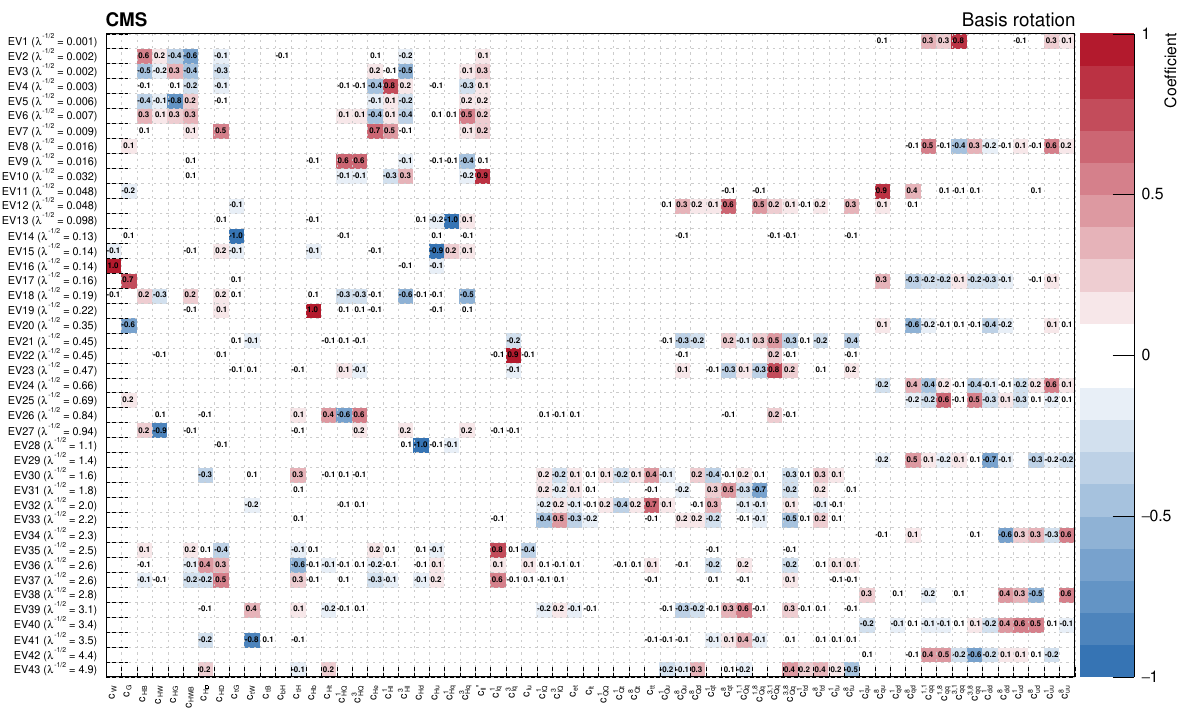}
   		\caption{Rotation matrix obtained by performing the PCA on the Hessian matrix of the full set of measurements. Only matrix coefficients with absolute value ${\geq}0.05$ are displayed.}\label{fig:PCA_withttX}
      \end{minipage}}
\end{figure*}

The rotation matrix, obtained by performing the PCA on the Hessian matrix of the full set of measurements, is visualized in Fig.~\ref{fig:PCA_withttX}. A total of \nEVsConstr{} eigenvectors with $1/\sqrt{\lambda}<5$ are retained in the analysis, which are linear combinations of \nWCsConstr{} WCs. The remaining 21 eigenvectors are fixed to their SM value (0), as the analysis is not sufficiently sensitive to these directions to be able to consider all \nWCsConstr{} parameters in a combined fit.

\begin{figure*}[htbp]
   	\centering
      \rotatebox{-90}{
      \begin{minipage}{\textheight}
   		\includegraphics[width=\textwidth]{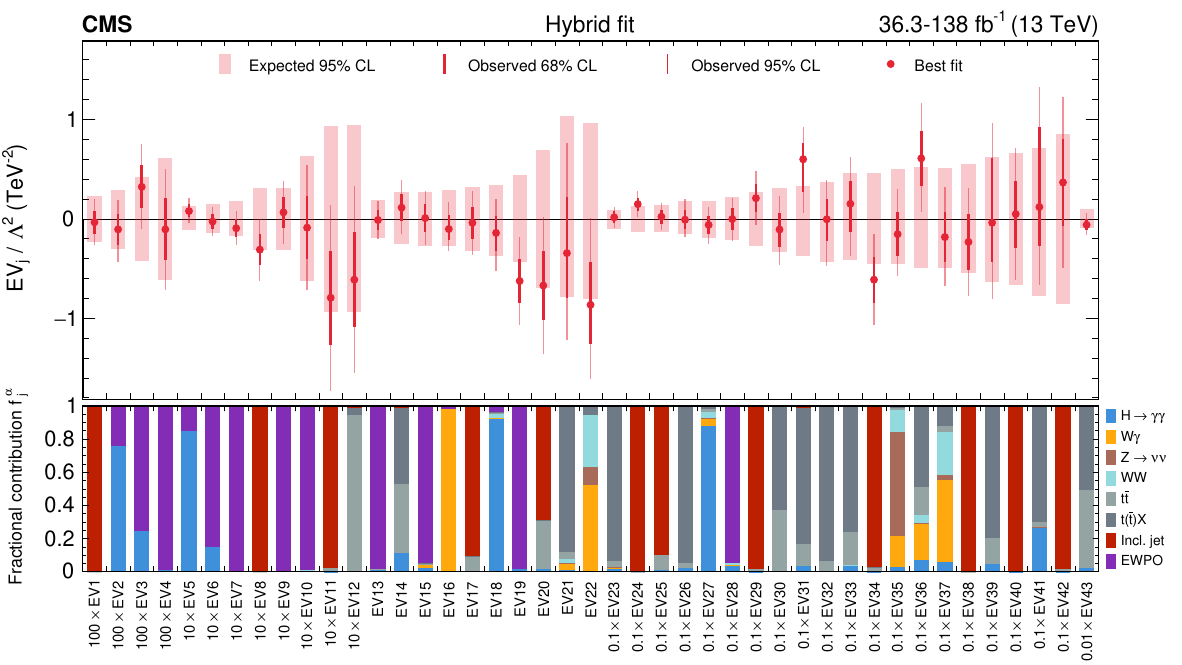}
   		\caption{Constraints on linear combinations of WCs, for the hybrid fit including the full set of measurements. The shaded areas correspond to the expected 95\% confidence intervals, the thick and thin bars to the observed 68\% and 95\% confidence intervals, respectively. The lower panel shows the contribution of different input measurements to the total constraints. The constraints are scaled by powers of 10 to ensure the constraints on all \nEVsConstr{} eigenvectors can be visualized on the same $y$ axis scale.}\label{fig:hybrid_plus_rotated}
      \end{minipage}}
\end{figure*}

Constraints on the linear combinations of WCs are shown in Fig.~\ref{fig:hybrid_plus_rotated} and Table~\ref{tab:hybrid_plus_rotated}. In this fit, all linear combinations of WCs are varied simultaneously. The 95\% confidence intervals on the \nEVsConstr{} eigenvector directions are in the range $\pm 10\TeV^{-2}$ for the least constrained direction, to $\pm 0.002\TeV^{-2}$ for the most constrained direction. The $p$-value for the compatibility with the SM (all WCs equal to 0) is 2.5\%. The deviation from the SM is mostly driven by the inclusive jet measurement; when excluding it from the combination, the SM compatibility $p$-value is 27\%. The level of agreement between the inclusive jet measurement and the SM prediction is known to be particularly sensitive to the PDFs. It has been demonstrated in Ref.~\cite{CMS:2021yzl} that a full PDF fit to the inclusive jet data yields good agreement with the SM prediction across all rapidity bins. A simultaneous PDF and SMEFT fit, \eg as in Refs.~\cite{Greljo_2021,Kassabov_2023}, with the full set of input measurements included here, is not currently feasible. Developing the technical capabilities for such an approach will be an important focus in the future.

\begin{figure*}[htbp]
   	\centering
      \rotatebox{-90}{
      \begin{minipage}{\textheight}
   		\includegraphics[width=\textwidth]{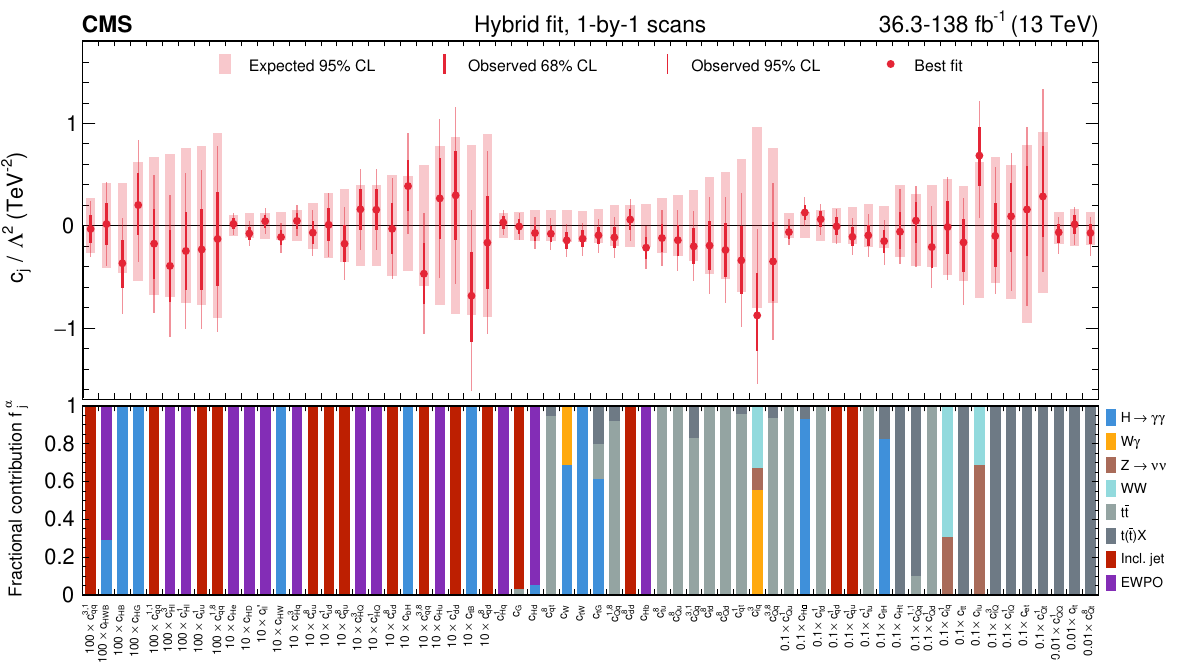}
   		\caption{Constraints on individual WCs, for the hybrid fit including the full set of measurements. The constraints for each WC are obtained keeping the other coefficients fixed to 0. The shaded areas correspond to the expected 95\% confidence intervals, the thick and thin bars to the observed 68\% and 95\% confidence intervals, respectively. The lower panel shows the contribution of different input measurements to the total constraints. The constraints are scaled by powers of 10 to ensure the constraints on all \nWCsConstr{} WCs can be visualized on the same $y$ axis scale.}\label{fig:hybrid_plus_individual}
      \end{minipage}}
\end{figure*}

Figure~\ref{fig:hybrid_plus_individual} and Tables~\ref{tab:hybrid_plus_individual_top} and~\ref{tab:hybrid_plus_individual_bottom} show the constraints on \nWCsConstr{} individual WCs, obtained when fixing all other WCs to 0. The 95\% confidence intervals on $c_j/\Lambda^2$ range from around $\pm 20\TeV^{-2}$ for the loosest constraint, \ctl, to $\pm 0.003\TeV^{-2}$ for the tightest constraint, \cjj{3}{1}. The WCs constrained mainly by the \ttbar measurement have best fit values below zero; this is because these operators generally have positive interference terms ($A_{\alpha,j}^i > 0$), and the differential cross sections measured in the input analysis are below the SM prediction (cf.~Fig.~\ref{fig:param_effect3}).

{\tolerance=800
The breakdown of contributions from the different measurements to each of the constraints is also shown. This breakdown is evaluated by considering the symmetrized 68\% confidence interval for the parameter $c_j/\Lambda^2$ evaluated with each measurement $\alpha$, $\sigma^{\alpha}_{c_j}$. The contribution of a measurement $\alpha$ to the constraint on $c_j/\Lambda^2$ is then defined as
\begin{equation}
   f^{\alpha}_j = \frac{\bigl( 1/\sigma^{\alpha}_{c_j} \bigr)^2}{\sum_{\beta}\bigl( 1/\sigma^{\beta}_{c_j} \bigr)^2}
   = \frac{H_{jj}^{\alpha}}{\sum_{\beta} H_{jj}^{\beta}} .
\end{equation}
Several operators receive significant constraints from multiple analyses. Through the combination of the \ttbar cross section measurements and the dedicated \ttX EFT analysis, we obtain stronger constraints on the two-heavy-two-light-quark coefficients. Improved constraints are found on, for example, \cqt{8}, \cqj{1}{8}, and \cqj{3}{8}, in comparison with those in Refs.~\cite{ATLAS:2022mlu,Celada:2024mcf}. The combination of the $\WGam$ and $\HtoGamGam$ data yields an improved constraint on \cw~with respect to any single-analysis result. For example, the linear-only sensitivity is approximately 45\% higher with respect to the $\WGam$ result of Ref.~\cite{CMS:2021cxr}.
For several operators, the inclusion of the EWPO measurements provides significantly stronger constraints than would be obtained from the CMS data alone, for example, on \chl{3}, \chq{3}, \chdd{}, and \cllprime.
In comparison with the global combinations from Refs.~\cite{Ellis_2021,bartocci2023global}, the combination presented here constrains a larger number of WCs simultaneously. However, as a result of differences in the input data sets and EFT scheme choices, it is not possible to perform a direct comparison of these constraints.
\par}

\begin{figure*}[bt]
   \centering
   \includegraphics[width=\textwidth]{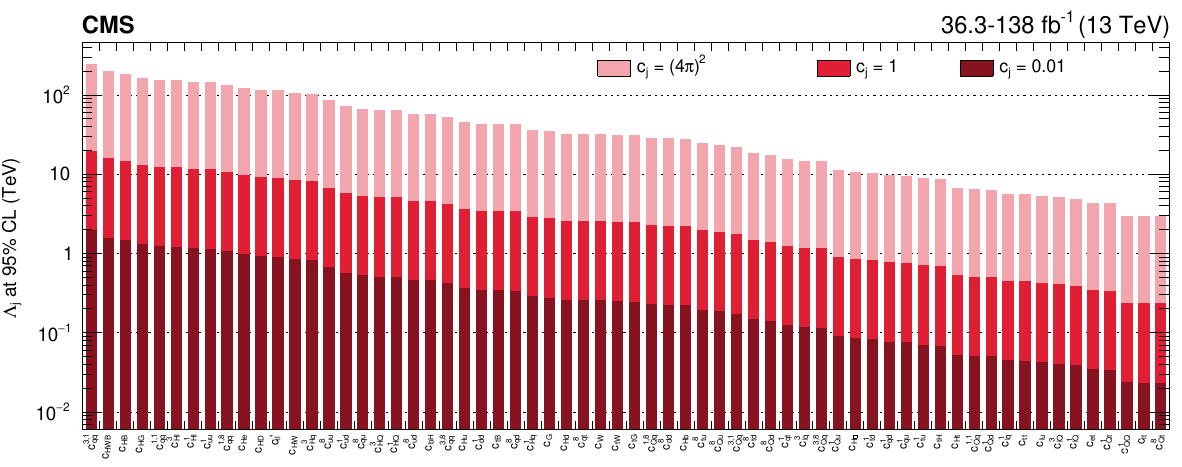}
   \caption{The 95\% \CL lower limits on the scales $\Lambda_j$ for the indicated values of the WCs $c_j$.}\label{fig:hybrid_plus_npscale_individual}
\end{figure*}

We also translate the obtained constraints into 95\% confidence level (\CL) lower limits on the scale of BSM physics $\Lambda_j$, as shown in Fig.~\ref{fig:hybrid_plus_npscale_individual}. These lower limits are evaluated from the expected constraints on $c_j/\Lambda^2$. By setting $c_j$ to specific values, the expected constraint is converted to a lower limit on the energy scale, for given values of the WC in question.
The WC can take a broad range of possible values, depending on whether it arises from weakly- or strongly-coupled BSM physics, at tree level or at loop level. For the loop expansion to converge, the condition $c_j < (4\pi)^2$ must be satisfied~\cite{Degrande_2013}. In line with the conventions used in other EFT interpretations, such as in Ref.~\cite{Ellis_2021}, we display the lower limits on the scale $\Lambda_j$ for $c_j = 0.01$, 1, and $(4\pi)^2$.

The WCs can be matched to parameters in UV-complete BSM models, allowing the constraints on $c_j/\Lambda^2$ to be interpreted as limits on these parameters. Ref.~\cite{de_Blas_2018} provides a general overview of the BSM models that can be constrained by these WCs.

\begin{figure*}[htbp]
   	\centering
      \rotatebox{-90}{
      \begin{minipage}{\textheight}
   		\includegraphics[width=\textwidth]{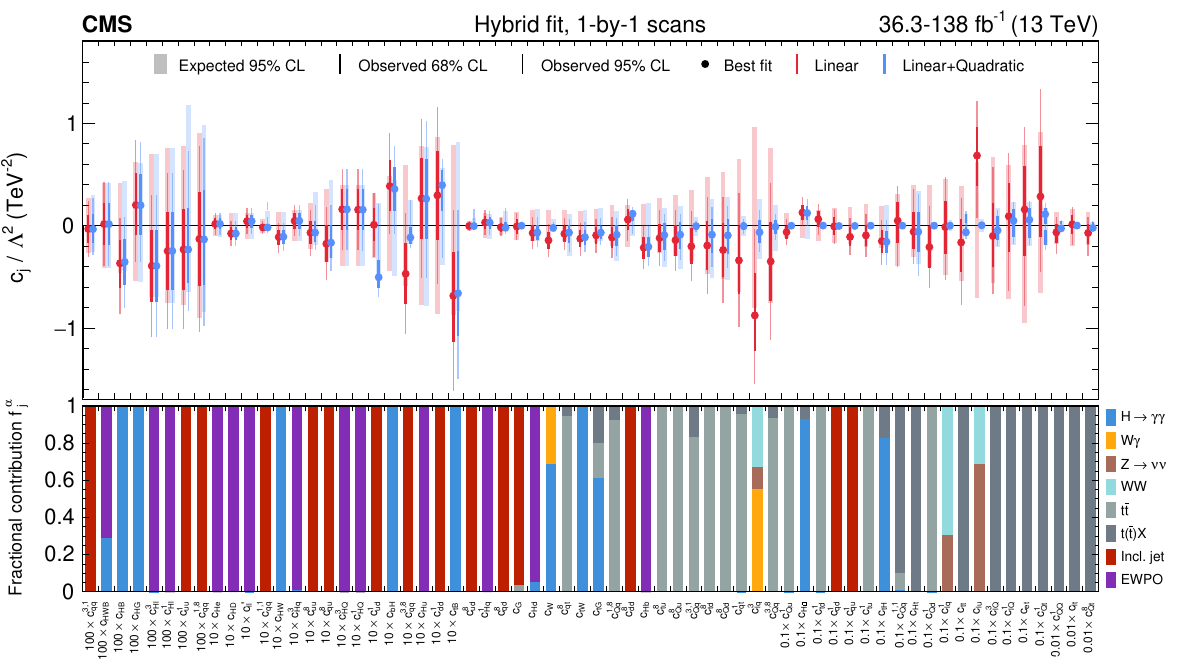}
   		\caption{Constraints on individual WCs, showing both the constraints considering only linear terms in the SMEFT parameterization and those considering both linear and quadratic terms. The constraints for each WC are obtained keeping the other coefficients fixed to 0. The shaded areas correspond to the expected 95\% confidence intervals, the thick and thin bars to the observed 68\% and 95\% confidence intervals, respectively. The constraints are scaled by powers of 10 to ensure the constraints on all \nWCsConstr{} WCs can be visualized on the same $y$ axis scale.}\label{fig:hybrid_plus_linplusquad}
      \end{minipage}}
\end{figure*}

{\tolerance=800
In Fig.~\ref{fig:hybrid_plus_linplusquad}, the constraints on the individual WCs when using the parameterizations up to linear order ($\propto\Lambda^{-2}$) are compared with those using parameterizations up to quadratic order ($\propto\Lambda^{-4}$). For some of the WCs with the loosest constraints, including the quadratic terms significantly improves these constraints. This shows that for these operators, contributions at $\mathcal{O}(\Lambda^{-4})$ are important. That is, the sensitivity to these operators is such that the BSM contributions dominate. For the WCs that are more tightly constrained, the addition of the quadratic terms generally only leads to small changes. This indicates that in these cases terms of the order of $\Lambda^{-2}$, \ie those corresponding to the SM-BSM interference, dominate the sensitivity. It has also been observed~\cite{Bernlochner:2022oiw} that confidence intervals derived using the asymptotic approximation, as used here, may over- or under-cover when quadratic terms are included in the EFT parameterization. The coverage has been verified for the corresponding fits with the simplified likelihood of Appendix~\ref{sec:simplified}, using pseudodata samples to construct the test statistic distributions, and is found to be in reasonable agreement with the target coverage for all parameters. The studies performed are described in more detail in Appendix~\ref{sec:coverage_tests}.
\par}

\section{Summary}\label{sec:summary}
A standard model effective field theory (SMEFT) interpretation of data collected by the CMS experiment has been presented. This combined interpretation is based on a simultaneous fit of seven sets of CMS measurements that probe Higgs boson, electroweak vector boson, top quark, and multijet production, and also incorporates measurements of electroweak precision observables.
These input measurements were chosen to obtain sensitivity to a broad set of SMEFT operators. Out of 129 operators in the SMEFT basis considered in this paper, the combined interpretation constrains \nWCsConstr{} Wilson coefficients (WCs) individually. The constraints are provided for both linear-only and linear-plus-quadratic parameterizations. Simultaneous constraints are set on \nEVsConstr{} linear combinations of WCs.
In the fit that constrains the linear combinations of WCs, the $p$-value for the compatibility with the standard model is 2.5\%. When excluding the inclusive jet measurement from the combination, the $p$-value is 27\%.
The 95\% confidence intervals range from around $\pm0.002$ to $\pm10\TeV^{-2}$ for the constraints on the linear combinations of WCs, whereas for the individual WCs the constraints range from $\pm0.003$ to $\pm 20\TeV^{-2}$. These constraints are also translated into lower limits on the probed energy scale of new physics $\Lambda$, for given values of the WCs. This combined interpretation yields improved constraints with respect to single-analysis results from CMS.

\begin{acknowledgments}
   We congratulate our colleagues in the CERN accelerator departments for the excellent performance of the LHC and thank the technical and administrative staffs at CERN and at other CMS institutes for their contributions to the success of the CMS effort. In addition, we gratefully acknowledge the computing centres and personnel of the Worldwide LHC Computing Grid and other centres for delivering so effectively the computing infrastructure essential to our analyses. Finally, we acknowledge the enduring support for the construction and operation of the LHC, the CMS detector, and the supporting computing infrastructure provided by the following funding agencies: SC (Armenia), BMBWF and FWF (Austria); FNRS and FWO (Belgium); CNPq, CAPES, FAPERJ, FAPERGS, and FAPESP (Brazil); MES and BNSF (Bulgaria); CERN; CAS, MoST, and NSFC (China); MINCIENCIAS (Colombia); MSES and CSF (Croatia); RIF (Cyprus); SENESCYT (Ecuador); ERC PRG, RVTT3 and MoER TK202 (Estonia); Academy of Finland, MEC, and HIP (Finland); CEA and CNRS/IN2P3 (France); SRNSF (Georgia); BMBF, DFG, and HGF (Germany); GSRI (Greece); NKFIH (Hungary); DAE and DST (India); IPM (Iran); SFI (Ireland); INFN (Italy); MSIP and NRF (Republic of Korea); MES (Latvia); LMTLT (Lithuania); MOE and UM (Malaysia); BUAP, CINVESTAV, CONACYT, LNS, SEP, and UASLP-FAI (Mexico); MOS (Montenegro); MBIE (New Zealand); PAEC (Pakistan); MES and NSC (Poland); FCT (Portugal); MESTD (Serbia); MICIU/AEI and PCTI (Spain); MOSTR (Sri Lanka); Swiss Funding Agencies (Switzerland); MST (Taipei); MHESI and NSTDA (Thailand); TUBITAK and TENMAK (Turkey); NASU (Ukraine); STFC (United Kingdom); DOE and NSF (USA).
   
   \hyphenation{Rachada-pisek} Individuals have received support from the Marie-Curie programme and the European Research Council and Horizon 2020 Grant, contract Nos.\ 675440, 724704, 752730, 758316, 765710, 824093, 101115353, 101002207, and COST Action CA16108 (European Union); the Leventis Foundation; the Alfred P.\ Sloan Foundation; the Alexander von Humboldt Foundation; the Science Committee, project no. 22rl-037 (Armenia); the Fonds pour la Formation \`a la Recherche dans l'Industrie et dans l'Agriculture (FRIA-Belgium); the Beijing Municipal Science \& Technology Commission, No. Z191100007219010 and Fundamental Research Funds for the Central Universities (China); the Ministry of Education, Youth and Sports (MEYS) of the Czech Republic; the Shota Rustaveli National Science Foundation, grant FR-22-985 (Georgia); the Deutsche Forschungsgemeinschaft (DFG), among others, under Germany's Excellence Strategy -- EXC 2121 ``Quantum Universe" -- 390833306, and under project number 400140256 - GRK2497; the Hellenic Foundation for Research and Innovation (HFRI), Project Number 2288 (Greece); the Hungarian Academy of Sciences, the New National Excellence Program - \'UNKP, the NKFIH research grants K 131991, K 133046, K 138136, K 143460, K 143477, K 146913, K 146914, K 147048, 2020-2.2.1-ED-2021-00181, TKP2021-NKTA-64, and 2021-4.1.2-NEMZ\_KI-2024-00036 (Hungary); the Council of Science and Industrial Research, India; ICSC -- National Research Centre for High Performance Computing, Big Data and Quantum Computing and FAIR -- Future Artificial Intelligence Research, funded by the NextGenerationEU program (Italy); the Latvian Council of Science; the Ministry of Education and Science, project no. 2022/WK/14, and the National Science Center, contracts Opus 2021/41/B/ST2/01369 and 2021/43/B/ST2/01552 (Poland); the Funda\c{c}\~ao para a Ci\^encia e a Tecnologia, grant CEECIND/01334/2018 (Portugal); the National Priorities Research Program by Qatar National Research Fund; MICIU/AEI/10.13039/501100011033, ERDF/EU, "European Union NextGenerationEU/PRTR", and Programa Severo Ochoa del Principado de Asturias (Spain); the Chulalongkorn Academic into Its 2nd Century Project Advancement Project, and the National Science, Research and Innovation Fund via the Program Management Unit for Human Resources \& Institutional Development, Research and Innovation, grant B39G670016 (Thailand); the Kavli Foundation; the Nvidia Corporation; the SuperMicro Corporation; the Welch Foundation, contract C-1845; and the Weston Havens Foundation (USA).
\end{acknowledgments}

\bibliography{auto_generated}

\clearpage

\appendix
\numberwithin{figure}{section}
\numberwithin{table}{section}

\section{Validation of the simplified likelihood approach}\label{sec:validation_simplified}

To verify the consistency of the simplified likelihood approach with the experimental likelihood approach, we present results with a reduced set of input measurements, once incorporated into the simplified likelihood of Eq.~(\ref{eqn:likelihood_simpl}), $\mathcal{L}^\text{simpl}$, and once into the experimental likelihood of Eq.~(\ref{eqn:likelihood_expt}), $\mathcal{L}^\text{expt}$. The input measurements used in this validation are \HtoGamGam, \WGam, \ZtoNuNu, and \WW, the only four measurements that can be implemented in either configuration.

\begin{figure*}[hpb]
   	\centering
   	\includegraphics[width=\textwidth]{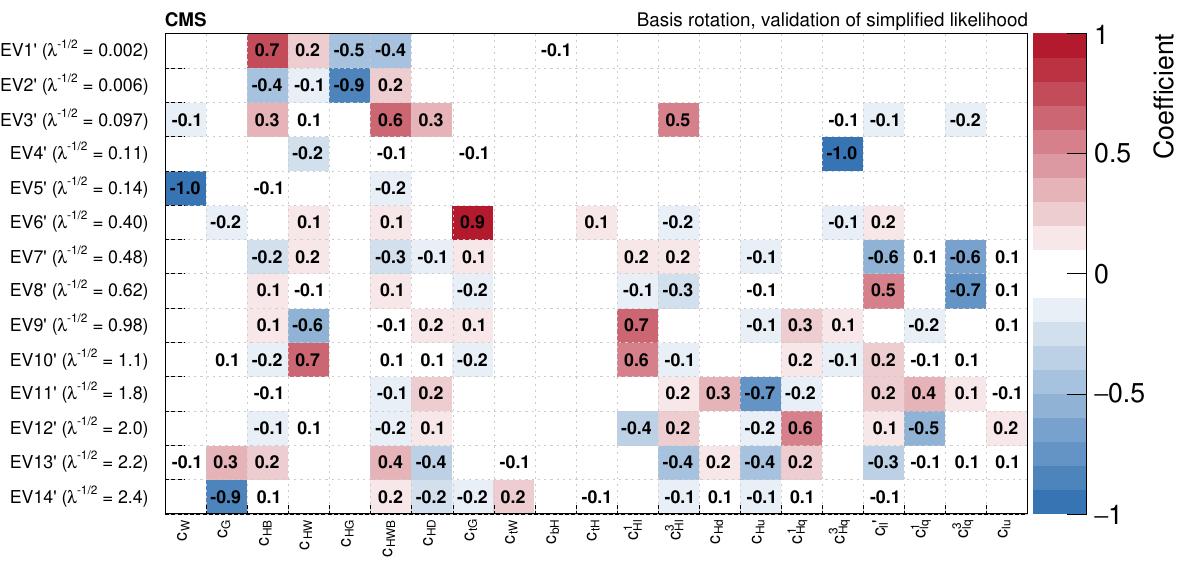}
   	\caption{Rotation matrix obtained by performing the PCA on the Hessian matrix of a reduced set of input measurements, \HtoGamGam, \WGam, \ZtoNuNu, and $\WW$. Only matrix coefficients with absolute value $\geq$0.05 are displayed.}
   	\label{fig:PCA_EW_H}
\end{figure*}

\begin{figure*}[hp]
   	\centering
   	\includegraphics[width=0.93\textwidth]{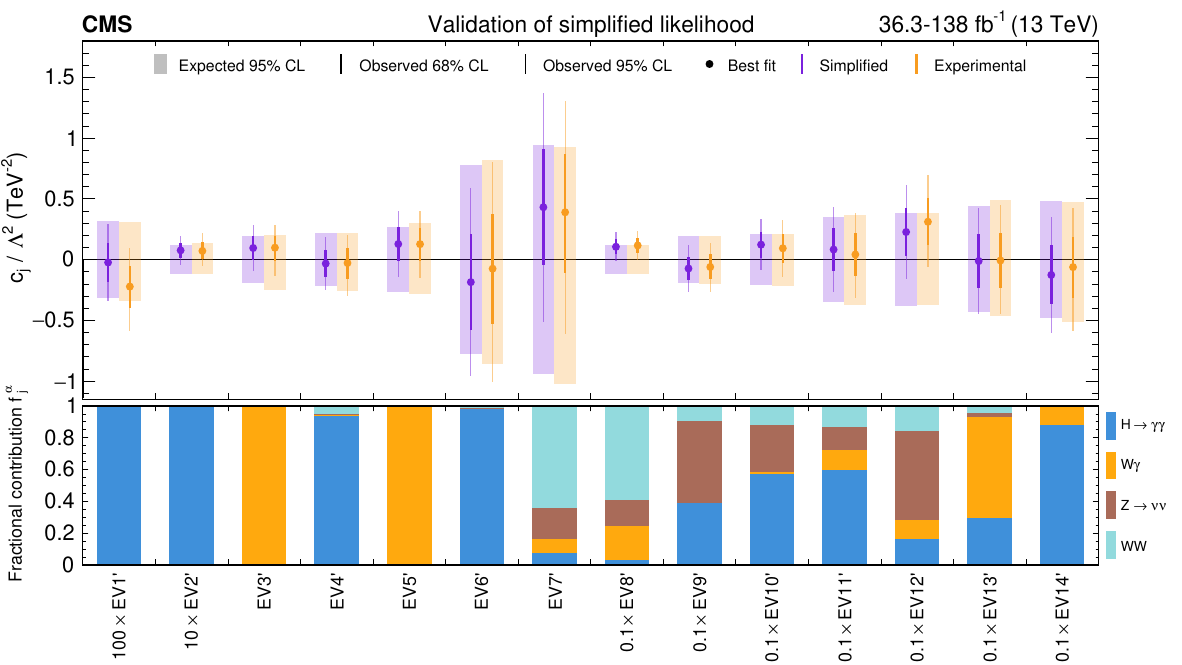}
   	\caption{Constraints on linear combinations of WCs, using a reduced set of input measurements, \HtoGamGam, \WGam, \ZtoNuNu, and \WW. The shaded areas correspond to the expected 95\% confidence intervals, the thick and thin bars to the observed 68\% and 95\% confidence intervals, respectively. The lower panel shows the contribution of different input measurements to the total constraints. The constraints are scaled by powers of 10 to ensure the constraints on all eigenvectors can be visualized on the same $y$ axis scale.}
   	\label{fig:results_EW_H_rotated}
\end{figure*}

\begin{figure*}[hp]
   	\centering
   	\includegraphics[width=0.93\textwidth]{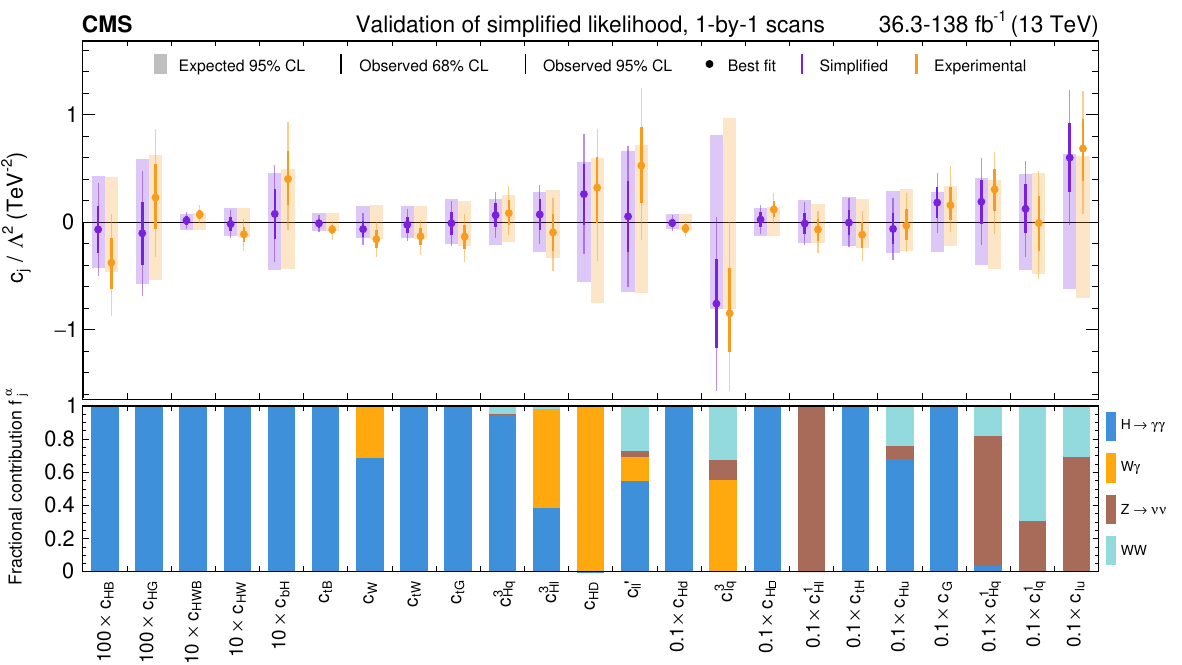}
   	\caption{Constraints on individual WCs, using a reduced set of input measurements, \HtoGamGam, \WGam, \ZtoNuNu, and \WW. The constraints for each WC are obtained keeping the other coefficients fixed to 0. The shaded areas correspond to the expected 95\% confidence intervals, the thick and thin bars to the observed 68\% and 95\% confidence intervals, respectively. The lower panel shows the contribution of different input measurements to the total constraints. The constraints are scaled by powers of 10 to ensure the constraints on all WCs can be visualized on the same $y$ axis scale.}
   	\label{fig:results_EW_H_individual}
\end{figure*}

The rotation matrix obtained by performing the PCA on this reduced set of input measurements is shown in Fig.~\ref{fig:PCA_EW_H}. In total, 14 eigenvectors are retained, the remaining 9 are fixed to their SM value (0). Constraints on these linear combinations of the WCs are shown in Fig.~\ref{fig:results_EW_H_rotated}, with constraints on the individual WCs shown in Fig.~\ref{fig:results_EW_H_individual}.

We observe that the differences in the constraints on the parameters of interest are typically small. The largest differences are observed for parameters constrained by the \HtoGamGam measurement, where the simplified likelihood does not approximate the experimental one as well as in the other analyses. Nonetheless, the simplified likelihood is a good approximation of the experimental likelihood in cases where the experimental likelihood model is not available, or when a computationally light, approximate model is more convenient.

\ifthenelse{\boolean{cms@external}}{}{\clearpage}

\section{Simplified likelihood fit}\label{sec:simplified}

Results are presented in an additional configuration that incorporates all input measurements, apart from the \ttX measurement, into the simplified likelihood of Eq.~(\ref{eqn:likelihood_simpl}), $\mathcal{L}^\text{simpl}$. This simplified likelihood model, based on a multivariate Gaussian approximation, is computationally light, which makes it suitable to perform additional studies. In Appendix~\ref{sec:coverage_tests}, it is used to study the coverage of confidence intervals derived using the asymptotic approximation when $\mathcal{O}(c_j^2/\Lambda^4)$ terms are included in the SMEFT parameterization. All information needed to construct the simplified likelihood model is provided in the HEPData record for this analysis~\cite{HEPData}.

In the simplified likelihood model, the Hessian matrix $H$ of the combined measurement is constructed directly from the block-diagonal covariance matrix, $V = V_\text{ex} + V_\text{th}$, and the SMEFT parameterization, as
\begin{equation}
\label{eqn:hessian_simplified}
	H = \mathcal{A}^T V^{-1} \mathcal{A} ,
\end{equation}
where the entries of the matrix $\mathcal{A}$ are the linear terms $A_{\alpha,j}^i$ of the SMEFT parameterization of Eq.~(\ref{eqn:crosssec_EFT}), with the rows running over the measurements bins and the columns over the WCs.

\begin{figure*}[hp]
   \centering
   \rotatebox{-90}{
   \begin{minipage}{\textheight}
   		\includegraphics[width=\textwidth]{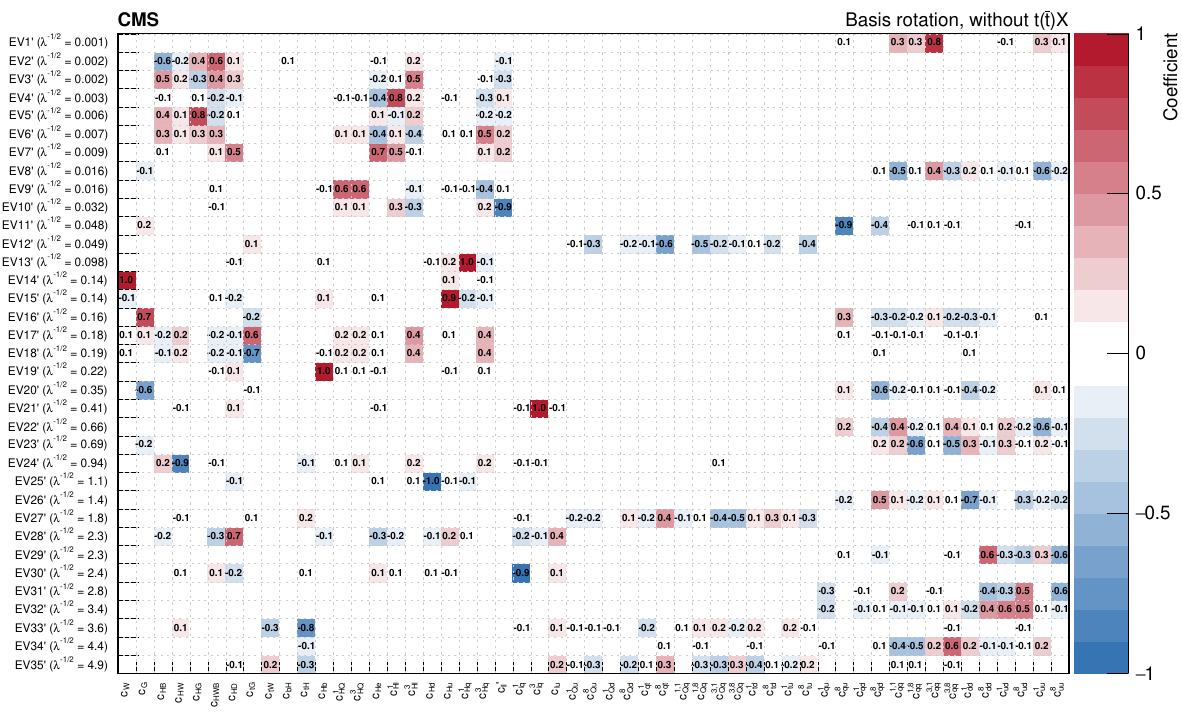}
   		\caption{Rotation matrix obtained by performing the PCA on the Hessian matrix of a reduced set of measurements, excluding the \ttX measurement. Only matrix coefficients with absolute value $\geq$0.05 are displayed.}\label{fig:PCA_nottX}
   \end{minipage}}
\end{figure*}

The matrix $H$ is diagonalized to obtain a set of linear combinations of WCs, $\mathrm{EV}_j'$, that are orthogonal to each other and can be constrained in a simultaneous fit:
\begin{equation}
\label{eqn:pca_simplified}
	H = \mathcal{R}^T \Lambda \mathcal{R} , \quad \mathrm{EV}_j' = \mathcal{R}^{\,jk} c_k .
\end{equation}
The rotation matrix we obtain is shown in Fig.~\ref{fig:PCA_nottX}. There are 35 eigenvectors to which the analysis is deemed sensitive enough for them to be incorporated in the combined fit.

\begin{figure*}[hp]
   	\centering
      \rotatebox{-90}{
      \begin{minipage}{\textheight}
   		\includegraphics[width=\textwidth]{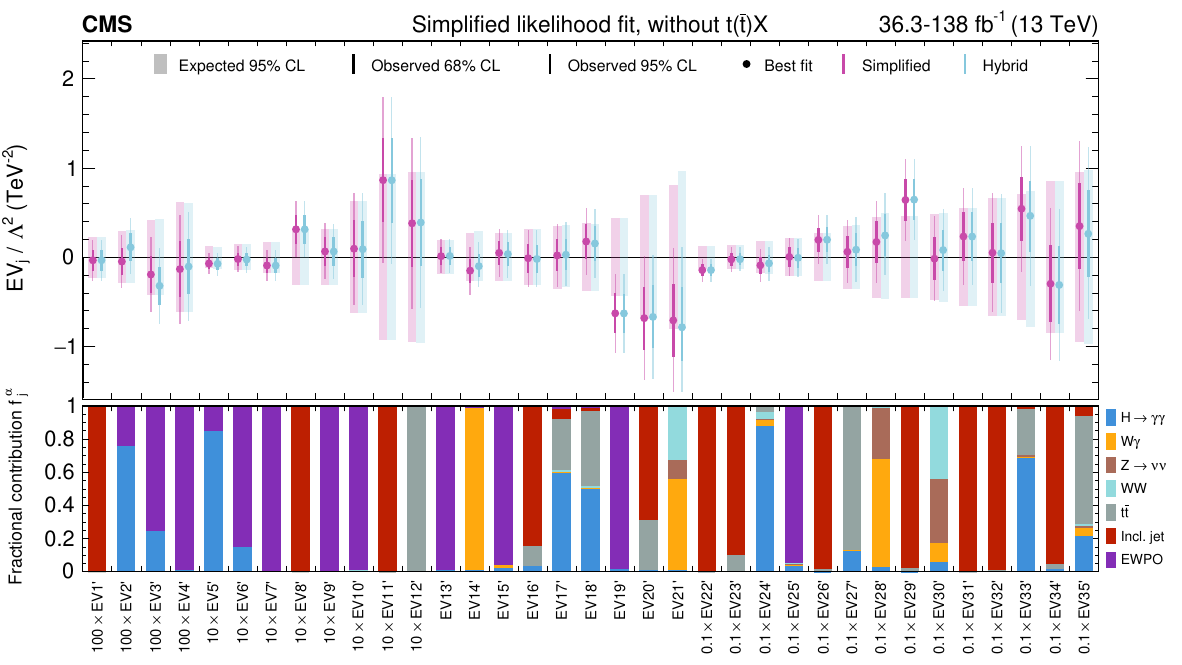}
   		\caption{Constraints on linear combinations of WCs, from the simplified and hybrid likelihood fits, excluding the \ttX analysis. The shaded areas correspond to the expected 95\% confidence intervals, the thick and thin bars to the observed 68\% and 95\% confidence intervals, respectively. The lower panel shows the contribution of different input measurements to the total constraints. The constraints are scaled by powers of 10 to ensure the constraints on all eigenvectors can be visualized on the same $y$ axis scale.}\label{fig:results_nottX}
      \end{minipage}}
\end{figure*}

\begin{figure*}[hp]
   	\centering
      \rotatebox{-90}{
      \begin{minipage}{\textheight}
   		\includegraphics[width=\textwidth]{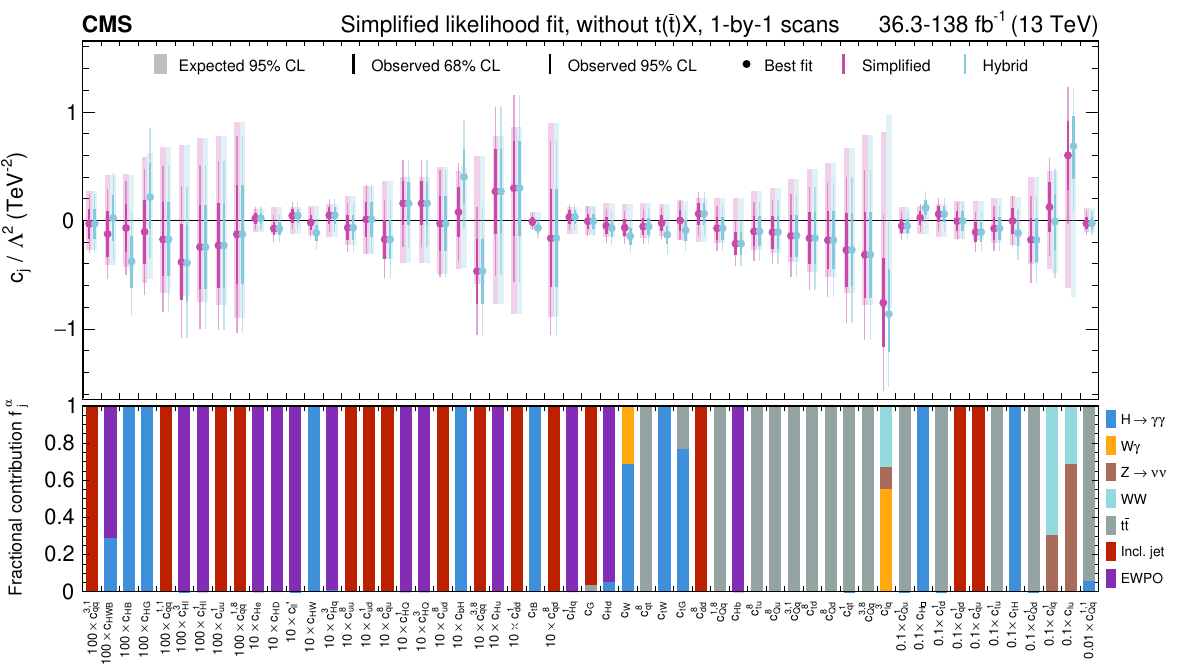}
   		\caption{Constraints on individual WCs, from the simplified and hybrid likelihood fits, excluding the \ttX analysis. The constraints for each WC are obtained keeping the other coefficients fixed to 0. The shaded areas correspond to the expected 95\% confidence intervals, the thick and thin bars to the observed 68\% and 95\% confidence intervals, respectively. The lower panel shows the contribution of different input measurements to the total constraints. The constraints are scaled by powers of 10 to ensure the constraints on all WCs can be visualized on the same $y$ axis scale.}\label{fig:results_nottX_individual}
      \end{minipage}}
\end{figure*}

The constraints on linear combinations of the WCs are shown in Fig.~\ref{fig:results_nottX}, with the constraints on the individual WCs shown in Fig.~\ref{fig:results_nottX_individual}. The constraints, in terms of the 95\% confidence interval, range from $\pm10\TeV^{-2}$ for the most loosely constrained eigenvector, to $\pm0.002\TeV^{-2}$ for the tightest constraint. For the individual WCs, the loosest constraint is on $\cqj{1}{1}/\Lambda^2$; the 95\% confidence interval is around $\pm 11\TeV^{-2}$. The tightest constraint, with a 95\% confidence interval of around $\pm 0.003\TeV^{-2}$, is found on $\cjj{3}{1}/\Lambda^2$.

The $p$-value for the compatibility with the SM (all WCs equal to 0) is 5.2\%. When excluding the inclusive jet measurement from the combination, the $p$-value is 54\%.

\ifthenelse{\boolean{cms@external}}{}{\clearpage}

\section{Coverage with quadratic terms in the parameterization}\label{sec:coverage_tests}
It has been shown in Ref.~\cite{Bernlochner:2022oiw} that, in the case where quadratic terms are included in the EFT parameterization, the confidence intervals that are constructed using the asymptotic approximation may not have the specified frequentist coverage. The impact of this effect on the analysis presented here has been studied using the simplified likelihood of Appendix.~\ref{sec:simplified}. Only the coverage of the constraints on the individual WCs was investigated. As the simplified likelihood, in which the \ttX analysis is not incorporated, was employed for this study, the WCs that are only constrained by the \ttX analysis are omitted.

This study is performed by generating pseudo-experiments and evaluating the test statistic (Eq.~(\ref{eqn:plr})) for each pseudodata set. Based on the distribution of test statistic values that is obtained, critical values $q^{68\%}$ and $q^{95\%}$ are extracted. These are defined for each value of $c_\text{inj}$ such that
\begin{equation}
   \begin{aligned}
   &\int_{q^{68\%}}^{+\infty} f(q_{c_{\text{inj}}}|c_{\text{inj}})dq_{c_{\text{inj}}} = 0.32, \\
   &\int_{q^{95\%}}^{+\infty} f(q_{c_{\text{inj}}}|c_{\text{inj}})dq_{c_{\text{inj}}} = 0.05,
   \end{aligned}
\end{equation}
where $f(q_{c_{\text{inj}}}|c_{\text{inj}})$ represents the sampling distribution of the test statistic, at the value $c_{\text{inj}}$ for which the pseudodata set is generated.

This procedure is repeated for a range of values of each WC, keeping the other WCs fixed to 0. The range of values for each WC is chosen to be wider than the asymptotically determined observed 95\% confidence intervals. This ensures both the 68\% and 95\% confidence intervals can be extracted by finding the intercept of the likelihood scan with the curve obtained by interpolating between the sets of points $q^{68\%}$ and $q^{95\%}$, even in the case where the asymptotic intervals under-cover. In all cases, 10000 pseudo-experiments per WC value are generated, and the model is fit to each pseudodata set. The test statistic value $q(c_{\text{inj}})$ is then evaluated.

\begin{figure*}[hp]
   \centering
   \includegraphics[width=0.48\textwidth]{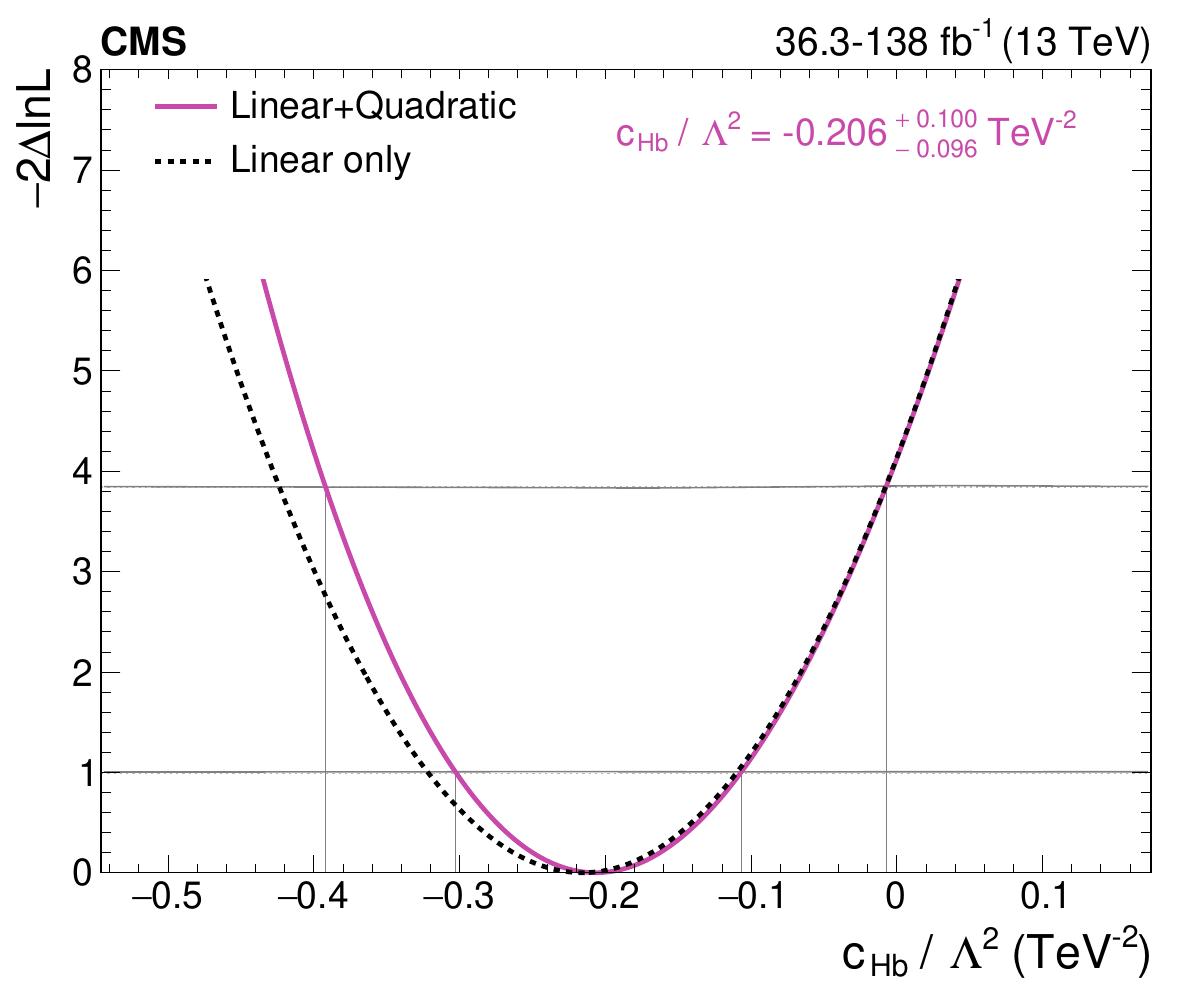}
   \includegraphics[width=0.48\textwidth]{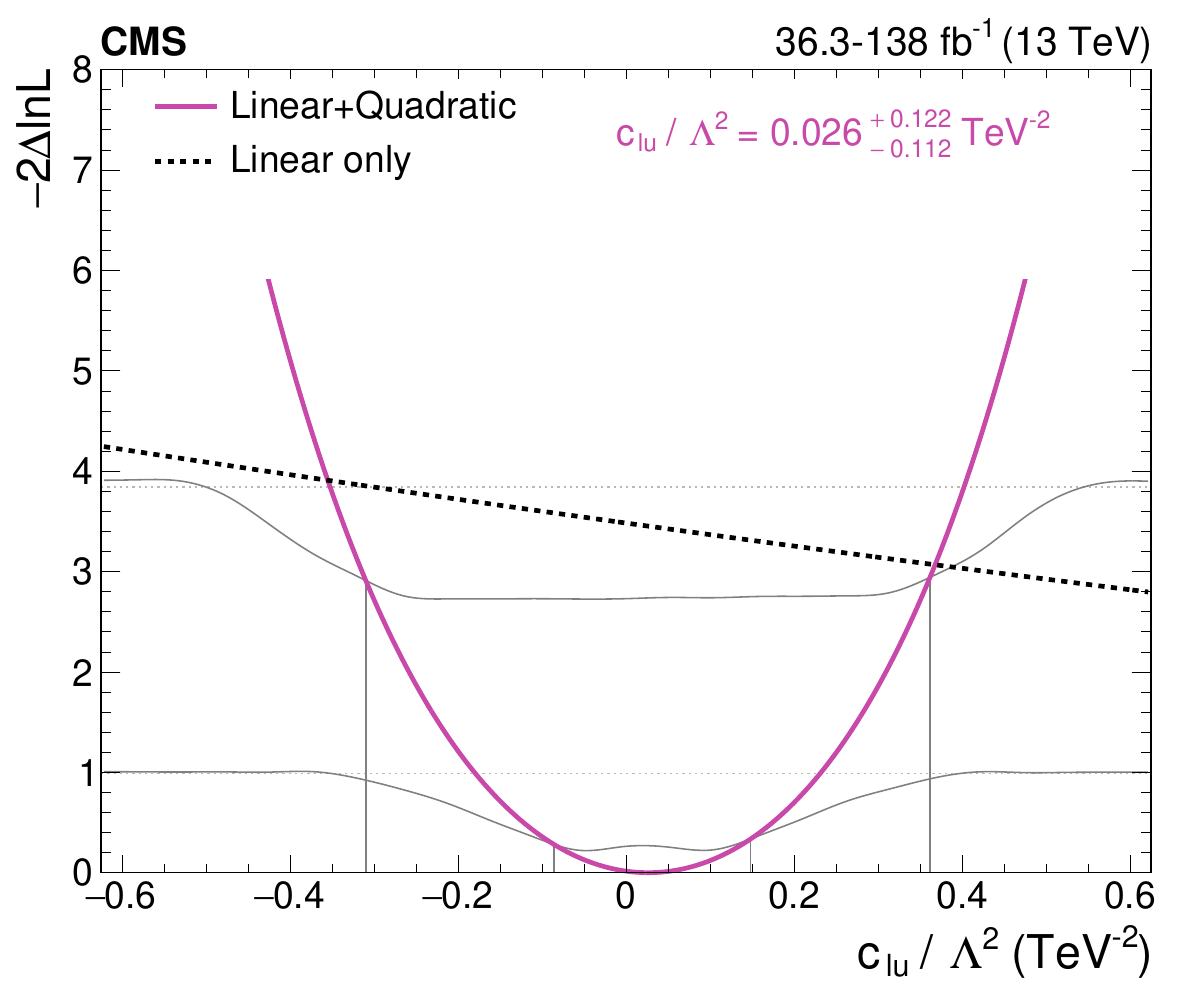}~\\
   \includegraphics[width=0.48\textwidth]{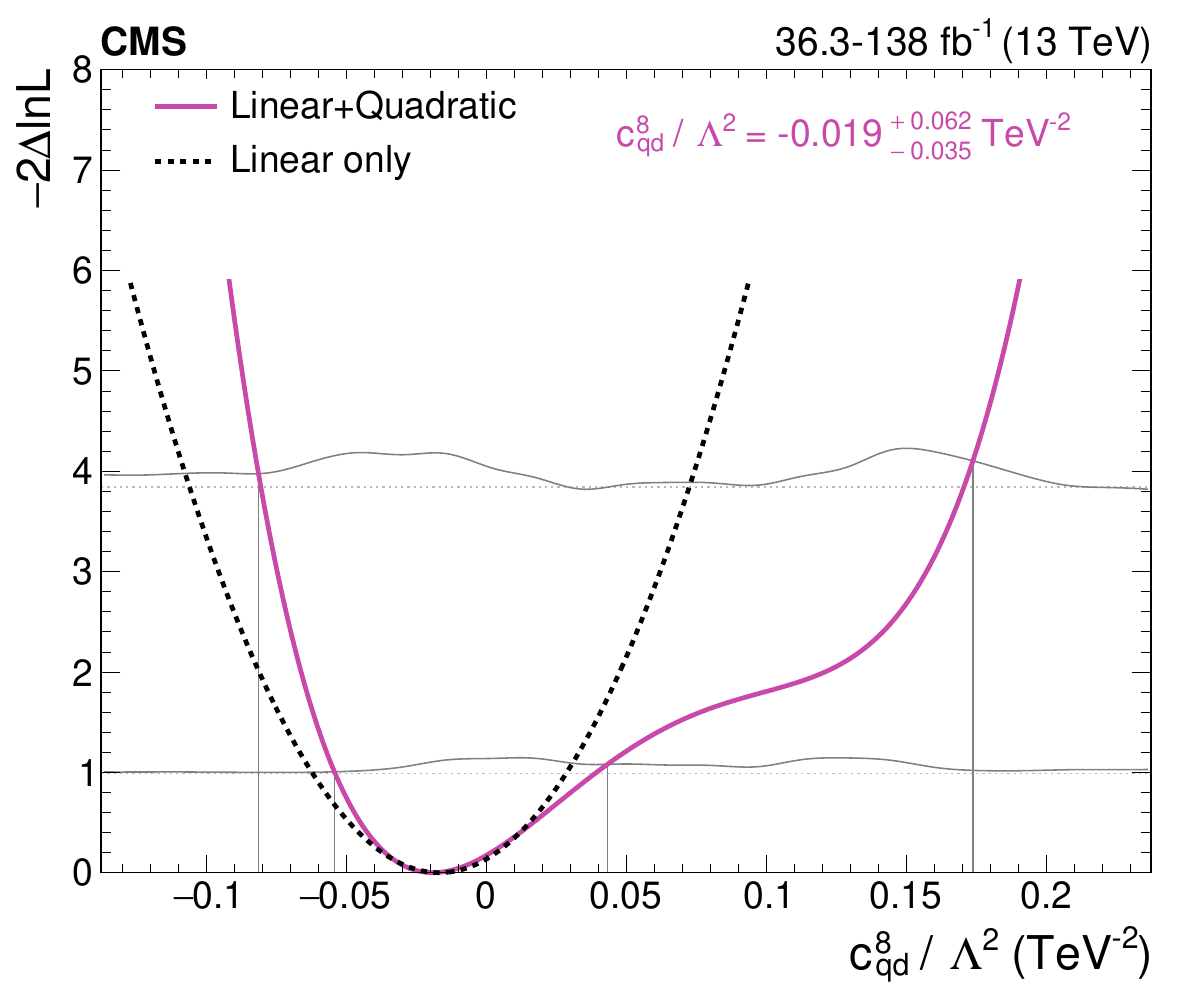}
   \caption{Observed likelihood scans for $\chbq$, where linear terms dominate (upper left); $\clu$, where quadratic terms dominate (upper right); and $\cjd{8}$, where quadratic and linear terms both contribute (lower). The results with quadratic terms included in the parameterization (solid purple line) and with linear terms only (dashed black line) are shown. The solid grey lines indicate the sets of points $q^{68\%}$ (lower line) and $q^{95\%}$ (upper line), whereas the dashed grey lines denote the test statistic values used to determine the 68\% and 95\% confidence intervals in the asymptotic approximation. The confidence intervals shown on the figures are the 68\% confidence intervals extracted from the intersection of the linear-plus-quadratic curve with the $q^{68\%}$ line.}\label{fig:scans_coverage}
\end{figure*}

{\tolerance=800
Likelihood scans, with $q^{68\%}$, $q^{95\%}$, their interpolations, and the extracted intervals, are shown for three WCs in Fig.~\ref{fig:scans_coverage}. In all cases, likelihood scans are shown for the case where quadratic terms are included in the parameterization, as well as for the linear-only parameterization.
In Fig.~\ref{fig:scans_coverage} (upper left), this result is shown for a WC for which the linear terms are dominant. In this case, the lines traced by the points $q^{68\%}$ and $q^{95\%}$ converge to the asymptotically used values, and the asymptotic intervals cover. Figure~\ref{fig:scans_coverage} (upper right) shows the likelihood for an example of a WC for which the quadratic terms dominate. The interpolations of the points $q^{68\%}$ and $q^{95\%}$ do not yield a straight line, and in this case the asymptotically extracted intervals over-cover. Finally, Fig.~\ref{fig:scans_coverage} (lower) shows the likelihood for a case where the linear and quadratic terms have similar-sized contributions. In this case, the asymptotic interval under-covers by a small amount.
\par}

\begin{figure*}[hp]
   	\centering
      \rotatebox{-90}{
      \begin{minipage}{\textheight}
   		\includegraphics[width=\textwidth]{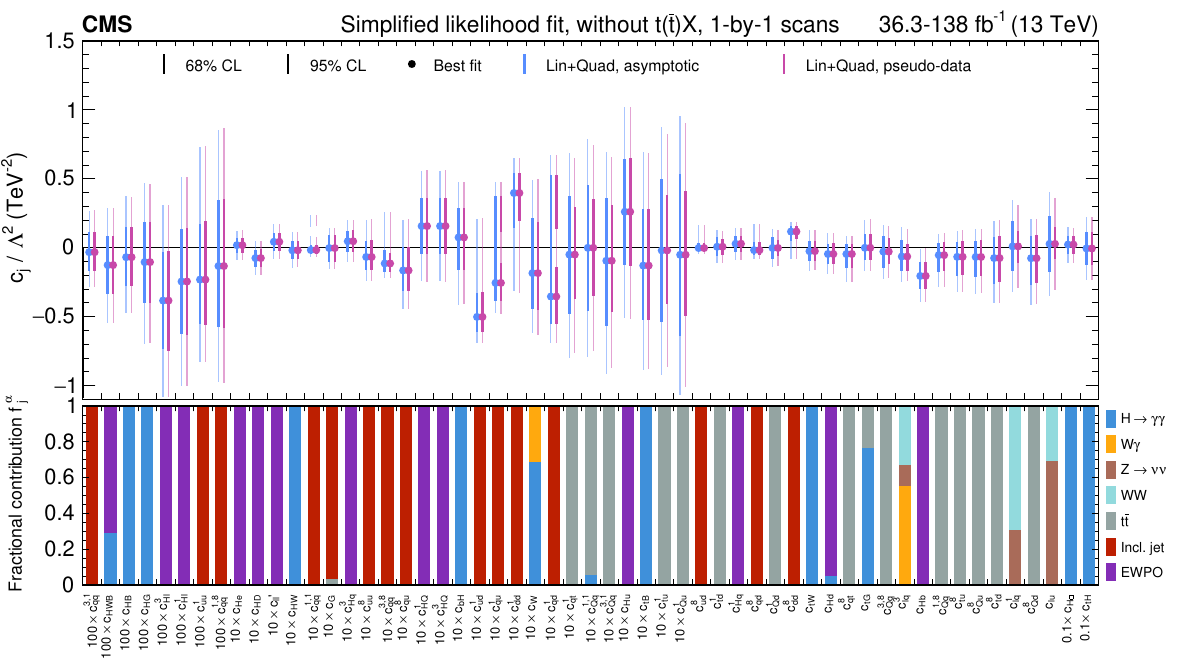}
   		\caption{Summary of best fit values and confidence intervals extracted with the asymptotic approximation (blue lines) and with the pseudo-experiment-based method described in this appendix (purple lines). The constraints are scaled by powers of 10 to ensure the constraints on all WCs can be visualized on the same $y$ axis scale. The intervals are generally compatible with each other, with only some small differences visible.}\label{fig:coverage_summary}
      \end{minipage}}
\end{figure*}

For 26 out of the 55 WCs that we study, the asymptotic approximation is valid and the coverage of the 68\% and 95\% confidence intervals is correct. For 9 WCs there is slight under-coverage, for 13 WCs there is slight over-coverage. In the remaining cases, the interval extracted using $q^{68\%}$ and $q^{95\%}$ would be marginally wider on one side of the best fit value, and narrower on the other side. A summary of the confidence intervals extracted with the asymptotic approximation, and using the method described in this appendix, is shown in Fig.~\ref{fig:coverage_summary}. In the cases where the asymptotic approximation is not valid, only small differences with the intervals computed using pseudo-experiments are observed.

\section{Relative effect of linear combinations of Wilson coefficients}\label
{sec:ev_impacts}

\begin{figure*}[hp]
   	\centering
      \rotatebox{-90}{
      \begin{minipage}{\textheight}
   		\includegraphics[width=\textwidth]{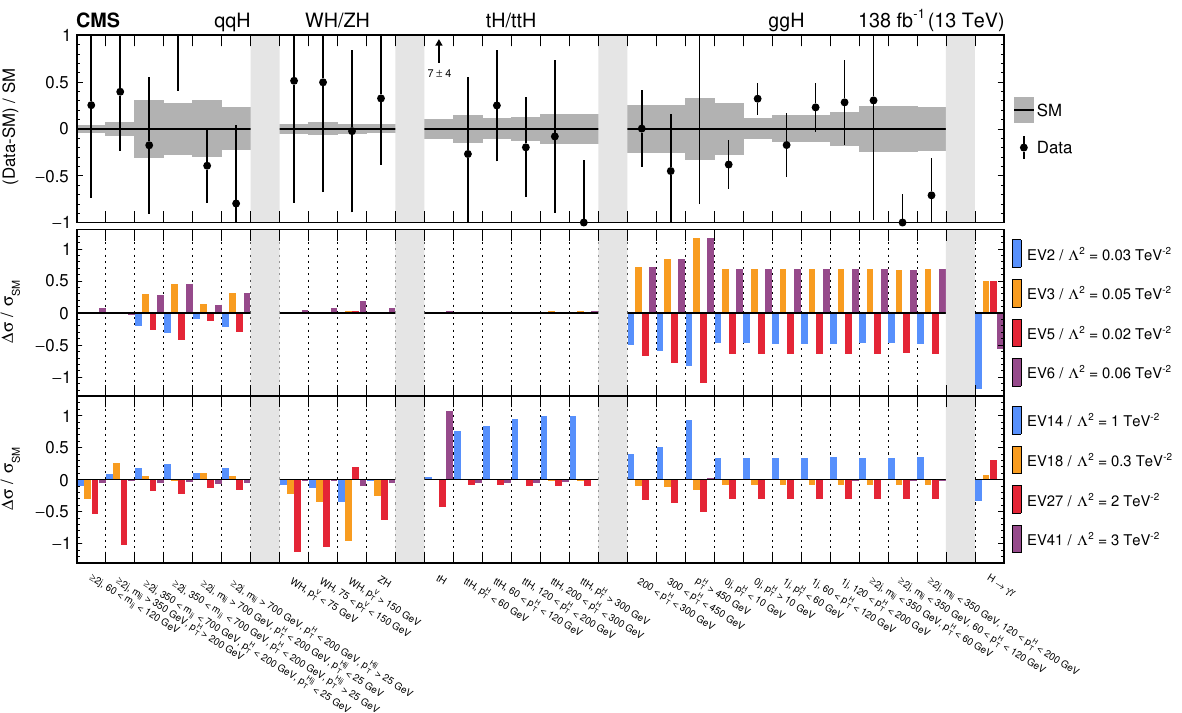}
   		\caption{Relative effect of the linear combinations of WCs that affect the Higgs STXS cross sections and the \HtoGamGam branching fraction. The parameters $\mathrm{EV}_j/\Lambda^2$ are set to different values to ensure the effect of all linear combinations can be visualized on the same $y$ axis scale. The upper panel shows the measured values and their uncertainties relative to the predictions in the SM. As these are measurements of the cross sections times branching fraction, no measurement is displayed in the rightmost bin (labelled ``\HtoGamGam'').}\label{fig:param_effect_pca1}
      \end{minipage}}
\end{figure*}

\begin{figure*}[hp]
   	\centering
      \rotatebox{-90}{
      \begin{minipage}{\textheight}
   		\includegraphics[width=\textwidth]{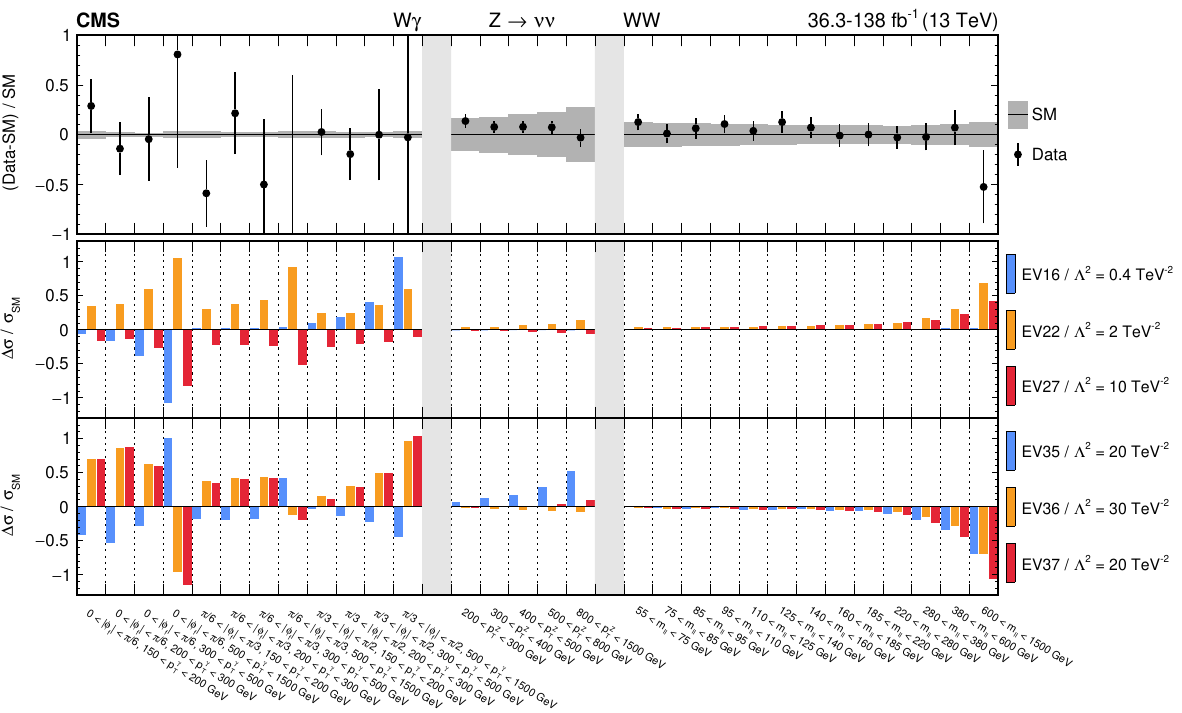}
   		\caption{Relative effect of the linear combinations of WCs that affect the $\WGam$, $\ZtoNuNu$, and $\WW$ differential cross sections. The parameters $\mathrm{EV}_j/\Lambda^2$ are set to different values to ensure the effect of all linear combinations can be visualized on the same $y$ axis scale. The upper panel shows the measured values and their uncertainties relative to the predictions in the SM.}\label{fig:param_effect_pca2}
      \end{minipage}}
\end{figure*}

\begin{figure*}[hp]
   	\centering
      \rotatebox{-90}{
      \begin{minipage}{\textheight}
   		\includegraphics[width=\textwidth]{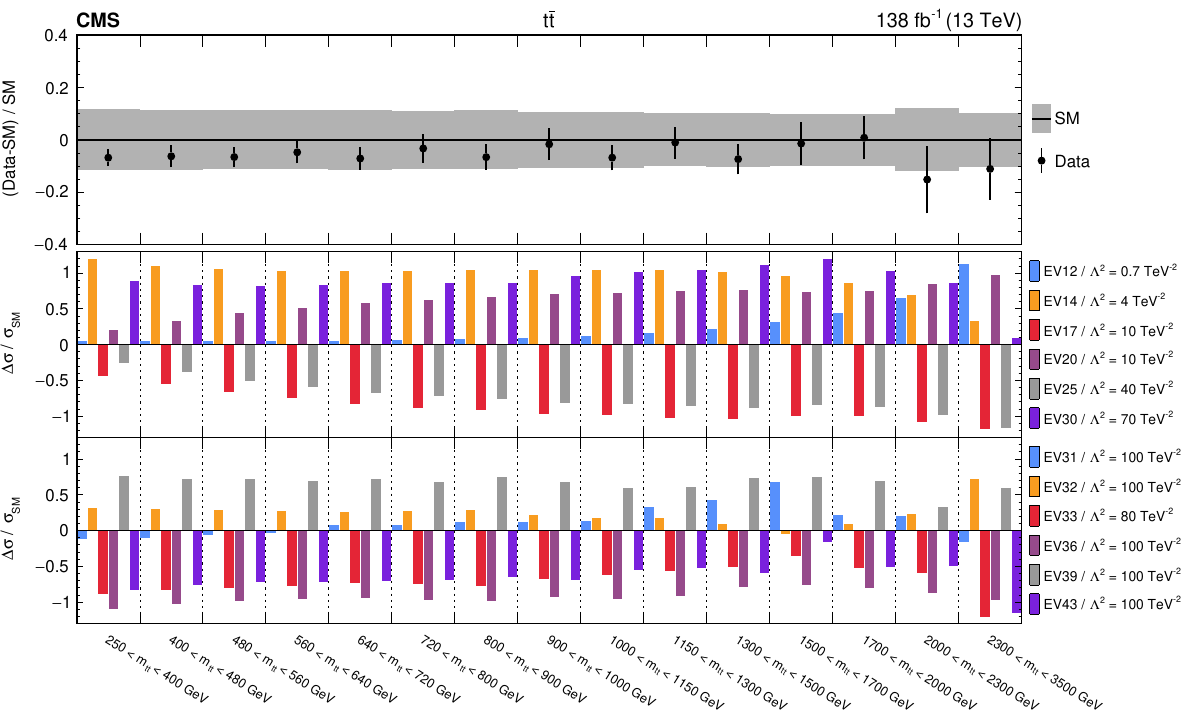}
   		\caption{Relative effect of the linear combinations of WCs that affect the \ttbar differential cross sections. The parameters $\mathrm{EV}_j/\Lambda^2$ are set to different values to ensure the effect of all linear combinations can be visualized on the same $y$ axis scale. The upper panel shows the measured values and their uncertainties relative to the predictions in the SM.}\label{fig:param_effect_pca3}
      \end{minipage}}
\end{figure*}

\begin{figure*}[hp]
   	\centering
      \rotatebox{-90}{
      \begin{minipage}{\textheight}
   		\includegraphics[width=\textwidth]{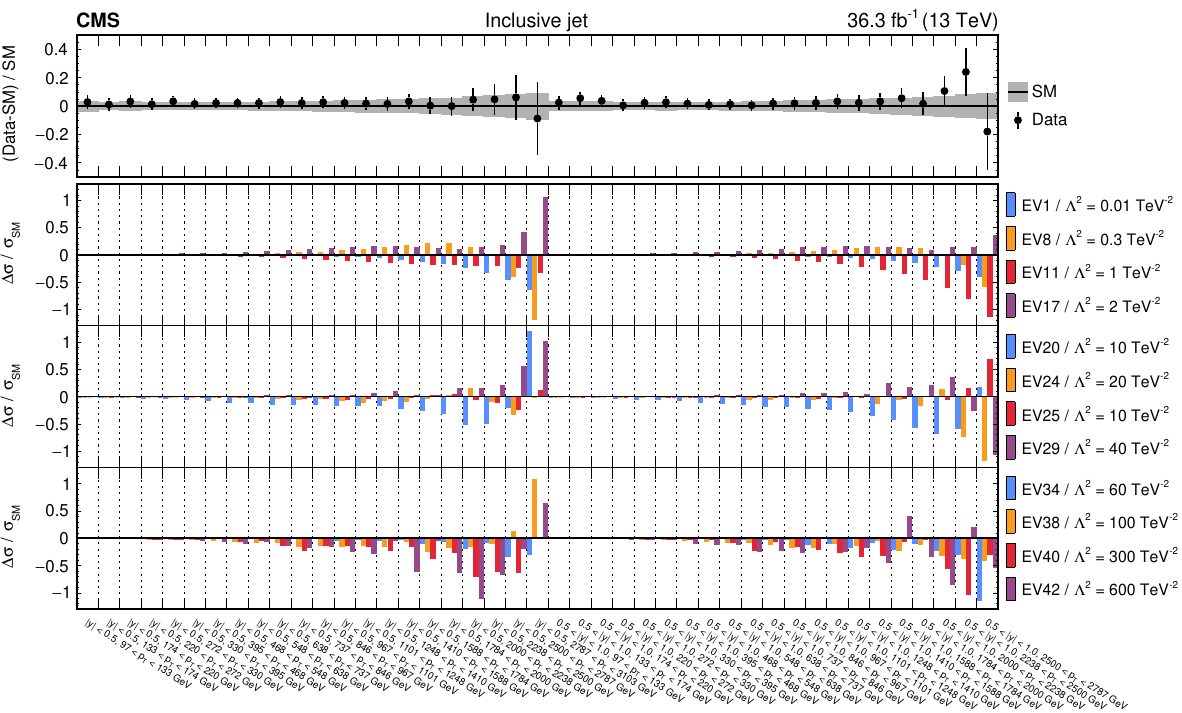}
   		\caption{Relative effect of the linear combinations of WCs that affect the inclusive jet differential cross sections in the rapidity bins $(0, 0.5)$ and $(0.5, 1)$. The parameters $\mathrm{EV}_j/\Lambda^2$ are set to different values to ensure the effect of all linear combinations can be visualized on the same $y$ axis scale. The upper panel shows the measured values and their uncertainties relative to the predictions in the SM.}\label{fig:param_effect_pca4}
      \end{minipage}}
\end{figure*}

\begin{figure*}[hp]
   	\centering
      \rotatebox{-90}{
      \begin{minipage}{\textheight}
   		\includegraphics[width=\textwidth]{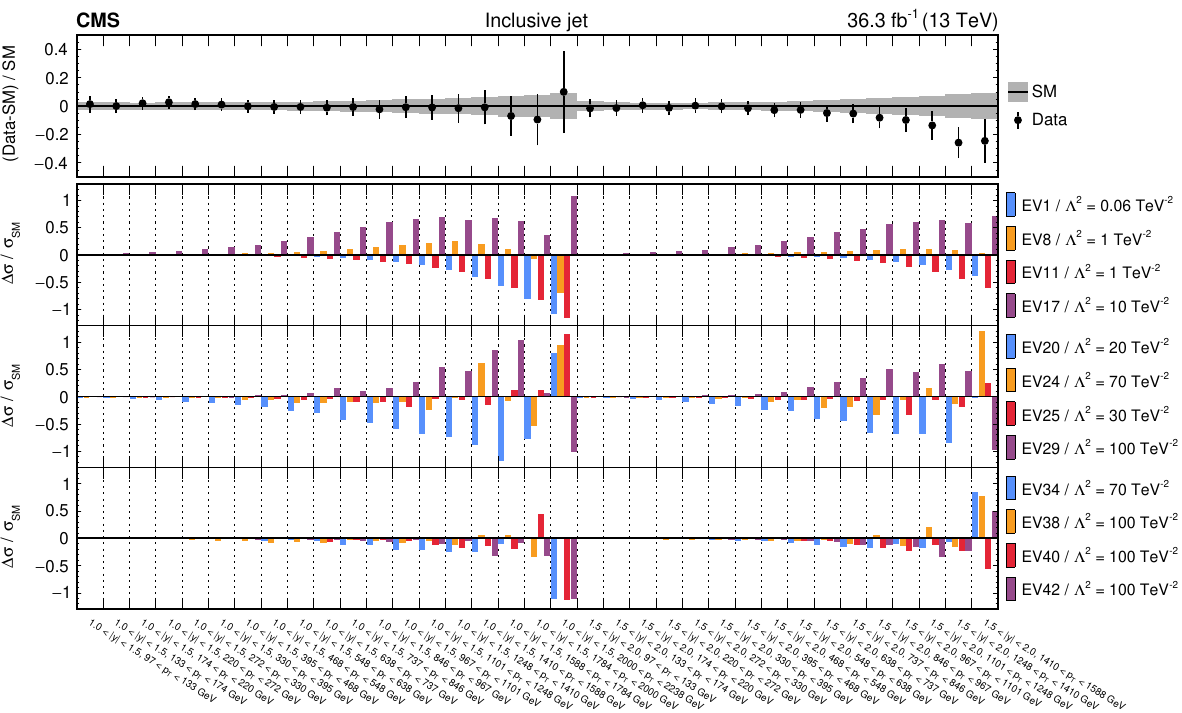}
   		\caption{Relative effect of the linear combinations of WCs that affect the inclusive jet differential cross sections in the rapidity bins $(1, 1.5)$ and $(1.5, 2)$. The parameters $\mathrm{EV}_j/\Lambda^2$ are set to different values to ensure the effect of all linear combinations can be visualized on the same $y$ axis scale. The upper panel shows the measured values and their uncertainties relative to the predictions in the SM.}\label{fig:param_effect_pca5}
      \end{minipage}}
\end{figure*}

\begin{figure*}[hp]
   	\centering
      \rotatebox{-90}{
      \begin{minipage}{\textheight}
   		\includegraphics[width=\textwidth]{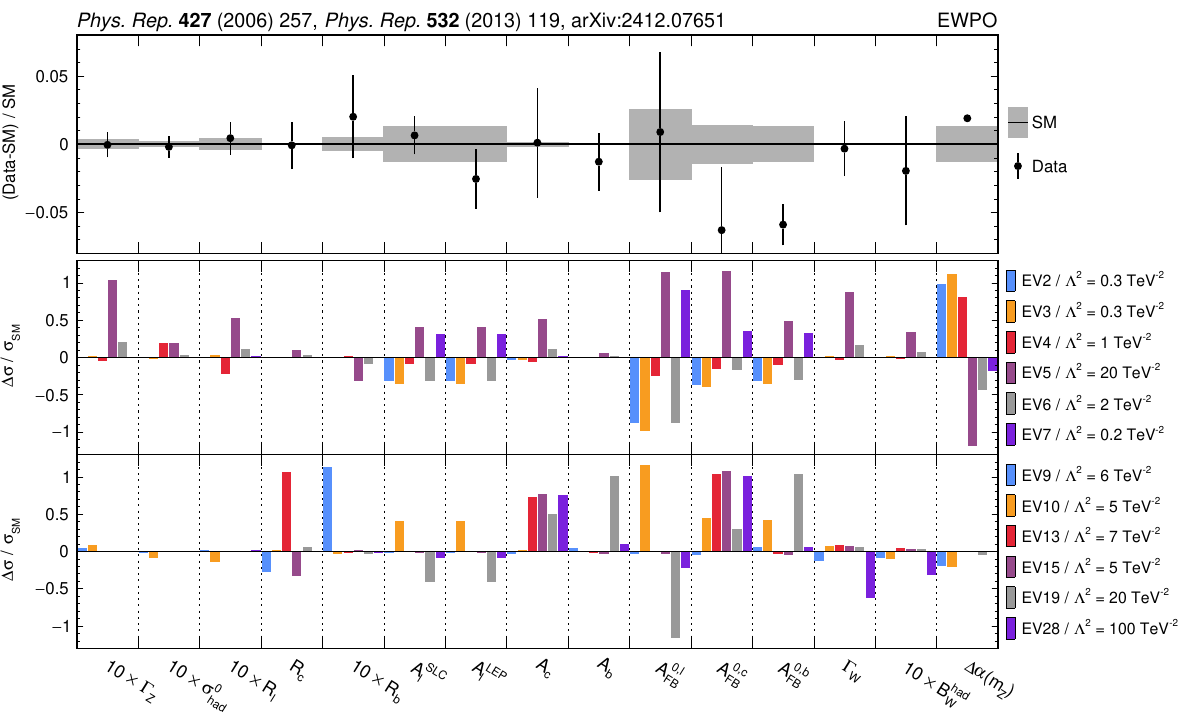}
   		\caption{Relative effect of the linear combinations of WCs that affect the EWPO~\cite{EWPO:ex1, EWPO:ex2, ewpd4lhc}. The parameters $\mathrm{EV}_j/\Lambda^2$ are set to different values to ensure the effect of all linear combinations can be visualized on the same $y$ axis scale. The upper panel shows the measured values and their uncertainties relative to the predictions in the SM.}\label{fig:param_effect_pca6}
      \end{minipage}}
\end{figure*}

The relative effect of the linear combinations of Wilson coefficients on the \HtoGamGam, \WGam, \ZtoNuNu, \WW, \ttbar, and inclusive jet  cross sections and the EWPO is shown in Figs.~\ref{fig:param_effect_pca1}--\ref{fig:param_effect_pca6}. This quantity is computed as the change in the cross section, relative to the SM expectation, for the parameter values indicated in the legend. The upper panels in these figures show the measured values of the cross sections with respect to the SM predictions.

\clearpage
\section{Numerical results}\label{sec:numerical_results}

\begin{table}[!h]
  \topcaption{Expected and observed 95\% \CL limits on linear combinations of WCs from the hybrid fit with the full set of input measurements, in units of $\TeV^{-2}$.}\label{tab:hybrid_plus_rotated}
  \centering
  \specialTableA{
  \begin{tabular}{cccc}
  \multirow{2}{*}{\shortstack{POI}} & \multirow{2}{*}{\shortstack{Expected\\(linear)}} & \multirow{2}{*}{\shortstack{Observed\\(linear)}} & \multirow{2}{*}{\shortstack{Best fit\\(linear)}} \\
    &  &  &  \\
  \hline
  EV1$/\Lambda^2$ & $[-0.002, 0.002]$ & $[-0.003, 0.002]$ & $-0.000$ \\
  EV2$/\Lambda^2$ & $[-0.003, 0.003]$ & $[-0.004, 0.002]$ & $-0.001$ \\
  EV3$/\Lambda^2$ & $[-0.004, 0.004]$ & $[-0.001, 0.007]$ & $+0.003$ \\
  EV4$/\Lambda^2$ & $[-0.006, 0.006]$ & $[-0.007, 0.005]$ & $-0.001$ \\
  EV5$/\Lambda^2$ & $[-0.012, 0.013]$ & $[-0.005, 0.021]$ & $+0.008$ \\
  EV6$/\Lambda^2$ & $[-0.015, 0.014]$ & $[-0.018, 0.012]$ & $-0.003$ \\
  EV7$/\Lambda^2$ & $[-0.017, 0.017]$ & $[-0.027, 0.008]$ & $-0.010$ \\
  EV8$/\Lambda^2$ & $[-0.031, 0.031]$ & $[-0.062, 0.000]$ & $-0.031$ \\
  EV9$/\Lambda^2$ & $[-0.031, 0.031]$ & $[-0.025, 0.038]$ & $+0.006$ \\
  EV10$/\Lambda^2$ & $[-0.063, 0.063]$ & $[-0.072, 0.054]$ & $-0.009$ \\
  EV11$/\Lambda^2$ & $[-0.093, 0.093]$ & $[-0.173, 0.014]$ & $-0.079$ \\
  EV12$/\Lambda^2$ & $[-0.094, 0.094]$ & $[-0.155, 0.032]$ & $-0.061$ \\
  EV13$/\Lambda^2$ & $[-0.19, 0.19]$ & $[-0.21, 0.18]$ & $-0.01$ \\
  EV14$/\Lambda^2$ & $[-0.25, 0.27]$ & $[-0.15, 0.39]$ & $+0.11$ \\
  EV15$/\Lambda^2$ & $[-0.27, 0.27]$ & $[-0.27, 0.28]$ & $+0.01$ \\
  EV16$/\Lambda^2$ & $[-0.27, 0.29]$ & $[-0.33, 0.17]$ & $-0.10$ \\
  EV17$/\Lambda^2$ & $[-0.32, 0.32]$ & $[-0.36, 0.28]$ & $-0.04$ \\
  EV18$/\Lambda^2$ & $[-0.37, 0.34]$ & $[-0.52, 0.19]$ & $-0.14$ \\
  EV19$/\Lambda^2$ & $[-0.44, 0.44]$ & $[-1.07, -0.19]$ & $-0.63$ \\
  EV20$/\Lambda^2$ & $[-0.69, 0.69]$ & $[-1.37, 0.02]$ & $-0.67$ \\
  EV21$/\Lambda^2$ & $[-0.79, 1.03]$ & $[-1.22, 0.76]$ & $-0.35$ \\
  EV22$/\Lambda^2$ & $[-0.81, 0.96]$ & $[-1.62, 0.01]$ & $-0.87$ \\
  EV23$/\Lambda^2$ & $[-0.99, 0.91]$ & $[-0.96, 1.22]$ & $+0.14$ \\
  EV24$/\Lambda^2$ & $[-1.3, 1.3]$ & $[0.2, 2.8]$ & $+1.5$ \\
  EV25$/\Lambda^2$ & $[-1.4, 1.4]$ & $[-1.2, 1.6]$ & $+0.2$ \\
  EV26$/\Lambda^2$ & $[-1.5, 1.8]$ & $[-1.9, 2.0]$ & $-0.1$ \\
  EV27$/\Lambda^2$ & $[-1.9, 1.8]$ & $[-2.6, 1.2]$ & $-0.6$ \\
  EV28$/\Lambda^2$ & $[-2.2, 2.2]$ & $[-2.2, 2.1]$ & $-0.0$ \\
  EV29$/\Lambda^2$ & $[-2.7, 2.7]$ & $[-0.7, 4.8]$ & $+2.1$ \\
  EV30$/\Lambda^2$ & $[-3.4, 3.1]$ & $[-4.6, 2.3]$ & $-1.1$ \\
  EV31$/\Lambda^2$ & $[-3.7, 3.3]$ & $[0.6, 9.2]$ & $+6.0$ \\
  EV32$/\Lambda^2$ & $[-4.4, 3.7]$ & $[-4.7, 3.9]$ & $-0.1$ \\
  EV33$/\Lambda^2$ & $[-4.1, 4.6]$ & $[-3.7, 6.3]$ & $+1.5$ \\
  EV34$/\Lambda^2$ & $[-4.6, 4.6]$ & $[-10.7, -1.6]$ & $-6.1$ \\
  EV35$/\Lambda^2$ & $[-4.6, 5.0]$ & $[-5.8, 2.9]$ & $-1.6$ \\
  EV36$/\Lambda^2$ & $[-5.0, 5.2]$ & $[0.7, 11.6]$ & $+6.1$ \\
  EV37$/\Lambda^2$ & $[-4.9, 5.1]$ & $[-6.7, 3.2]$ & $-1.8$ \\
  EV38$/\Lambda^2$ & $[-5.5, 5.5]$ & $[-7.8, 3.1]$ & $-2.3$ \\
  EV39$/\Lambda^2$ & $[-6.3, 6.2]$ & $[-8.1, 9.7]$ & $-0.4$ \\
  EV40$/\Lambda^2$ & $[-6.7, 6.6]$ & $[-6.2, 7.1]$ & $+0.5$ \\
  EV41$/\Lambda^2$ & $[-7.7, 7.2]$ & $[-6.7, 13.3]$ & $+1.2$ \\
  EV42$/\Lambda^2$ & $[-8.6, 8.6]$ & $[-4.9, 12.2]$ & $+3.7$ \\
  EV43$/\Lambda^2$ & $[-9.6, 9.5]$ & $[-16.6, 5.7]$ & $-6.1$
\end{tabular}}
\end{table}

\begin{table*}[htb]
   \topcaption{Expected and observed individual 95\% \CL limits on WCs from the hybrid fit with the full set of input measurements, in units of $\TeV^{-2}$. This table shows the 32 WCs with the strongest expected constraints, when considering the fit with linear terms only.}\label{tab:hybrid_plus_individual_top}
   \centering
   \cmsTable{
   \begin{tabular}{ccccccc}
      \multirow{2}{*}{\shortstack{POI}} & \multirow{2}{*}{\shortstack{Expected\\(linear)}} & \multirow{2}{*}{\shortstack{Observed\\(linear)}} & \multirow{2}{*}{\shortstack{Best fit\\(linear)}} & \multirow{2}{*}{\shortstack{Expected\\(lin.+quad.)}} & \multirow{2}{*}{\shortstack{Observed\\(lin.+quad.)}} & \multirow{2}{*}{\shortstack{Best fit\\(lin.+quad.)}} \\
      &  &  &  &  &  &  \\
    \hline
  \cjj{3}{1}$/\Lambda^2$ & $[-0.003, 0.003]$ & $[-0.003, 0.002]$ & $-0.000$ & $[-0.003, 0.003]$ & $[-0.003, 0.003]$ & $-0.000$ \\
  \chwb$/\Lambda^2$ & $[-0.004, 0.004]$ & $[-0.004, 0.004]$ & $+0.000$ & $[-0.004, 0.004]$ & $[-0.004, 0.004]$ & $+0.000$ \\
  \chb$/\Lambda^2$ & $[-0.005, 0.004]$ & $[-0.009, 0.001]$ & $-0.004$ & $[-0.004, 0.004]$ & $[-0.008, 0.001]$ & $-0.004$ \\
  \chg$/\Lambda^2$ & $[-0.005, 0.006]$ & $[-0.004, 0.008]$ & $+0.002$ & $[-0.005, 0.006]$ & $[-0.004, 0.008]$ & $+0.002$ \\
  \cjj{1}{1}$/\Lambda^2$ & $[-0.007, 0.007]$ & $[-0.008, 0.005]$ & $-0.002$ & $[-0.005, 0.014]$ & $[-0.007, 0.008]$ & $-0.002$ \\
  \chl{3}$/\Lambda^2$ & $[-0.007, 0.007]$ & $[-0.011, 0.003]$ & $-0.004$ & $[-0.007, 0.007]$ & $[-0.011, 0.003]$ & $-0.004$ \\
  \chl{1}$/\Lambda^2$ & $[-0.008, 0.008]$ & $[-0.010, 0.005]$ & $-0.002$ & $[-0.008, 0.008]$ & $[-0.010, 0.005]$ & $-0.002$ \\
  \cuu{1}$/\Lambda^2$ & $[-0.008, 0.008]$ & $[-0.010, 0.005]$ & $-0.002$ & $[-0.006, 0.012]$ & $[-0.008, 0.007]$ & $-0.002$ \\
  \cjj{1}{8}$/\Lambda^2$ & $[-0.009, 0.009]$ & $[-0.010, 0.008]$ & $-0.001$ & $[-0.008, 0.010]$ & $[-0.010, 0.009]$ & $-0.001$ \\
  \che$/\Lambda^2$ & $[-0.011, 0.011]$ & $[-0.009, 0.012]$ & $+0.002$ & $[-0.011, 0.011]$ & $[-0.009, 0.012]$ & $+0.002$ \\
  \chdd$/\Lambda^2$ & $[-0.012, 0.012]$ & $[-0.020, 0.005]$ & $-0.008$ & $[-0.012, 0.012]$ & $[-0.020, 0.005]$ & $-0.008$ \\
  \cllprime$/\Lambda^2$ & $[-0.013, 0.013]$ & $[-0.008, 0.017]$ & $+0.004$ & $[-0.013, 0.013]$ & $[-0.008, 0.017]$ & $+0.004$ \\
  \chw$/\Lambda^2$ & $[-0.014, 0.013]$ & $[-0.027, 0.003]$ & $-0.011$ & $[-0.014, 0.013]$ & $[-0.025, 0.003]$ & $-0.011$ \\
  \chj{3}$/\Lambda^2$ & $[-0.015, 0.015]$ & $[-0.010, 0.020]$ & $+0.005$ & $[-0.015, 0.015]$ & $[-0.010, 0.020]$ & $+0.005$ \\
  \cuu{8}$/\Lambda^2$ & $[-0.022, 0.022]$ & $[-0.029, 0.016]$ & $-0.007$ & $[-0.019, 0.033]$ & $[-0.024, 0.020]$ & $-0.007$ \\
  \cud{1}$/\Lambda^2$ & $[-0.032, 0.032]$ & $[-0.031, 0.032]$ & $+0.001$ & $[-0.054, 0.022]$ & $[-0.070, 0.021]$ & $-0.050$ \\
  \cju{8}$/\Lambda^2$ & $[-0.036, 0.036]$ & $[-0.054, 0.018]$ & $-0.018$ & $[-0.031, 0.044]$ & $[-0.045, 0.020]$ & $-0.017$ \\
  \chq{3}$/\Lambda^2$ & $[-0.040, 0.040]$ & $[-0.024, 0.056]$ & $+0.016$ & $[-0.040, 0.040]$ & $[-0.024, 0.056]$ & $+0.016$ \\
  \chq{1}$/\Lambda^2$ & $[-0.040, 0.040]$ & $[-0.024, 0.055]$ & $+0.016$ & $[-0.040, 0.040]$ & $[-0.024, 0.055]$ & $+0.016$ \\
  \cbhre$/\Lambda^2$ & $[-0.044, 0.049]$ & $[-0.008, 0.091]$ & $+0.039$ & $[-0.049, 0.045]$ & $[-0.008, 0.078]$ & $+0.036$ \\
  \cud{8}$/\Lambda^2$ & $[-0.049, 0.049]$ & $[-0.052, 0.046]$ & $-0.003$ & $[-0.041, 0.075]$ & $[-0.045, 0.160]$ & $-0.004$ \\
  \cjj{3}{8}$/\Lambda^2$ & $[-0.059, 0.059]$ & $[-0.106, 0.012]$ & $-0.047$ & $[-0.018, 0.025]$ & $[-0.022, 0.025]$ & $-0.012$ \\
  \chu$/\Lambda^2$ & $[-0.078, 0.077]$ & $[-0.051, 0.104]$ & $+0.027$ & $[-0.078, 0.077]$ & $[-0.051, 0.102]$ & $+0.026$ \\
  \ctbre$/\Lambda^2$ & $[-0.087, 0.078]$ & $[-0.161, 0.015]$ & $-0.068$ & $[-0.083, 0.082]$ & $[-0.149, 0.015]$ & $-0.066$ \\
  \cdd{1}$/\Lambda^2$ & $[-0.086, 0.086]$ & $[-0.057, 0.116]$ & $+0.030$ & $[-0.038, 0.056]$ & $[-0.031, 0.065]$ & $+0.040$ \\
  \cjd{8}$/\Lambda^2$ & $[-0.089, 0.089]$ & $[-0.106, 0.073]$ & $-0.017$ & $[-0.066, 0.138]$ & $[-0.080, 0.171]$ & $-0.019$ \\
  \chj{1}$/\Lambda^2$ & $[-0.12, 0.12]$ & $[-0.09, 0.15]$ & $+0.03$ & $[-0.13, 0.12]$ & $[-0.09, 0.14]$ & $+0.03$ \\
  \cg$/\Lambda^2$ & $[-0.14, 0.14]$ & $[-0.14, 0.13]$ & $-0.01$ & $[-0.017, 0.015]$ & $[-0.016, 0.014]$ & $-0.000$ \\
  \ctwre$/\Lambda^2$ & $[-0.16, 0.15]$ & $[-0.30, 0.03]$ & $-0.13$ & $[-0.15, 0.15]$ & $[-0.26, 0.03]$ & $-0.11$ \\
  \ctj{8}$/\Lambda^2$ & $[-0.15, 0.15]$ & $[-0.23, 0.08]$ & $-0.08$ & $[-0.30, 0.12]$ & $[-0.26, 0.07]$ & $-0.07$ \\
  \chd$/\Lambda^2$ & $[-0.15, 0.16]$ & $[-0.23, 0.08]$ & $-0.07$ & $[-0.15, 0.16]$ & $[-0.21, 0.08]$ & $-0.07$ \\
  \cw$/\Lambda^2$ & $[-0.16, 0.15]$ & $[-0.30, 0.01]$ & $-0.14$ & $[-0.061, 0.061]$ & $[-0.066, 0.037]$ & $-0.022$
 \end{tabular}}
 \end{table*}

 \begin{table*}[htb]
   \topcaption{Expected and observed individual 95\% \CL limits on WCs from the hybrid fit with the full set of input measurements, in units of $\TeV^{-2}$. This table shows the 32 WCs with the weakest expected constraints, when considering the fit with linear terms only.}\label{tab:hybrid_plus_individual_bottom}
   \centering
   \cmsTable{
   \begin{tabular}{ccccccc}
      \multirow{2}{*}{\shortstack{POI}} & \multirow{2}{*}{\shortstack{Expected\\(linear)}} & \multirow{2}{*}{\shortstack{Observed\\(linear)}} & \multirow{2}{*}{\shortstack{Best fit\\(linear)}} & \multirow{2}{*}{\shortstack{Expected\\(lin.+quad.)}} & \multirow{2}{*}{\shortstack{Observed\\(lin.+quad.)}} & \multirow{2}{*}{\shortstack{Best fit\\(lin.+quad.)}} \\
      &  &  &  &  &  &  \\
      \hline
  \ctgre$/\Lambda^2$ & $[-0.17, 0.16]$ & $[-0.27, 0.07]$ & $-0.09$ & $[-0.15, 0.17]$ & $[-0.19, 0.07]$ & $-0.07$ \\
  \cqj{1}{8}$/\Lambda^2$ & $[-0.20, 0.20]$ & $[-0.31, 0.08]$ & $-0.12$ & $[-0.35, 0.14]$ & $[-0.29, 0.08]$ & $-0.09$ \\
  \cdd{8}$/\Lambda^2$ & $[-0.20, 0.20]$ & $[-0.14, 0.26]$ & $+0.06$ & $[-0.10, 0.16]$ & $[-0.08, 0.18]$ & $+0.12$ \\
  \chbq$/\Lambda^2$ & $[-0.21, 0.21]$ & $[-0.42, -0.01]$ & $-0.22$ & $[-0.20, 0.22]$ & $[-0.39, -0.01]$ & $-0.21$ \\
  \ctu{8}$/\Lambda^2$ & $[-0.27, 0.27]$ & $[-0.39, 0.15]$ & $-0.12$ & $[-0.40, 0.18]$ & $[-0.34, 0.11]$ & $-0.09$ \\
  \cqu{8}$/\Lambda^2$ & $[-0.30, 0.30]$ & $[-0.44, 0.15]$ & $-0.14$ & $[-0.42, 0.19]$ & $[-0.34, 0.12]$ & $-0.09$ \\
  \cqj{3}{1}$/\Lambda^2$ & $[-0.35, 0.35]$ & $[-0.54, 0.15]$ & $-0.20$ & $[-0.102, 0.084]$ & $[-0.062, 0.049]$ & $-0.006$ \\
  \ctd{8}$/\Lambda^2$ & $[-0.47, 0.47]$ & $[-0.67, 0.28]$ & $-0.19$ & $[-0.51, 0.26]$ & $[-0.41, 0.18]$ & $-0.09$ \\
  \cqd{8}$/\Lambda^2$ & $[-0.52, 0.52]$ & $[-0.76, 0.28]$ & $-0.24$ & $[-0.53, 0.28]$ & $[-0.42, 0.19]$ & $-0.10$ \\
  \ctj{1}$/\Lambda^2$ & $[-0.65, 0.65]$ & $[-0.99, 0.32]$ & $-0.34$ & $[-0.103, 0.090]$ & $[-0.078, 0.066]$ & $-0.006$ \\
  \cqj{3}{8}$/\Lambda^2$ & $[-0.76, 0.75]$ & $[-1.11, 0.41]$ & $-0.35$ & $[-0.24, 0.20]$ & $[-0.14, 0.12]$ & $-0.01$ \\
  \clj{3}$/\Lambda^2$ & $[-0.81, 0.97]$ & $[-1.54, -0.04]$ & $-0.88$ & $[-0.32, 0.26]$ & $[-0.26, 0.16]$ & $-0.07$ \\
  \cqu{1}$/\Lambda^2$ & $[-1.2, 1.2]$ & $[-1.9, 0.6]$ & $-0.6$ & $[-0.14, 0.13]$ & $[-0.10, 0.09]$ & $-0.01$ \\
  \chbox$/\Lambda^2$ & $[-1.2, 1.4]$ & $[-0.1, 2.8]$ & $+1.3$ & $[-1.3, 1.3]$ & $[-0.1, 2.6]$ & $+1.2$ \\
  \ctd{1}$/\Lambda^2$ & $[-1.5, 1.5]$ & $[-0.9, 2.1]$ & $+0.6$ & $[-0.15, 0.17]$ & $[-0.11, 0.13]$ & $+0.01$ \\
  \cjd{1}$/\Lambda^2$ & $[-1.7, 1.7]$ & $[-1.8, 1.7]$ & $-0.1$ & $[-0.060, 0.059]$ & $[-0.069, 0.067]$ & $-0.036$ \\
  \cju{1}$/\Lambda^2$ & $[-1.8, 1.8]$ & $[-2.9, 0.7]$ & $-1.1$ & $[-0.040, 0.040]$ & $[-0.048, 0.047]$ & $-0.026$ \\
  \cthre$/\Lambda^2$ & $[-2.2, 1.9]$ & $[-3.8, 0.6]$ & $-1.5$ & $[-2.1, 2.0]$ & $[-3.9, 0.5]$ & $-1.6$ \\
  \ctu{1}$/\Lambda^2$ & $[-2.1, 2.1]$ & $[-3.0, 1.1]$ & $-1.0$ & $[-0.12, 0.11]$ & $[-0.091, 0.086]$ & $-0.003$ \\
  \cqj{1}{1}$/\Lambda^2$ & $[-3.9, 3.0]$ & $[-4.0, 3.9]$ & $+0.5$ & $[-0.10, 0.10]$ & $[-0.075, 0.075]$ & $+0.000$ \\
  \cht$/\Lambda^2$ & $[-3.1, 4.0]$ & $[-3.8, 3.6]$ & $-0.6$ & $[-3.8, 3.4]$ & $[-4.9, 3.1]$ & $-0.6$ \\
  \cqd{1}$/\Lambda^2$ & $[-4.0, 4.0]$ & $[-6.1, 1.9]$ & $-2.1$ & $[-0.18, 0.17]$ & $[-0.14, 0.13]$ & $-0.00$ \\
  \ctt$/\Lambda^2$ & $[-5.4, 3.9]$ & $[-7.8, 2.7]$ & $-1.6$ & $[-1.0, 1.1]$ & $[-1.3, 1.4]$ & $-0.7$ \\
  \clj{1}$/\Lambda^2$ & $[-4.8, 4.6]$ & $[-5.3, 4.7]$ & $-0.1$ & $[-0.54, 0.56]$ & $[-0.36, 0.36]$ & $-0.00$ \\
  \cql{3}$/\Lambda^2$ & $[-5.6, 6.7]$ & $[-6.7, 5.7]$ & $-1.0$ & $[-2.1, 1.7]$ & $[-2.0, 1.6]$ & $-0.5$ \\
  \cql{1}$/\Lambda^2$ & $[-7.2, 5.9]$ & $[-6.4, 7.1]$ & $+0.9$ & $[-1.9, 2.3]$ & $[-1.8, 2.3]$ & $+0.5$ \\
  \clu$/\Lambda^2$ & $[-7.0, 6.2]$ & $[0.8, 12.2]$ & $+6.8$ & $[-0.60, 0.65]$ & $[-0.37, 0.44]$ & $+0.03$ \\
  \cqt{1}$/\Lambda^2$ & $[-6.6, 9.1]$ & $[-4.5, 13.4]$ & $+2.8$ & $[-1.9, 1.7]$ & $[-2.4, 2.2]$ & $+1.1$ \\
  \cte$/\Lambda^2$ & $[-9.5, 7.8]$ & $[-7.9, 9.6]$ & $+1.6$ & $[-2.0, 2.4]$ & $[-1.9, 2.4]$ & $+0.6$ \\
  \cqq{1}$/\Lambda^2$ & $[-18.2, 13.1]$ & $[-27.9, 8.4]$ & $-6.5$ & $[-3.9, 4.6]$ & $[-5.2, 5.8]$ & $-2.7$ \\
  \cqt{8}$/\Lambda^2$ & $[-19.1, 13.8]$ & $[-29.3, 8.4]$ & $-7.1$ & $[-3.4, 4.0]$ & $[-4.4, 4.9]$ & $-2.2$ \\
  \ctl$/\Lambda^2$ & $[-20.0, 16.6]$ & $[-19.1, 18.5]$ & $+1.3$ & $[-2.1, 2.2]$ & $[-2.0, 2.2]$ & $+0.1$
 \end{tabular}}
 \end{table*}

\cleardoublepage \section{The CMS Collaboration \label{app:collab}}\begin{sloppypar}\hyphenpenalty=5000\widowpenalty=500\clubpenalty=5000
\cmsinstitute{Yerevan Physics Institute, Yerevan, Armenia}
{\tolerance=6000
V.~Chekhovsky, A.~Hayrapetyan, V.~Makarenko\cmsorcid{0000-0002-8406-8605}, A.~Tumasyan\cmsAuthorMark{1}\cmsorcid{0009-0000-0684-6742}
\par}
\cmsinstitute{Institut f\"{u}r Hochenergiephysik, Vienna, Austria}
{\tolerance=6000
W.~Adam\cmsorcid{0000-0001-9099-4341}, J.W.~Andrejkovic, L.~Benato\cmsorcid{0000-0001-5135-7489}, T.~Bergauer\cmsorcid{0000-0002-5786-0293}, K.~Damanakis\cmsorcid{0000-0001-5389-2872}, M.~Dragicevic\cmsorcid{0000-0003-1967-6783}, C.~Giordano, P.S.~Hussain\cmsorcid{0000-0002-4825-5278}, M.~Jeitler\cmsAuthorMark{2}\cmsorcid{0000-0002-5141-9560}, N.~Krammer\cmsorcid{0000-0002-0548-0985}, A.~Li\cmsorcid{0000-0002-4547-116X}, D.~Liko\cmsorcid{0000-0002-3380-473X}, I.~Mikulec\cmsorcid{0000-0003-0385-2746}, J.~Schieck\cmsAuthorMark{2}\cmsorcid{0000-0002-1058-8093}, R.~Sch\"{o}fbeck\cmsAuthorMark{2}\cmsorcid{0000-0002-2332-8784}, D.~Schwarz\cmsorcid{0000-0002-3821-7331}, M.~Sonawane\cmsorcid{0000-0003-0510-7010}, W.~Waltenberger\cmsorcid{0000-0002-6215-7228}, C.-E.~Wulz\cmsAuthorMark{2}\cmsorcid{0000-0001-9226-5812}
% ...
\par}
\vskip\cmsinstskip
\dag:~Deceased\\
$^{1}$Also at Yerevan State University, Yerevan, Armenia\\
$^{2}$Also at TU Wien, Vienna, Austria\\
% ...
\end{sloppypar}
\end{document}